\documentclass[aps,
               prd,
               preprint,
               longbibliography,
               ]{revtex4-1}

\makeatletter
\DeclareRobustCommand*{\bfseries}{%
  \not@math@alphabet\bfseries\mathbf
  \fontseries\bfdefault\selectfont
  \boldmath
}
\makeatother

\usepackage{tikz}
\usepackage[amssymb,textstyle]{SIunits}
\usepackage{amsmath,amssymb}
\usepackage[update,prepend]{epstopdf}
\usepackage{relsize}
\usepackage{microtype}
\usepackage{setspace}
\usepackage{atlasphysics} 

\usepackage{bm}
\usepackage{booktabs}
\usepackage{xspace}
\usepackage{multirow}
\usepackage{bigstrut}
\usepackage{subfigure}
\usepackage{hyperref}
\usepackage{placeins}

\makeatletter
\def\@bibitemShut{}
\makeatother

\graphicspath{{./}}

\newcommand{\Pythia}{PYTHIA\xspace}
\newcommand{\Phojet}{PHOJET\xspace}
\newcommand{\Herwig}{HERWIG\xspace}
\newcommand{\Jimmy}{JIMMY\xspace}
\newcommand{\HerwigJimmy}{HERWIG+JIMMY\xspace}

\newcommand{\effvtx}{\ensuremath{\epsilon_\text{vtx}}\xspace}
\newcommand{\efftrig}{\ensuremath{\epsilon_\text{trig}}\xspace}
\newcommand{\efftrk}{\ensuremath{\epsilon_\text{trk}}\xspace}
\newcommand{\effldtrk}{\ensuremath{\epsilon_\text{ld\,trk}}\xspace}
\newcommand{\effbin}{\ensuremath{\epsilon_\text{bin}}\xspace}
\newcommand{\Nselbs}{\ensuremath{N_\text{sel}^\text{BS}}\xspace}
\newcommand{\ptlead}{\ensuremath{\pt^\text{lead}}\xspace}
\newcommand{\ptmean}{\ensuremath{\langle\pt\rangle}\xspace}
\newcommand{\ptsum}{\ensuremath{\sum\mspace{-0.8mu}\pt}\xspace}
\newcommand{\etamod}{\ensuremath{|\eta\mspace{0.2mu}|}\xspace}

\newcommand{\Nchg}{\ensuremath{N_\text{ch}}\xspace}
\newcommand{\dNchgdetadphi}{\ensuremath {\langle \mathrm{d}^2N_\text{ch}/\mathrm{d}\eta\,\mathrm{d}\phi\rangle} \xspace}
\newcommand{\dpTsumdetadphi}{\ensuremath{\langle \mathrm{d}^2\!\sum\!\pt/\mathrm{d}\eta\,\mathrm{d}\phi \rangle}\xspace}

\newcommand{\FigureRef}[2][]{Figure#1~\ref{#2}\xspace}
\newcommand{\FigRef}[2][]{Fig#1.~\ref{#2}\xspace}
\newcommand{\SecRef}[2][]{Section#1~\ref{#2}\xspace}
\newcommand{\TabRef}[2][]{Table#1~\ref{#2}\xspace}

\begin{document}


\title{Measurement of underlying event characteristics using\\ charged particles
in $pp$ collisions at $\sqrt{s} = \unit{900}{\GeV}$ and \unit{7}{\TeV}\\ with the ATLAS detector}
\date{\today}
\author{G. Aad \textit{et al.}}\thanks{Full author list given at the end of the article in Appendix~\ref{app:ATLASColl}.}
\collaboration{The ATLAS Collaboration}

\begin{abstract}%
  Measurements of charged particle distributions, sensitive to the underlying event, have been performed
  with the ATLAS detector at the LHC. The measurements are based on data collected using a
  minimum-bias trigger to select proton--proton collisions at center-of-mass energies of
  \unit{900}{\GeV} and \unit{7}{\TeV}.
  The ``underlying event'' is defined as those aspects of a hadronic interaction
  attributed not to the hard scattering process, but rather to the accompanying
  interactions of the rest of the proton. Three regions are defined
  in azimuthal angle with respect to
  the highest transverse momentum charged particle in the event, such that the region
  transverse to the dominant momentum-flow is most sensitive to the underlying
  event. In each of these regions, distributions of the charged particle multiplicity,
  transverse momentum density, and average \pt are measured.
  The data show generally higher underlying event activity than that predicted by
  Monte Carlo models tuned to pre-LHC data.%
\end{abstract}%

\pacs{12.38.-t,13.75.-n}
\maketitle


\section{Introduction}

To perform precise Standard Model measurements or search for new physics
phenomena at hadron colliders, it is essential to have a good understanding not
only of the short-distance ``hard'' scattering process, but also of the
accompanying interactions of the rest of the proton -- collectively termed the
``underlying event'' (UE). It is impossible to uniquely separate the UE from the
hard scattering process on an event-by-event basis.  However, observables can be
measured which are sensitive to its properties.


The UE may involve contributions from both hard and soft physics, where ``soft''
refers to interactions with low \pT transfer between the scattering particles.
Soft interactions cannot reliably be calculated with perturbative QCD methods, and
are generally described in the context of different phenomenological models, usually implemented
in Monte Carlo (MC) event generators. These models contain many parameters whose
values are not \textit{a priori} known. Therefore, to obtain insight into the nature of
soft QCD processes and to optimize the description of UE contributions for
studies of hard-process physics such as hadronic jet observables, the model parameters
must be fitted to experimental data.

Measurements of primary charged particle multiplicities have been performed in
``minimum bias'' (MB) events at the LHC~\cite{Aamodt:2010pp,Aamodt:2010ft,Khachatryan:2010xs,
  Collaboration:2010rd, Collaboration:2010ir}. Such inclusive studies
provide important constraints on soft hadron-interaction models.
However, observables constructed for the study of the UE measure
the structure of hadronic events in a different way, focusing on the correlation
of soft-process features to one another and to those of the hardest
processes in the event.
UE observables have been measured in $p\bar{p}$ collisions in dijet and Drell-Yan
events at CDF in Run~I~\cite{PhysRevD.70.072002} and
Run~II~\cite{PhysRevD.82.034001} at center-of-mass energies of
$\sqrt{s}=\unit{1.8}{\TeV}$ and \unit{1.96}{\TeV} respectively, and in $pp$ collisions at
$\sqrt{s}=\unit{900}{\GeV}$ in a detector-specific study by CMS~\cite{Khachatryan:2010pv}.

This paper reports the measurement of UE observables, performed with the ATLAS
detector~\cite{:2008zzm} at the LHC using proton--proton collisions at
center-of-mass energies of \unit{900}{\GeV} and \unit{7}{\TeV}. The UE
observables are constructed from primary charged particles in the pseudorapidity
range $\etamod < 2.5$, whose transverse momentum component\footnote{The ATLAS
  reference system is a Cartesian right-handed coordinate system, with the
  nominal collision point at the origin. The anti-clockwise beam direction
  defines the positive $z$-axis, while the positive $x$-axis is defined as
  pointing from the collision point to the center of the LHC ring and the
  positive $y$-axis points upwards. The azimuthal angle $\phi$ is measured
  around the beam axis, and the polar angle $\theta$ is the angle measured with
  respect to~the $z$-axis. The pseudorapidity is given by $\eta = -\ln
  \tan\mspace{-0.1mu}( \theta/2 )$. Transverse momentum is defined relative to the beam axis.} is separately required to be $\pt >
\unit{100}{\MeV}$ or $\pt > \unit{500}{\MeV}$. Primary charged particles are
defined as those with a mean proper lifetime $\tau \gtrsim 0.3 \times
\unit{\ensuremath{10^{-10}}}{\second}$, either directly produced in $pp$ interactions or in
the decay of particles with a shorter lifetime.
%
At the detector level, charged particles are observed as tracks in the inner tracking system.
The direction of the track with the largest \pt in the event -- referred to as
the ``leading'' track -- is used to define regions of the $\eta$--$\phi$ plane
which have different sensitivities to the UE.
The axis given by the leading track is well-defined for all events, and is highly
correlated with the axis of the hard scattering in high-\pt events.
A single track is used as opposed to a jet or the decay products of a massive gauge
boson, as it allows significant results to be derived with limited luminosity
and avoids the systematic measurement complexities of alignment with more
complex objects.

\begin{figure}[tbp]
  \begin{center}
    \includegraphics{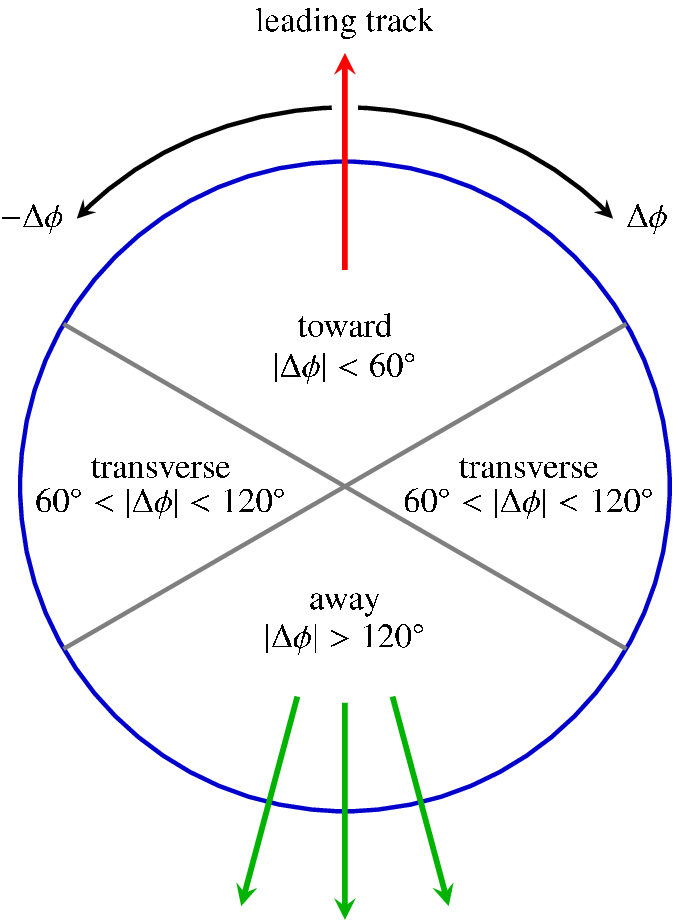}
    \caption{Definition of regions in the azimuthal angle with respect to the leading track.}
    \label{fig:ueregions}
  \end{center}
\end{figure}

As illustrated in \FigRef{fig:ueregions}, the azimuthal angular difference between charged tracks and the leading track,
$|\Delta\phi|=|\phi-\phi_\text{leading~track}|$, is used to define the following three azimuthal regions~\cite{PhysRevD.70.072002}:
\begin{itemize}
\item $|\Delta\phi| < 60^{\circ}$, the ``toward region'';
\item $60^{\circ} < |\Delta\phi| < 120^{\circ}$, the ``transverse region''; and
\item $|\Delta\phi| > 120^{\circ}$, the ``away region''.
\end{itemize}

\noindent
The transverse regions are most sensitive to the underlying event, since they
are generally perpendicular to the axis of hardest scattering and hence have the lowest
level of activity from this source. However, the hard scatter can of course also emit particles
perpendicular to the event axis: the regional division is not, and cannot be, an
exact filter.  The observables examined in this analysis are described in
\TabRef{tab:obs}. The detector level corresponds to the tracks passing the
selection criteria, and the particle level corresponds to true charged particles
in the event. The particle level can be compared directly with the QCD Monte
Carlo models at the generator level.

\begin{table}[p]

   \caption{Definition of the measured observables at particle and detector level. The particles and tracks are required to have $\pt> \unit{0.1}{\GeV}$ or $\unit{0.5}{\GeV}$ and $\etamod < 2.5$. Tracks are selected if they pass the criteria described in \SecRef{sec:selection}. The mean charged particle momentum $\langle \pt \rangle$ is constructed on an event-by-event basis and then averaged over the events.}

  \begin{center}
    \singlespacing
    \begin{tabular}{lll}
    \toprule
     \toprule
      Observable & Particle level & Detector level \\
     \midrule

     \ptlead & \multirow{3}{50mm}{Transverse momentum of the stable charged particle with maximum \pT in the event} & \multirow{3}{50mm}{Transverse momentum of the selected track with maximum \pT in the event}  \\
     &&\\
     &&\\
     &&\\
     $\etamod^\text{lead}$ & \multirow{2}{50mm}{$\etamod$ of the maximum \pt stable charged particle in the event} & \multirow{2}{50mm}{$\etamod$ of the maximum \pt selected track in the event}  \\
     &&\\
     &&\\
     \dNchgdetadphi & \multirow{2}{50mm}{Mean number of stable charged particles per unit $\eta$--$\phi$}  & \multirow{2}{50mm}{Mean number of selected tracks per unit $\eta$--$\phi$} \\
     &&\\
     &&\\
     \dpTsumdetadphi & \multirow{2}{50mm}{Mean scalar \pt sum of stable charged particles per unit $\eta$--$\phi$} & \multirow{2}{50mm}{Mean scalar \pt sum of selected tracks per unit $\eta$--$\phi$} \\
     &&\\
     &&\\
     \multirow{2}{50mm}{Standard deviation of $\mathrm{d}^2N_\text{ch}/\mathrm{d}\eta\,\mathrm{d}\phi$} & \multirow{3}{50mm}{Standard deviation of number of stable charged particles per unit $\eta$--$\phi$}  & \multirow{2}{50mm}{Standard deviation of number of selected tracks per unit $\eta$--$\phi$}  \\
     &&\\
     &&\\
     &&\\
     \multirow{2}{50mm}{Standard deviation of $\mathrm{d}^2\!\sum\!\pt/\mathrm{d}\eta\,\mathrm{d}\phi$} & \multirow{3}{50mm}{Standard deviation of scalar \pt sum of stable charged particles per unit $\eta$--$\phi$} & \multirow{3}{50mm}{Standard deviation of scalar \pt sum of selected tracks per unit $\eta$--$\phi$} \\
     &&\\
     &&\\
     &&\\
     $\langle \pt \rangle$ & \multirow{3}{50mm}{Average \pt of stable charged particles (at least 1 charged particle is required)} & \multirow{2}{50mm}{Average \pt of selected tracks (at least 1 selected track is required)}  \\
     &&\\
     &&\\
     &&\\
     \multirow{2}{50mm}{Angular distribution of number~density} & \multirow{3}{50mm}{Number density of stable charged particles in intervals of $\Delta|\phi|$, measured relative to the leading charged particle} & \multirow{3}{50mm}{Number density of tracks in intervals of $\Delta|\phi|$, measured relative to the leading track}  \\
     &&\\
     &&\\
     &&\\
     &&\\
     \multirow{2}{50mm}{Angular distribution of \pt~density} & \multirow{3}{50mm}{\pt density of stable charged particles in the intervals of $\Delta|\phi|$, measured relative to the leading charged particle} & \multirow{3}{50mm}{\pt density of tracks in the intervals of $\Delta|\phi|$, measured relative to the leading track}  \\
     &&\\
     &&\\
     &&\\
    \bottomrule
    \bottomrule
    \end{tabular}
    \label{tab:obs}
  \end{center}

\end{table}

This paper is organized as follows: The ATLAS detector is described in \SecRef{sec:atlasdetector}.
In \SecRef{sec:mcmodels}, the QCD MC models used in this analysis are discussed.
\SecRef[s]{sec:selection}--\ref{sec:systematics} respectively describe the event selection, background contributions,
correction of the data back to particle level, and estimation of the systematic uncertainties.
The results are discussed in \SecRef{sec:results} and finally the conclusions are presented in \SecRef{sec:conclusions}.

\section{The ATLAS detector}
\label{sec:atlasdetector}

The ATLAS detector~\cite{:2008zzm} covers almost the whole solid angle around the collision point with layers of tracking detectors, calorimeters and muon chambers. It has been designed to study a wide range of physics topics at LHC energies.  For the measurements presented in this
paper, the trigger system and the tracking devices were of particular importance.

The ATLAS inner detector has full coverage in $\phi$ and covers the pseudorapidity range $\etamod~ < 2.5$.
It consists of a silicon pixel detector (pixel), a silicon strip detector namely the semiconductor tracker (SCT) and a straw-tube transition radiation tracker
(TRT). These detectors cover a 
radial distance from the interaction point of
50.5--\unit{150}{\mm}, 299--\unit{560}{\mm} and 563--\unit{1066}{\mm},
respectively, and are immersed in a 2~Tesla
axial magnetic field. The inner detector barrel (end-cap) parts consist of
3 ($2 \times 3$) pixel layers, 4 ($2 \times 9$) layers of double-sided silicon strip modules, and 73 ($2 \times 160$) layers of
TRT straw-tubes. These detectors have position resolutions of typically 10, 17 and
\unit{130}{\micro\metre} for the $r$--$\phi$ coordinate and (for the pixel and SCT) 115 and \unit{580}{\micro\metre} for the $r$--$z$ coordinate.
A track traversing the barrel would typically have 11 silicon hits (3 pixel clusters, and 8 strip clusters), and more than 30 straw-tube hits.

The ATLAS detector has a three-level trigger system:~level~1~(L1), level~2 (L2) and the event~filter~(EF).
For this measurement, the trigger relies on the beam pickup timing devices (BPTX) and
the minimum bias trigger scintillators (MBTS).
The BPTX are composed of electrostatic beam pick-ups attached to the beam pipe at a distance $z = \pm \unit{175}{\metre}$ from the center of the ATLAS detector.
The MBTS are mounted at each end of the detector in front of the liquid-argon endcap-calorimeter cryostats at $z = \pm \unit{3.56}{\metre}$ and are segmented into eight sectors in azimuth and two rings in pseudorapidity ($2.09 < \etamod < 2.82$ and $2.82 < \etamod < 3.84$).
Data were taken for this analysis using the single-arm MBTS trigger, formed from BPTX  and MBTS trigger signals.
The MBTS trigger was configured to require one hit above threshold from either side of the detector.  The MBTS trigger efficiency was studied with a separate pre-scaled L1 BPTX trigger, filtered to obtain inelastic interactions by inner detector requirements at L2 and EF.


\section{QCD Monte Carlo models}
\label{sec:mcmodels}

In scattering processes modeled by lowest-order perturbative QCD
two-to-two parton scatters, at sufficiently low \pt the partonic jet cross-section
exceeds that of the total hadronic cross-section.
This problem is resolved by allowing the possibility of multiple parton interactions (MPI) in
a given hadron-hadron interaction. In this picture, the ratio of the partonic jet cross-section to the
total cross-section is interpreted as the mean number of parton
interactions in such events. This idea is implemented in several Monte
Carlo event generators, and is usually complemented by phenomenological models
which continue to be developed. These include (non-exhaustively) further low \pt screening of the
partonic differential cross-section, use of phenomenological transverse hadronic-matter
distributions, reconfiguration of color string or cluster topologies
, saturation of parton densities at low-$x$, and connection to elastic scattering and cut-pomeron
models via the optical theorem.  Such models
typically contain several parameters, which may be tuned to data at different center-of-mass energies
and in various hadronic processes. MC tuning has been actively pursued in recent years,
and standard tunes are being iterated in response to early LHC data,
including those presented in ref.~\cite{Collaboration:2010ir}.

Samples of 10--20~million MC events were produced for single-diffractive,
double-diffractive and non-diffractive processes using the \Pythia~6.4.21
generator~\cite{Sjostrand:2006za} for collision energies of \unit{900}{\GeV} and \unit{7}{\TeV}.
The MC09~\cite{:atlasmc09} set of Tevatron-optimized parameters was used:
this employs the MRST LO*~\cite{Sherstnev:2007nd} parton density
functions (PDFs)~\footnote{The gluon density distribution is enhanced at low-$x$ in the modified LO* PDF with respect to the LO CTEQ5L/6L or MSTW2008LO PDFs.} and
the \Pythia \pt-ordered parton shower, and was tuned to describe underlying event
and minimum bias data at \unit{630}{\GeV}
and \unit{1.8}{\TeV}~\cite{Abe:1988yu} at CDF in $p\bar{p}$ collisions.
ATLAS~MC09 is the reference \Pythia tune throughout this paper, and
samples generated with this tune were used to calculate detector acceptances
and efficiencies to correct the data for detector effects. All events were processed through
the ATLAS detector simulation framework~\cite{2010arXiv1005.4568T}, which is
based on Geant4~\cite{Agostinelli:2002hh}. They were then reconstructed and
analyzed identically to the data. Particular attention was
devoted to the description in the simulation of the size and position of the
collision beam-spot and of the detailed detector conditions during the
data-taking runs.

For the purpose of comparing the present measurement to different
phenomenological models, several additional MC
samples were generated. For \Pythia, these were the
Perugia0~\cite{Skands:2009zm} tune, in which the soft-QCD part of the event is
tuned using only minimum bias data from the Tevatron and S$p\bar{p}$S colliders,
and the DW~\cite{CDFtuneA} \Pythia tune, which uses a
virtuality-ordered parton shower and an eikonal multiple scattering model including
impact-parameter correlations. This tune was constructed to describe CDF Run II
underlying event, dijet and Drell-Yan data. \Phojet ~\cite{phojet} and \Herwig
~\cite{Corcella:2002jc} were used as alternative models.  \Phojet describes
low-\pt physics using the two component Dual~Parton~Model~\cite{DPM1, DPM2},
which includes soft hadronic processes described by pomeron exchange and
semi-hard processes described by perturbative parton scattering; it relies on
\Pythia for the fragmentation of partons. The \Phojet versions used for this
study were shown to agree with previous
measurements~\cite{Abe:1988yu,Aaltonen:2009ne,Albajar:1989an,Abe:1989td}.
The \Phojet samples were also passed through full detector simulation for
systematic studies of acceptance and smearing corrections (unfolding).
\Herwig uses angular-ordered parton showers and a cluster hadronization model.
The UE is simulated using the \Jimmy package~\cite{Butterworth:1996zw} which, like
\Pythia, implements an eikonal multiple scattering model including impact-parameter
correlations. It does not contain any model of soft scatters. \HerwigJimmy was run with
the ATLAS~MC09 parameters~\cite{:atlasmc09}: these set a minimum partonic interaction \pt of
 \unit{3.0}{\GeV} at \unit{900}{\GeV} and \unit{5.2}{\GeV} at \unit{7}{\TeV}, and hence
agreement with data is not expected when the maximum track \pt is below this cut-off scale.

For \Pythia and \Phojet, non-diffractive, single-diffractive and
double-diffractive events were generated separately, and were mixed according to
the generator cross-sections to fully describe the inelastic scattering.
\Herwig does not contain any diffractive processes.

\section{Event and track selection}
\label{sec:selection}

All data used in this paper were taken during the LHC running periods with
stable beams and defined beam-spot values, between 6th--15th~December~2009 for the
analysis at $\sqrt{s} = \unit{900}{\GeV}$, and from 30th~March to 27th~April~2010
for the \unit{7}{\TeV} analysis.
The only operational requirement was that the MBTS trigger and all inner detector
subsystems were at nominal conditions.
During the December data taking period, more than 96\% of the pixel detector,
more than 99\% of the SCT and more than 98\% of the TRT was operational. These
efficiencies were higher in 2010.

To reduce the contribution from backgrounds and secondaries, as well as to
minimize the systematic uncertainties, the following criteria were imposed:
\begin{itemize}
\item the presence of a reconstructed primary vertex
  using at least two tracks, each with:
  \begin{itemize}
  \item $\pt > \unit{100}{\MeV}$;
  \item offline reconstruction within the inner detector,
    $\etamod < 2.5$;
  \item a transverse distance of closest approach with respect to the beam-spot (BS)
    position, $|d_\mathrm{0}^\mathrm{BS}|$, of less than
    \unit{4}{\mm};
  \item uncertainties on the transverse and longitudinal distances of closest
    approach of $\sigma(d_0^{BS}) ~<~$ \unit{5}{\mm} and $\sigma(z_0^{BS}) <$
    \unit{10}{\mm};
  \item at least one pixel hit, at least four SCT hits and at least six silicon
    hits in total.
  \end{itemize}

  Beam-spot information was used both in the track pre-selection and to constrain
  the fit during iterative vertex reconstruction, and vertices
  incompatible with the beam-spot were removed.  The vertices were ordered by the $\sum\pt^2$
  over the tracks assigned to the vertex, which is strongly correlated with the total number of associated tracks,
  with the highest-$\sum\pt^2$ vertex defined as the primary interaction vertex of the event.

  Events that had a second primary vertex with more than three tracks in the same bunch crossing were rejected.
  If the second vertex had three or fewer tracks, all tracks from the event that passed the selection were kept.
  After this cut, the fraction of events with more than one interaction in the same bunch crossing (referred to as pile-up)
  was found to be about $0.1\%$; the residual effect was thus neglected.
  At $\sqrt{s} = \unit{900}{\GeV}$, since the data were taken at the low luminosity period,
  the rate of pileup was even lower and was also neglected.

\item at least one track with:
  \begin{itemize}
  \item $\pt > \unit{1}{\GeV}$,
  \item a minimum of one pixel and six SCT hits\footnote{This is a more
      stringent requirement than the requirement of seven silicon hits at the track
      reconstruction step.};
  \item a hit in the innermost pixel layer (the b-layer), if the corresponding pixel module was active;
  \item transverse and weighted-longitudinal impact parameters with respect to
    the event-by-event primary vertex were required to be $|d_0| <
    \unit{1.5}{\mm}$ and $|z_0|\cdot \sin \theta < \unit{1.5}{\mm}$
    \footnote{The factor of $\sin\theta$ compensates for the $\sin\theta$ in the
      denominator of the uncertainty of $z_0$ derived from the measured distance
      of closest approach.};
  \item for tracks with $\pt > \unit{10}{\GeV}$, a $\chi^2$ probability of track
    fit $> 0.01$ was required in order to remove mismeasured tracks\footnote{A
      long non-Gaussian tail in the track momentum resolution, combined with the
      steeply falling \pt spectrum, leads to an observed migration of
      very-low-momentum particles to very high reconstructed \pt, which are
      referred to as mismeasured tracks.}.
  \end{itemize}

\end{itemize}

Only events with leading track $\pt >
\unit{1}{\GeV}$ were considered, in order to reject events where the leading
track selection can potentially introduce large systematic effects. This also has
the effect of further reducing the contribution from diffractive scattering processes.

Two separate analyses were performed, in which all the other tracks were required
to have either $\pt > \unit{100}{\MeV}$ or $\pt > \unit{500}{\MeV}$.  For
 $\pt >$ \unit{500}{\MeV} tracks, the silicon and impact parameter requirements were the
same as given earlier for tracks with $\pt > \unit{1}{\GeV}$.
For tracks with the lower \pt threshold, all other selection criteria were the same except that
only two, four or six SCT hits were required for tracks with \pt $\geq$ $100, 200$, \unit{300}{\MeV}, respectively.
Tracks with $\pt > \unit{500}{\MeV}$ are less prone than
lower-\pt tracks to inefficiencies and systematic uncertainties resulting from
interactions with the material inside the tracking volume.
Whenever possible, the tracks were extrapolated to include hits in the TRT.
Typically, 88\% of tracks inside the TRT acceptance ($\etamod< 2.0$) included a
TRT extension, which significantly improves the momentum resolution.

After these selections, for the \unit{500}{\MeV} (\unit{100}{\MeV}) analysis,
189,164 and 6,927,129 
events remained at \unit{900}{\GeV} and \unit{7}{\TeV} respectively, containing
1,478,900 (4,527,710) and 89,868,306 (209,118,594)
selected tracks and corresponding to integrated luminosities of
\unit{7}{\micro\reciprocal\barn} and \unit{168}{\micro\reciprocal\barn}, respectively.
For the MC models considered here, the contribution of diffractive events
to the underlying event observables was less than~$1\%$.

\section{Background contributions}
\label{sec:backgrounds}


\subsection{Backgrounds}
The amount of beam and non-beam (cosmic rays and detector noise) background
remaining after the full event selection was estimated using the number of pixel
hits which were not associated to a reconstructed track. This multiplicity
included unassigned hits from low-\pt looping tracks, but was dominated at
higher multiplicities by hits from charged particles produced in beam background
interactions.  The vertex requirement removed most of the beam background events
and the residual contribution from beam background events after this requirement
was below~0.1\%.  As the level of background was found to be very low, no
explicit background subtraction was performed.

\subsection{Fraction of secondary tracks}
The primary charged-particle multiplicities were measured from selected tracks
after correcting for the fractions of secondary and
poorly reconstructed tracks in the sample. The potential background from fake
tracks was found via MC studies to be less than 0.01\%.

Non-primary tracks predominantly arise from hadronic interactions, photon
conversions to positron-electron pairs, and decays of long-lived particles.
For \pt above \unit{500}{\MeV} the contribution from photon conversions is small, and
side-band regions of the transverse and longitudinal impact parameters from data
were used to find a scaling factor of 1.3 for the track yield in MC to get a
better agreement with the data. This is not the case at lower \pt.  A
separate fit to the tails of the $d_0$ distribution for primaries, non-primaries
from electrons and other non-primaries, was carried out in eight bins of
\unit{50}{\MeV} in the range $100 < \pt < \unit{500}{\MeV}$. The scaled MC was then used
to estimate the fraction of secondaries as a function of both \pt and \eta~in
the selected track sample, which is found to be at most 2\% for events in
both \unit{900}{\GeV} and \unit{7}{\TeV} collisions~\cite{Collaboration:2010rd,
  Collaboration:2010ir}. The systematic uncertainty on the secondaries is included in the
uncertainties due to tracking.

\section{Correction to particle level}
\label{sec:correction}

The data were corrected back to charged primary particle spectra satisfying the
event-level requirement of at least one primary charged particle within $\pt >
\unit{1}{\GeV}$ and $\etamod < 2.5$.  A two step correction process was used,
where first the event and track efficiency corrections were applied, then an
additional bin-by-bin unfolding was performed to account for possible bin
migrations and any remaining detector effects.


\subsection{Event-level correction}
Trigger and vertexing efficiencies were measured~\cite{Collaboration:2010ir} as a function of
the number of tracks, \Nselbs, passing all the track selection requirements except for the
primary vertex constraint. In this case the transverse impact parameter with
respect to the beam-spot~\cite{:atlasbs} was required to be less than \unit{1.8}{\mm}.
The event level corrections consisted of the following:

\begin{itemize}

\item The efficiency of the MBTS scintillator trigger, $\efftrig(\Nselbs)$ was
  determined from data using an orthogonal trigger. It consisted of a random
  trigger, requiring only that the event coincided with colliding bunches and
  had at least 4 pixel clusters and at least 4 SCT space points at L2. The
  trigger was found to be $\sim 97\%$ efficient for low-multiplicity events,
  and almost fully efficient otherwise. It showed no dependence on the \pt and
  pseudorapidity distributions of the selected tracks.

\item The vertex reconstruction efficiency, $\effvtx(\Nselbs, \langle \eta \rangle)$ was also measured
  in data, by taking the ratio of the number of triggered events with a reconstructed vertex
  to the total number of triggered events.  For events containing fewer than three
  selected tracks, the efficiency was found to depend on the projected
  separation along the beam axis of the two extrapolated
  tracks, $\Delta z_0^{BS}$. This efficiency amounted to approximately 90\% for the lowest bin of
  \Nselbs, rapidly rising to 100\%.

\item A correction factor, $\effldtrk(\efftrk)$ accounts for the probability that
  due to the tracking inefficiency none of the candidate leading tracks with
  $\pt > \unit{1}{\GeV}$ are reconstructed in an event, resulting in the event
  failing the selection criteria.
  A partial correction for this was provided by determining the probability that
  all possible reconstructed leading tracks would be missed for each event
  using the known tracking efficiencies, and then dividing the event weight by
  this probability. This process will in general yield an excessive correction,
  since the correct weight should be determined using the number and
  distributions of true charged particles with $\pt > \unit{1}{\GeV}$ and
  $\etamod < 2.5$ rather than the distributions of reconstructed tracks.  This
  leads to an over-estimation of the probability for the event to be omitted.
  Nevertheless, this correction represents a good estimate of the efficiency,
  given the efficiency estimate of tracks in each event. The efficiency was found to be $>
  98\%$ in low-\pt bins and almost $100\%$ in high-\pt bins.  The
  uncertainty for this correction is included as part of the tracking efficiency
  systematic uncertainty.  The correction was made with the expectation that the
  final unfolding in the form of bin-by-bin corrections will provide the small
  additional correction that is needed.

\end{itemize}

  The total correction applied to account for events lost due to the trigger,
  vertex, and tracking requirements (in bins of number of tracks with $\pt >
  \unit{0.5}{\GeV}$) is given by
  \begin{equation}
    \label{equation:event_weight}
    w_\text{ev} =
    \frac{1}{\efftrig(\Nselbs)} \cdot
    \frac{1}{\effvtx(\Nselbs, \langle \eta \rangle)} \cdot
    \frac{1}{\effldtrk(\efftrk)},
  \end{equation}
  where $\efftrig(\Nselbs)$, $\effvtx(\Nselbs, \langle \eta \rangle)$ and
  $\effldtrk(\efftrk)$ are the trigger, vertex reconstruction and leading track
  reconstruction efficiencies discussed earlier.

\subsection{Track-level correction}

The track-reconstruction efficiency in each bin of the $\pt$--$\eta$ kinematic
plane, was determined from simulation and defined as
\begin{equation}
  \effbin(\pt,\eta) =
  \frac{N^\text{matched}_\text{rec}(\pt,\eta)}{N_\text{gen}(\pt,\eta)},
\end{equation}
where $N^\text{matched}_\text{rec}(\pt,\eta)$ is the number of reconstructed
tracks in a given bin matched to a generated charged particle, and
$N_\text{gen}(\pt,\eta)$ is the number of generated particles in that bin.  The
matching between a generated particle and a reconstructed track was done using a
cone-matching algorithm in the $\eta$--$\phi$ plane and associating the particle
to the track with the smallest $R = \sqrt {(\Delta\phi)^2+(\Delta \eta
  )^2}$ within a cone of radius \mbox{$\Delta R < 0.15$}.  To reduce fake
matching, a common pixel hit between the reconstructed, simulated track and the
generated particle track in the Geant4 simulation was also required.
The efficiencies were slightly different between the datasets at the
two different center-of-mass energies
because of small differences in the configuration of the pixel and SCT detectors
between the 2009 and 2010 data-taking periods.

A weight,
\begin{equation}
  \label{equation:track_weight}
  w_\text{trk} =  \frac{1}{\effbin(\pt, \eta)} \cdot (1-f_\text{sec}(\pt)) \cdot (1-f_\text{fake}),
\end{equation}
was applied on a track-by-track basis to all track-level histograms. Here
$\effbin(\pt, \eta)$ is the track-reconstruction efficiency described earlier,
$f_\text{sec}$ is the fraction of secondaries, and $f_\text{fake}$ is the
fraction of fakes.

\subsection{Final unfolding step}

The efficiency corrections described so far do not account for bin-by-bin
migrations, nor for the possibility of not reconstructing the leading particle
in the event as the leading track (reorientation of an event).  To account for
these effects, an additional bin-by-bin unfolding was applied to all
distributions after applying the event- and track-level efficiency corrections
described above.

In this correction step, the unfolding factors were evaluated separately in each
bin for each observable listed in \TabRef{tab:obs},
\begin{equation}
  \text{U}_\text{bin} = \frac{\text{V}^\text{Gen}_\text{bin}}{\text{V}^\text{Reco,~eff~corr}_\text{bin}},
\end{equation}
where $\text{V}^\text{Gen}_\text{bin}$ and
$\text{V}^\text{Reco,~eff~corr}_\text{bin}$ respectively represent the generator
level MC value of the observable and the reconstructed MC value after applying
the event- and track-level efficiency corrections at each bin. The corrected
value for an observable is found by multiplying the measured value by
the corresponding unfolding factor.
%
This unfolding factor is within 5\% (10\%) of unity in the lowest-\pt bins for
the $\pt > \unit{100}{\MeV}$ (\unit{500}{\MeV}) analyses respectively, due to
the migration and reorientation effects, and very close to unity for higher-\pt
bins.


\section{Systematic uncertainties}
\label{sec:systematics}

A study of the systematic uncertainties was performed, and these
were propagated to the final distributions and added in quadrature to obtain a
total systematic uncertainty.

Systematic uncertainties from tracking efficiency were
studied~\cite{Collaboration:2010rd, Collaboration:2010ir}, and the largest were found to be due to the following:
\begin{itemize}

\item The material in the inner detector: the effect of material budget uncertainties
  in the inner detector was determined to affect the efficiency by a relative difference of
  $2\%$ in the barrel region, rising to over $7\%$ for $2.3 < \etamod < 2.5$, for
  tracks with $\pt > \unit{500}{\MeV}$.

\item Consequence of $\chi^2$ probability cut: the maximum difference between the
  fraction of events in data and MC which passed this cut was found to be
  10\%. This value was taken as a conservative estimate of the systematic
  uncertainty, applied to tracks with $\pT > \unit{10}{\GeV}$ only.

\end{itemize}
The systematic uncertainty from pile-up removal was estimated to be negligible.

The most common UE observable is a ``profile'' plot of the mean value of a
charged particle \pt or multiplicity observable as a function of the \pt of the
leading object in the event. Due to the steeply-falling \pt spectrum in minimum
bias events, the number of events in the low-\pt bins of these profiles is much
higher than in the higher-\pt bins, and so migration of the leading track from
the lower-\pt bins to higher ones is possible: this was accounted for in the
MC-based unfolding procedure. However, an additional systematic uncertainty was
included because more \ptlead migrations are expected in data than in the MC
detector modelling. This extra systematic contributes only to the region of the profiles with
$\ptlead > \unit{10}{\GeV}$, since a small fraction of highly mismeasured
leading tracks from the lowest \ptlead bin can still have a significant effect
upon the less-populated high-\ptlead bins. Since the greatest difference from
the \ptlead-profile values in $\ptlead > \unit{10}{\GeV}$ is seen in the first
\ptlead bin, a conservative systematic estimate was obtained by assuming all
migrations to come from the first bin.

The remaining contributions to the overall systematic uncertainty result from
the specific unfolding method used in this analysis.  The bin-by-bin
unfolding corrections are in general influenced by the number of charged
particles and their \pt distributions, so there is some dependence on the event
generator model. This introduces a second extra source of systematic uncertainty. In
order to estimate this uncertainty it is necessary to compare different plausible
event generation models, which deviate significantly from each other.
%
%
Between the various models and tunes already described, the maximal variation is
seen between \Pythia and \Phojet, and this difference is taken as a measure
of the uncertainty due to model-dependence. Where the \Phojet sample has sufficient
statistics, it is seen that beyond the statistical fluctuations the relative
difference between the required correction factors from \Phojet and \Pythia
are at most 4\% in the lowest-\pt bins, and 2\% everywhere else.

Since this uncertainty is independent of any efficiency systematics, it has been
summed in quadrature with the efficiency systematic uncertainty and the
statistical uncertainty.  In addition to the model-dependent uncertainty in the
bin-by-bin unfolding, there is also a statistical uncertainty due to the finite
size of the Monte Carlo sample. The statistical fluctuation of the \Pythia
unfolding factor is found to be negligible for low-\pt bins,
but rises to be a significant contribution in higher \pt bins.


\begin{figure}[tb]
  \begin{center}
    \includegraphics[width=.43\textwidth]{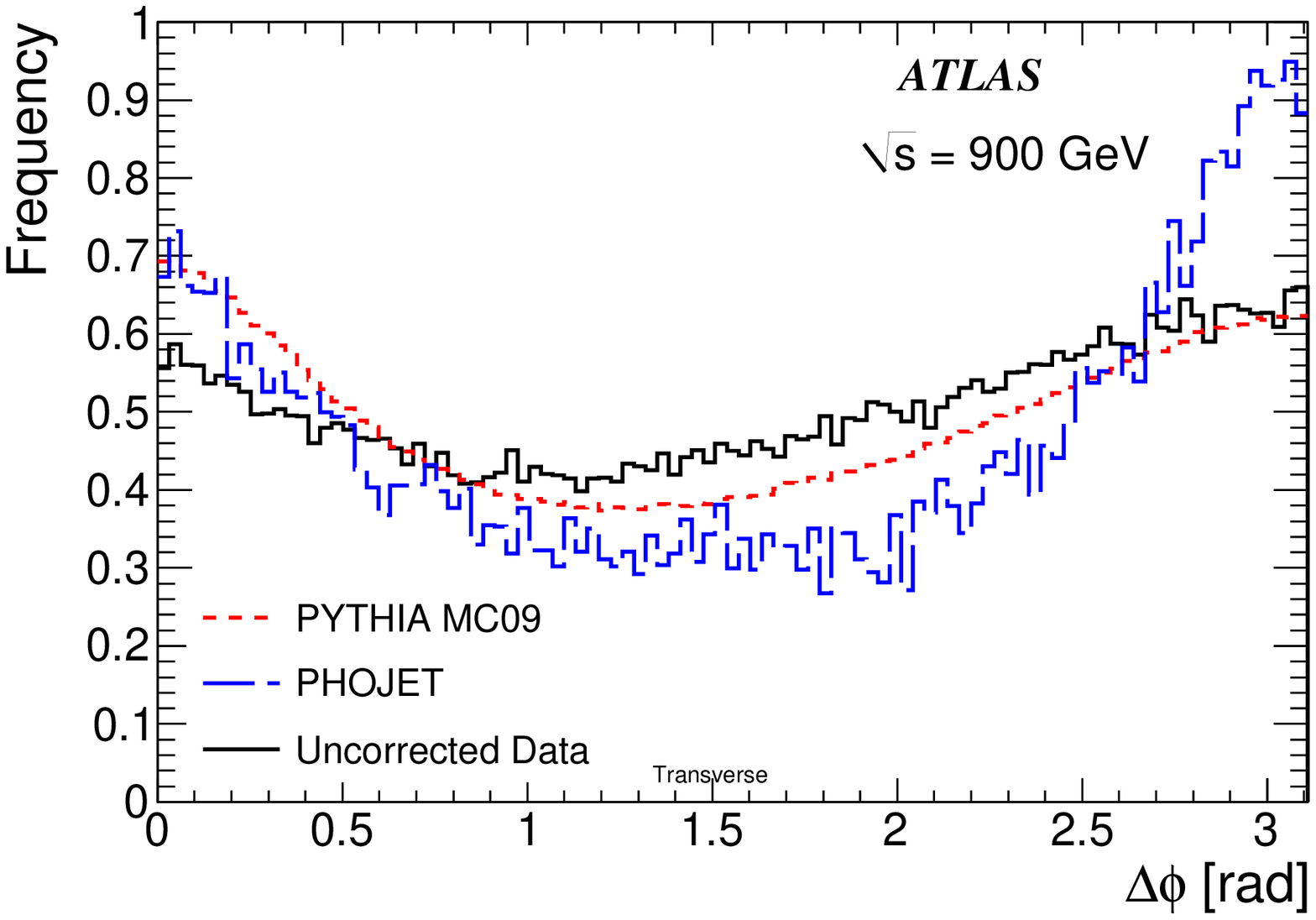} \hfill
    \includegraphics[width=.43\textwidth]{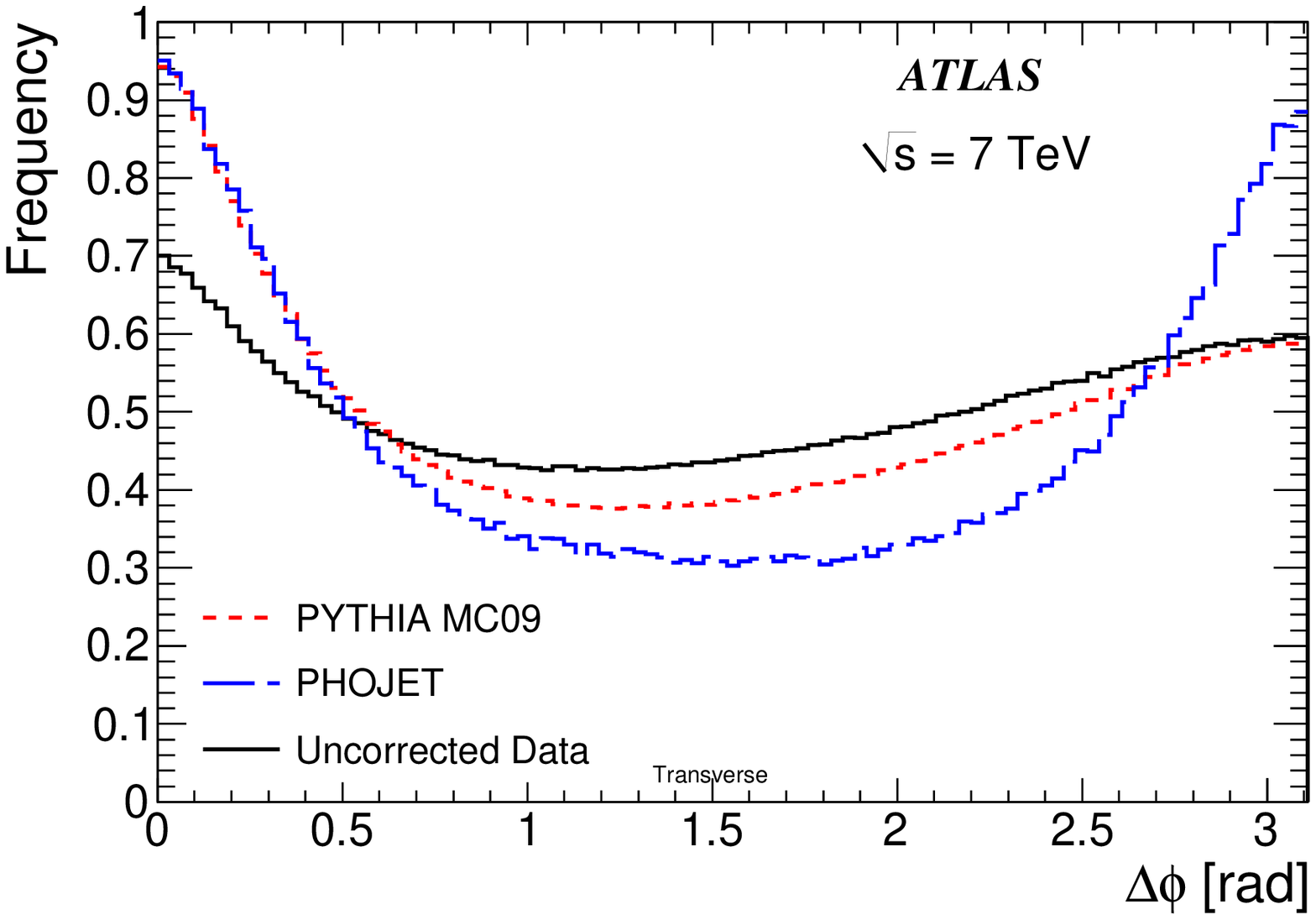}
  \end{center}
  \caption{Difference in $\phi$ between the leading and the sub-leading track in \Pythia, \Phojet and in uncorrected data. The left plot is for \unit{900}{\GeV} and the right is for \unit{7}{\TeV}. The MC curves are shown after the full detector simulation.}
  \label{fig:ldsld}
\end{figure}

The $|\Delta\phi|$ between the leading track and the track with
the second-highest \pt (the sub-leading track) is shown in \FigRef{fig:ldsld}.
It is seen to be most likely that the sub-leading charged particle lies in
either the true toward or the true away region, in which case there is
relatively little effect on the observables -- the transverse region is particularly
unaffected by a $\sim 180^\circ$ reorientation.  However, if the reconstructed
leading track lies in what should have been the transverse region, the effect
will be to reduce the densities in the toward and away regions, and to increase
the densities in the transverse region. The bin-by-bin unfolding derived from
the MC corrects for this effect, provided that it occurs with the frequency
of reorientation predicted by the MC simulation. \FigRef{fig:ldsld} is used to
estimate the relative frequency with which an event is reoriented such that the
true towards and away regions lie in the transverse region identified by the
reconstruction. Comparing the $|\Delta\phi|$ distribution in uncorrected data to
the same distributions (uncorrected and reconstructed) predicted by \Pythia and
\Phojet, it is seen that both generator models predict fewer event
reorientations of this type.  The final correction to the data uses bin-by-bin
unfolding factors that are derived from the \Pythia sample, so the relative
magnitude of the systematic uncertainty associated with this effect can again be
estimated by the difference of the \Pythia and \Phojet probabilities. This difference
is comparable with the difference between the data and \Pythia predictions.
The uncertainty is applied in both directions, reasonably assuming a symmetric effect,
so the difference in \Pythia and \Phojet corrections provides the systematic
uncertainty in the unfolding factor even though the \Phojet deviation from
\Pythia is in the opposite direction from the data.

\TabRef{tab:sysSummary} summarizes the various contributions to the systematic
uncertainties.

\begin{table}[hpb]
  \singlespacing
  \caption{Summary of systematic uncertainties, shown for the lowest-, intermediate- and highest-\pt bins.
    For the analysis with \unit{7}{\TeV} (\unit{900}{\GeV}) center-of-mass energy data, the lowest-\pt
    bin refers to \mbox{$\ptlead = 1.0-\unit{1.5}{\GeV}$}, the intermediate \pt bin refers to
    $\ptlead = 9-\unit{10}{\GeV}$ ($4-\unit{5}{\GeV}$), and the highest \pt bin refers to $\ptlead = 18-\unit{20}{\GeV}$ ($9-\unit{10}{\GeV}$).
    The uncertainties shown are from the transverse region charged $\sum\pt$ distribution,
    and all the other profiles are estimated to have comparable or less systematic uncertainty.
    Each uncertainty is given relative
    to the profile value at that stage in the correction sequence
    and they are an average over all of the phase-space values.
    In the cases where the uncertainties are different for
    \unit{900}{\GeV} and \unit{7}{\TeV} analysis, the \unit{900}{\GeV} value is shown in parentheses.}
  \begin{center}
    \begin{tabular}{lrrr}
      \toprule
      \toprule
      Leading charged particle bin & Lowest-\pt & Intermediate-\pt & Highest-\pt \\
      \midrule
      \multicolumn{4}{l}{ {\bf Systematic uncertainty on unfolding}}\\
      \Pythia/\Phojet difference ~ &  4\% & 2\% & 2\% \\
      \Pythia unfolding stat. uncertainty  &  $<0.1\%$  & 1\% (2\%) & 4\% (5\%)  \\
      \midrule
      \multicolumn{4}{l}{ {\bf Systematic uncertainties from efficiency corrections}}  \\
      Track reconstruction &  3\% &  4\% & 4\%  \\
      Leading track requirement &  1\% & $<0.1\%$ & $<0.1\%$ \\
      Trigger and vertex efficiency &  \multicolumn{3}{c}{------ \quad $< 0.1\%$ (everywhere) \quad ------} \\
      Total from efficiency corrections & 2.5\% & 4\% & 4\% \\
      \midrule
      \multicolumn{4}{l}{ {\bf Systematic uncertainty for bin migration}}  \\
      Bin migration due to mismeasured \pt & - & 2.5\% (0\%)& 5\% (0\%)\\
      \midrule
      {\bf Total systematic uncertainty}  & 4.5\%  & 4.5\% (5\%)  & 8\% (6.5\%)    \\
      \bottomrule
      \bottomrule
    \end{tabular}
    \label{tab:sysSummary}
  \end{center}
\end{table}

\clearpage

\section{Results and discussion}
\label{sec:results}

\subsection{Overview}

In this section, corrected distributions of underlying event
observables are compared to model predictions tuned to a wide range of measurements.
As described, the data have received minimally model-dependent corrections
to facilitate model comparisons. The transverse, toward and away regions each have
an area of $\Delta\phi \, \Delta\eta = 10\, \pi/3$ in $\eta$--$\phi$ space, so
the density of particles \dNchgdetadphi and transverse momentum sum \dpTsumdetadphi
are constructed by dividing the mean values by the corresponding area. The leading charged
particle is included in the toward region distributions, unless otherwise stated.

The data, corrected back to particle level in the transverse, toward and away regions
are compared with predictions by \Pythia with the ATLAS~MC09, DW, and Perugia0 tunes, by
\HerwigJimmy with the ATLAS~MC09 tune, and by \Phojet. The ratios of the MC
predictions to the data are shown at the bottom of these plots.
The error bars show the statistical
uncertainty while the shaded area shows the combined statistical and systematic
uncertainties.
For the higher values of leading charged particle \pt, the data statistics are limited,
so the distributions are shown only in the \pt range where sufficient statistics are
available.

\subsection{Charged particle multiplicity}
\label{sec:results:nch}

\begin{figure}[pbt]
 \begin{center}
   \includegraphics[width=.5\textwidth]{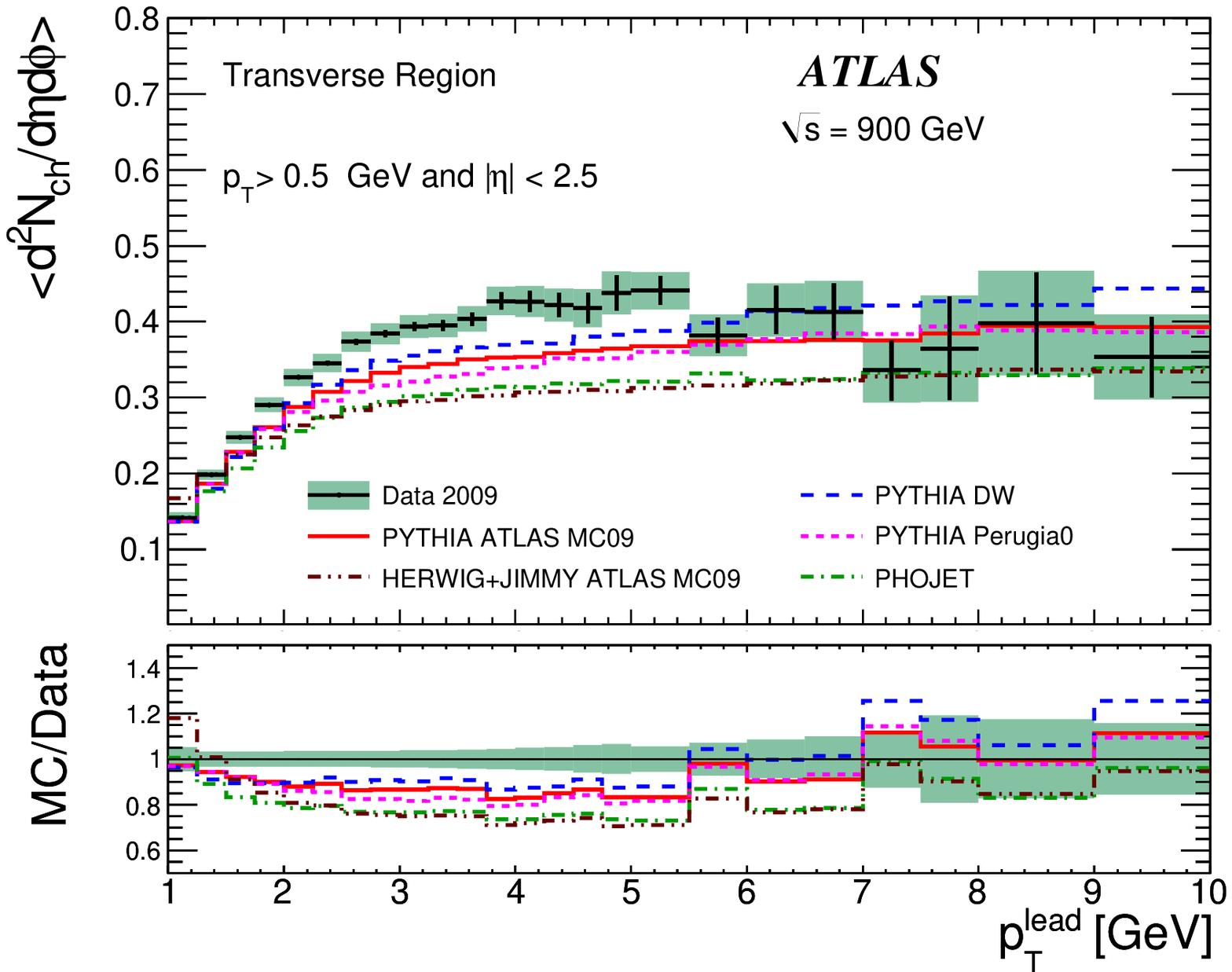}\hfill
   \includegraphics[width=.5\textwidth]{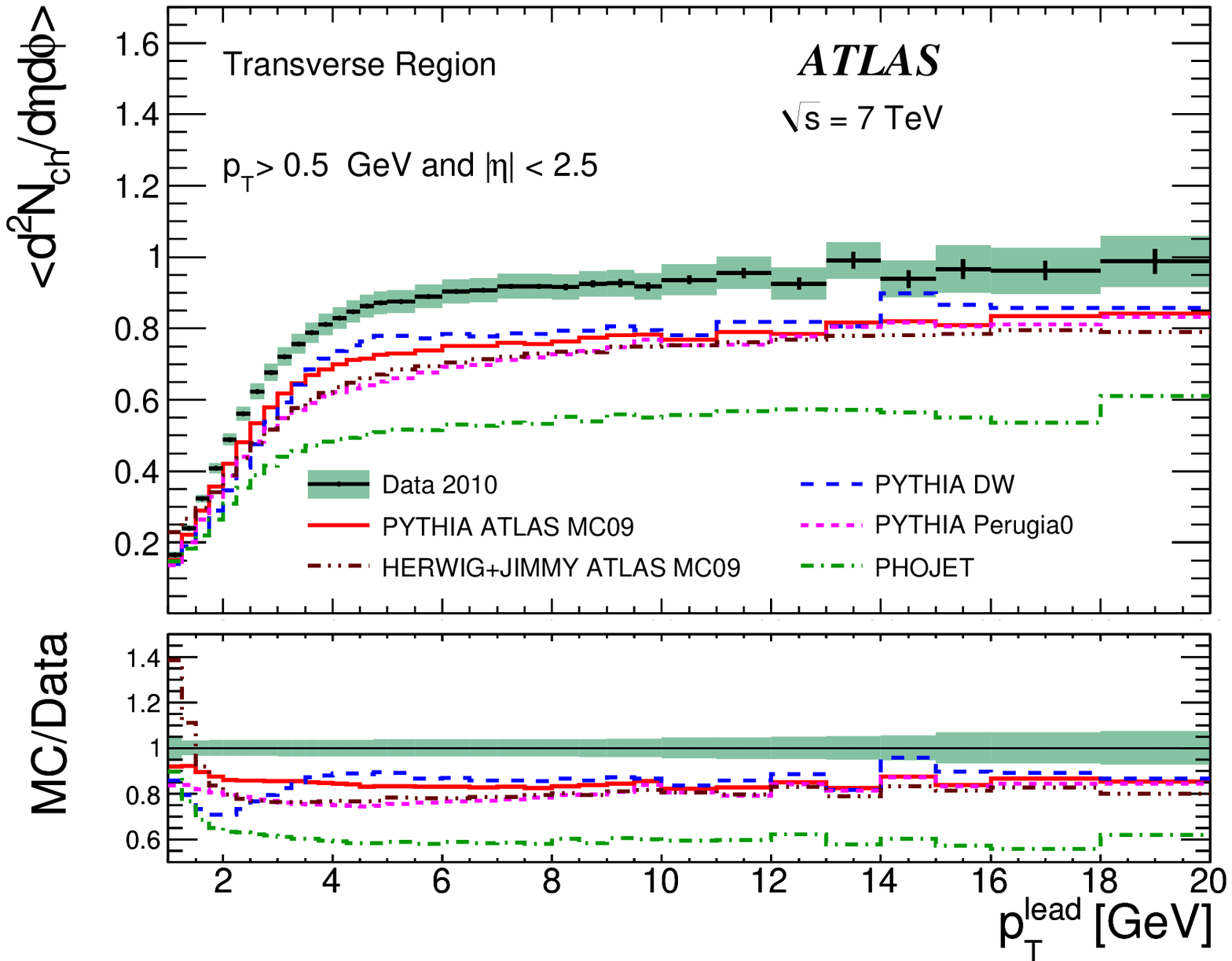}\hfill
   \includegraphics[width=.5\textwidth]{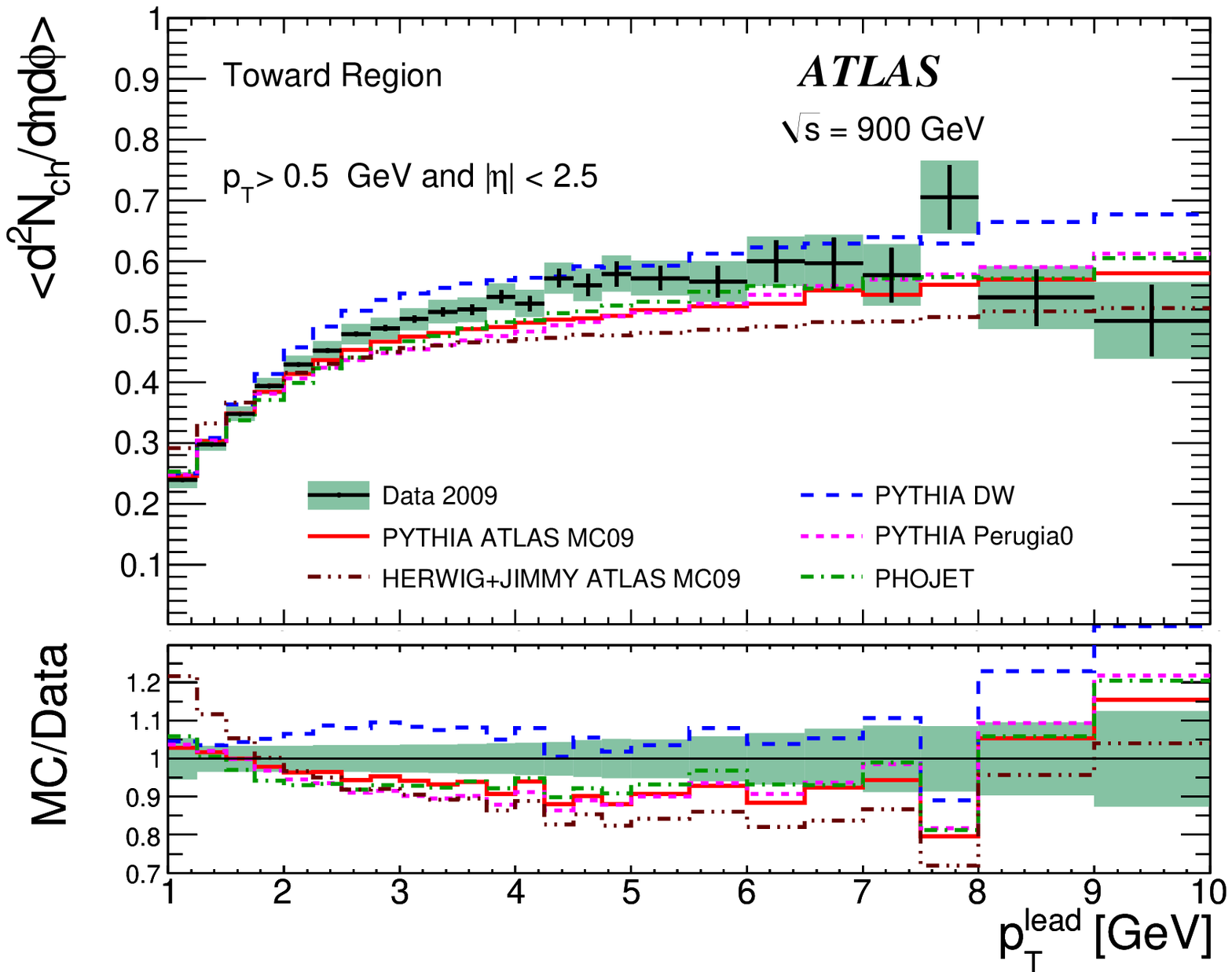}\hfill
   \includegraphics[width=.5\textwidth]{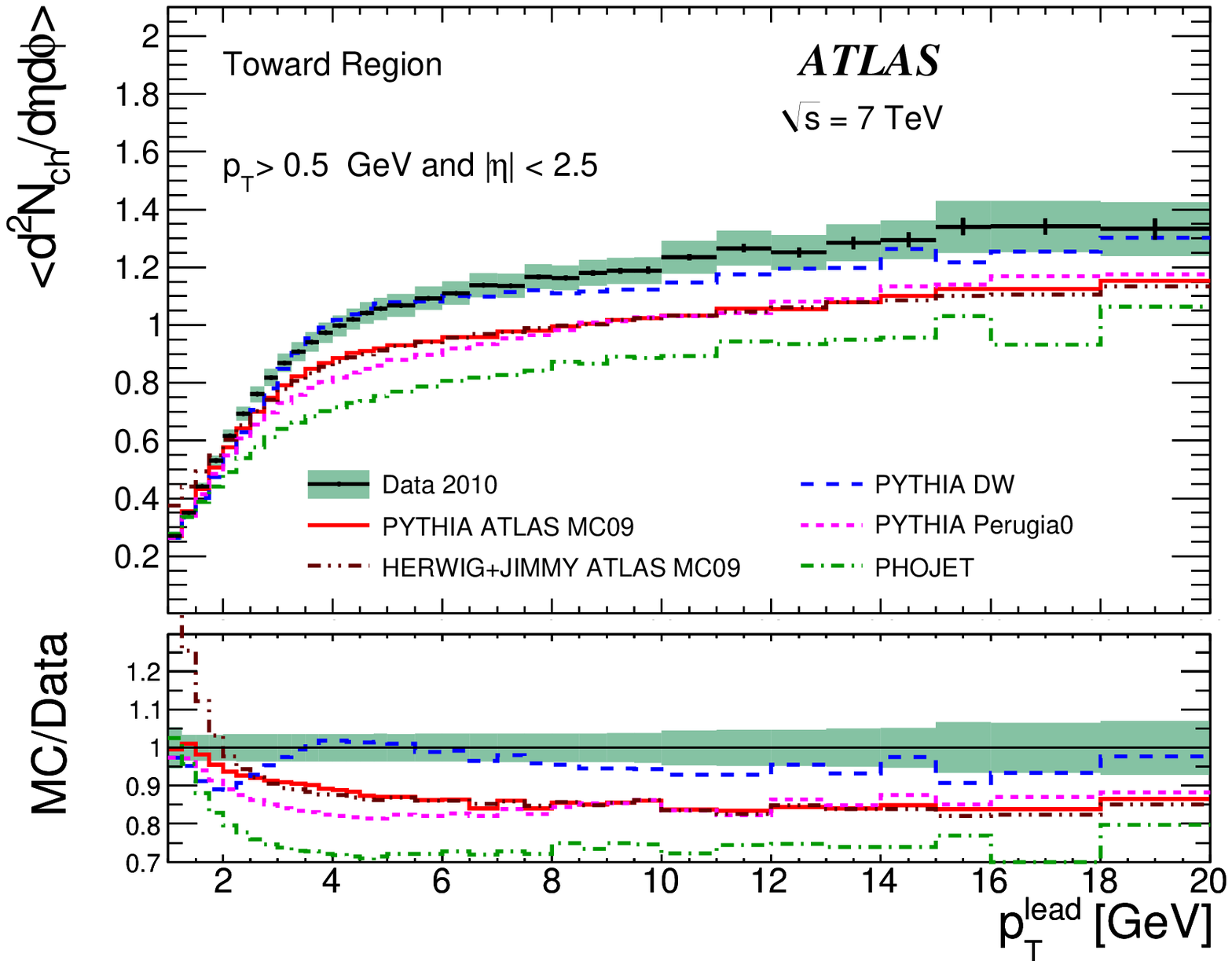}\hfill
   \includegraphics[width=.49\textwidth]{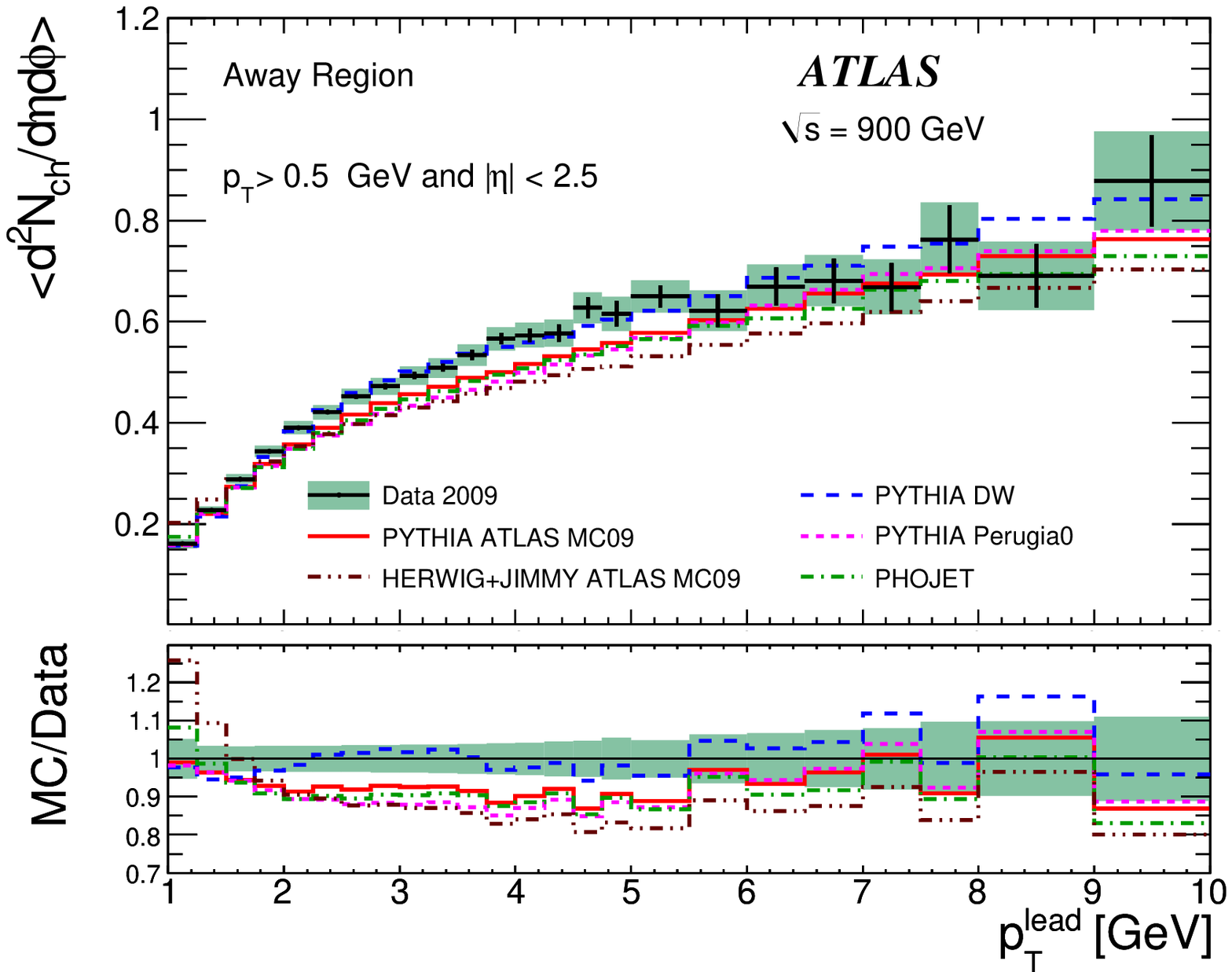} \hfill
   \includegraphics[width=.49\textwidth]{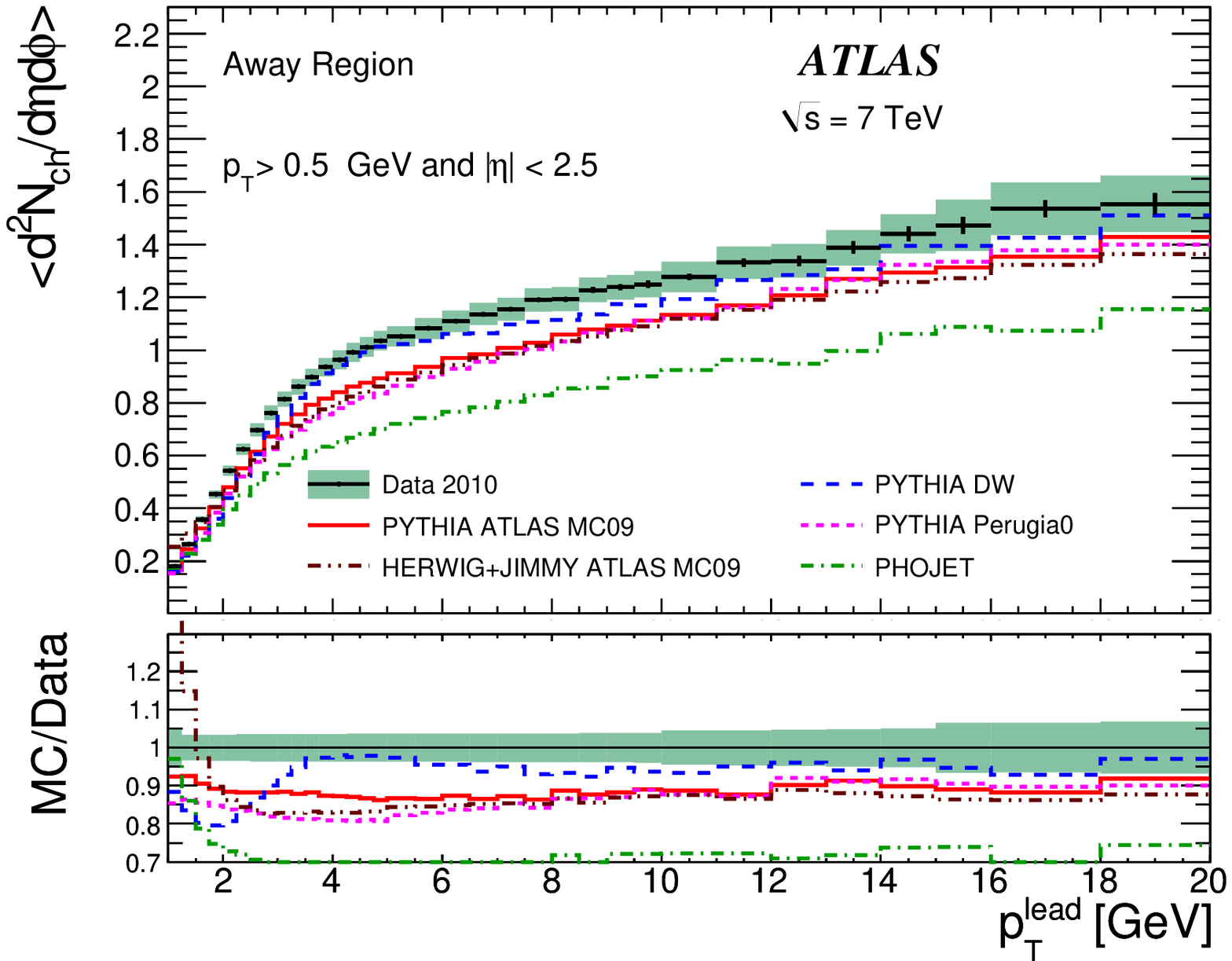}
 \end{center}

 \caption[]{ATLAS data at \unit{900}{\GeV} (left) and at \unit{7}{\TeV} (right)
   corrected back to particle level, showing the density of the charged
   particles \dNchgdetadphi with $\pt > \unit{0.5}{\GeV}$ and $\etamod < 2.5$, as
   a function of \ptlead.
   The data are compared with
   \Pythia ATLAS~MC09, DW and Perugia0 tunes, \Herwig+ \Jimmy ATLAS~MC09 tune, and \Phojet predictions.
   The top, middle and the bottom rows, respectively, show the transverse,
   toward and away regions defined by the leading charged particle.
   The error bars show the statistical uncertainty while the shaded area shows the
   combined statistical and systematic uncertainty.}
 \label{fig:nchg}
\end{figure}

The charged particle multiplicity density,
in the kinematic range $\pt > \unit{0.5}{\GeV}$ and $\etamod < 2.5$
is shown in \FigRef{fig:nchg} as a
function of \ptlead at $\sqrt{s} = \unit{900}{\GeV}$ and \unit{7}{\TeV}.


For the \unit{7}{\TeV} (\unit{900}{\GeV}) data, the average number of charged
particles in the transverse region doubles in going from $\ptlead =
\unit{2}{\GeV} (\unit{1.5}{\GeV})$ to \unit{5}{\GeV} (\unit{3}{\GeV}), and then
forms an approximately constant ``plateau'' for \mbox{$\ptlead > \unit{5}{\GeV}
  (\unit{3}{\GeV})$}.  If we assume the UE to be uniform in
azimuthal angle $\phi$ and pseudorapidity $\eta$, then for $\ptlead >
\unit{5}{\GeV} (\unit{3}{\GeV}) $, the charged particle density of 0.8 (0.4)
translates to about 5 (2.5) particles per unit $\eta$ (extrapolating to the full
$\phi$ space) on average per event, compared to the corresponding number of
$2.423 \pm 0.001~(\text{stat.}) \pm 0.042~(\text{syst.})~(1.343 \pm
0.004~(\text{stat.}) \pm 0.042~(\text{syst.}))$ obtained in the ATLAS
minimum bias measurement \cite{Collaboration:2010ir} with $\pt > \unit{500}{\MeV}$.

%

It can be concluded that the charged particle density in the underlying event,
for events with a leading charged particle in the plateau region (above approximately $3$ or
\unit{5}{\GeV} for the \unit{900}{\GeV} or \unit{7}{\TeV} data respectively), is
about a factor of two larger than the number of charged particles per unit
rapidity seen in the inclusive minimum bias spectrum. This is presumably due to
the selection effect for more momentum exchange in these events, and the
expected absence of diffractive contributions to the events which populate the
plateau region. Given that there is one hard scattering it is more probable to
have MPI, and hence, the underlying event has more activity than minimum bias.

All the pre-LHC MC tunes considered show at least 10--15\% lower activity than
the data in the 
transverse region plateau.
The \Pythia~DW tune is the closest model to data for the transverse
region, and in fact agrees well with the data in the toward and away regions.
The most significant difference between data and MC is seen for the \Phojet
generator, particularly at \unit{7}{\TeV}. The strong deviation of \HerwigJimmy
from the data at low-\ptlead is expected, as the \Jimmy model requires at least
one hard scattering and therefore is not expected to be applicable in this region.

The underlying event activity is seen to increase by a factor of approximately
two between the \unit{900}{\GeV} and \unit{7}{\TeV} data. This is roughly
consistent with the rate of increase predicted by MC models tuned to Tevatron
data.  The toward and away regions are dominated by jet-like activity, yielding
gradually rising number densities.  In contrast, the number density in the
transverse region appears to be independent of the energy scale defined by \ptlead
once it reaches the plateau. The \unit{900}{\GeV} and \unit{7}{\TeV} data show the same trend.

\subsection{Charged particle scalar \pt sum}
\label{sec:results:ptsum}

\begin{figure}[pbt]
 \begin{center}
   \includegraphics[width=.5\textwidth]{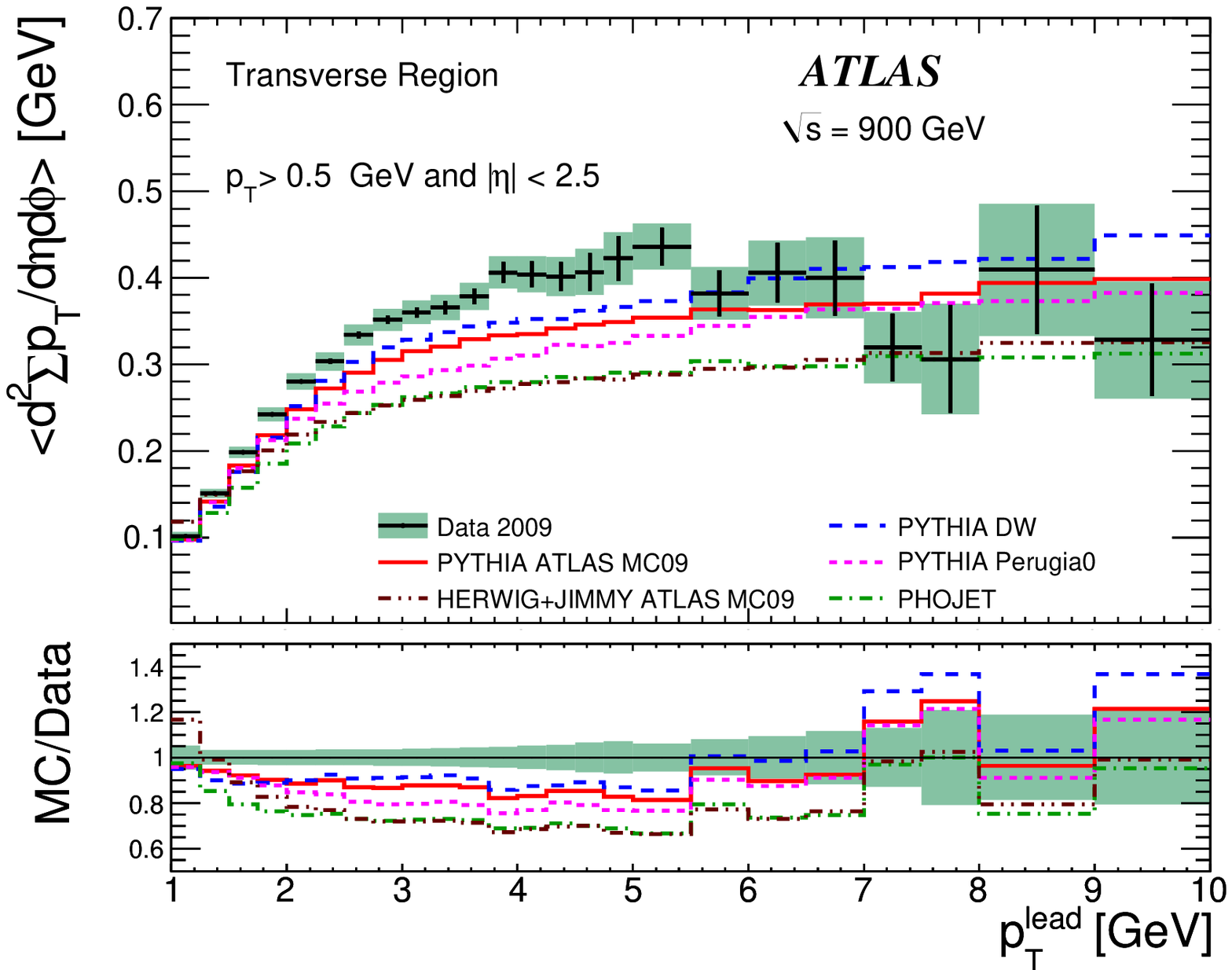}\hfill
   \includegraphics[width=.5\textwidth]{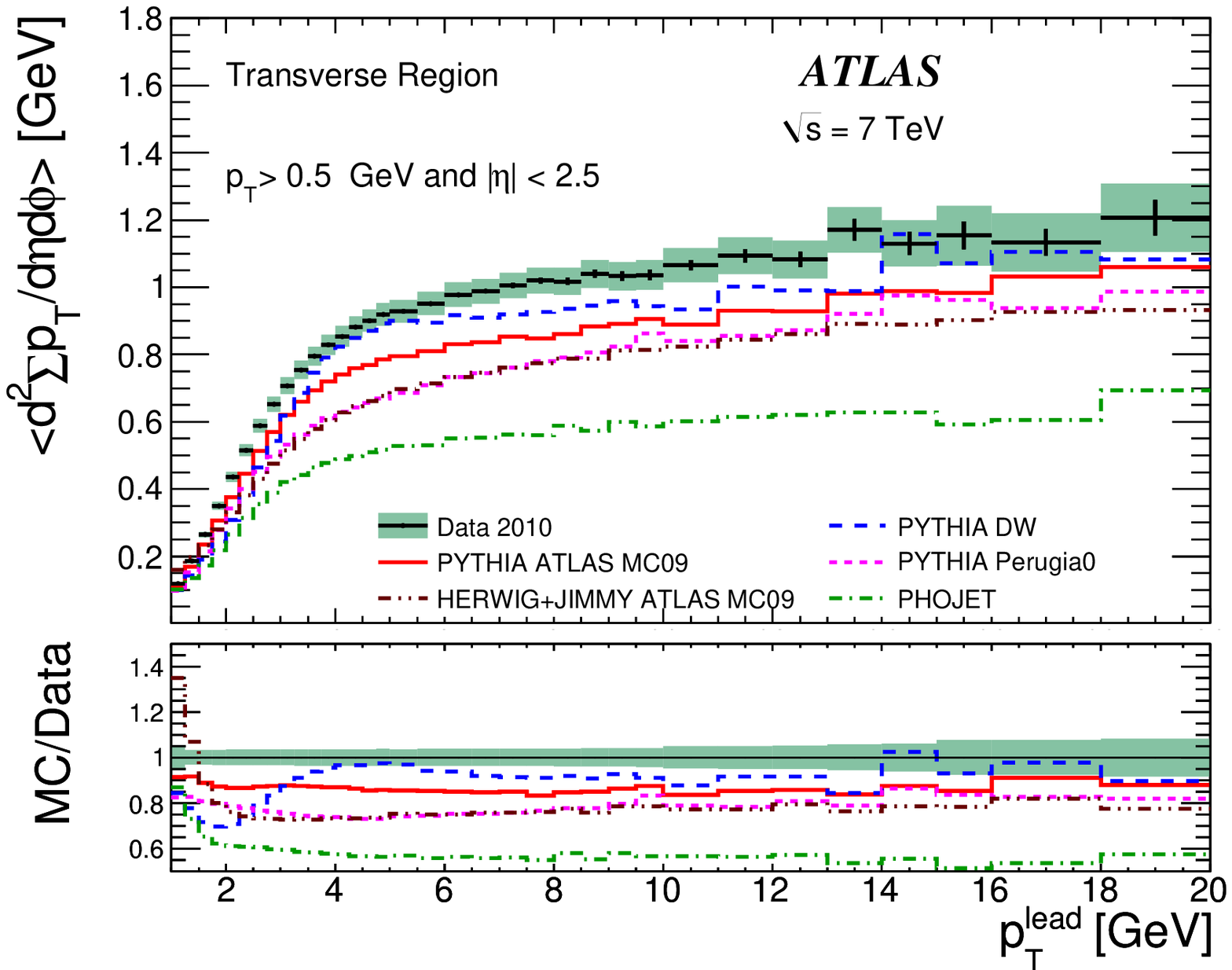}\hfill
   \includegraphics[width=.5\textwidth]{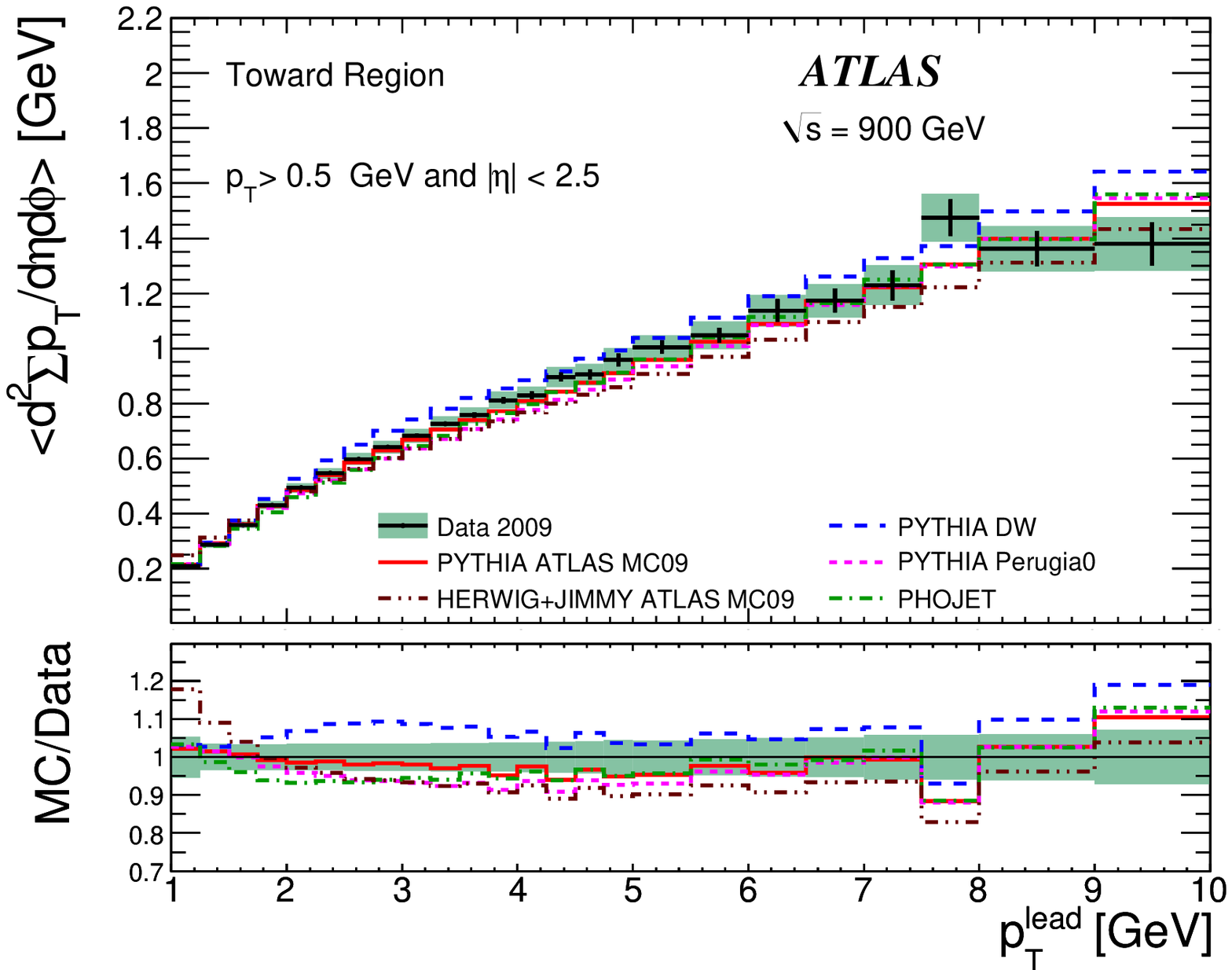}\hfill
   \includegraphics[width=.5\textwidth]{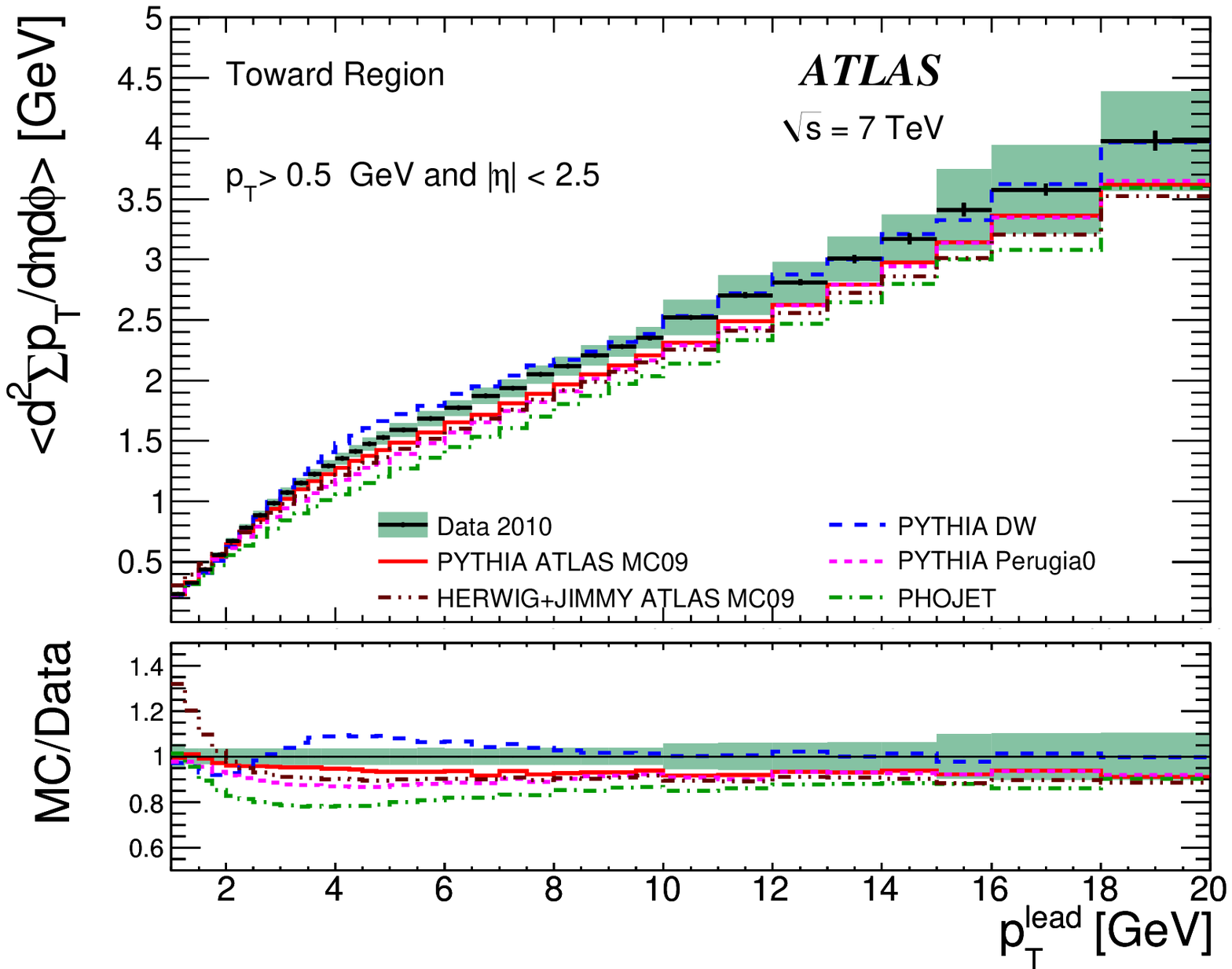}\hfill
   \includegraphics[width=.5\textwidth]{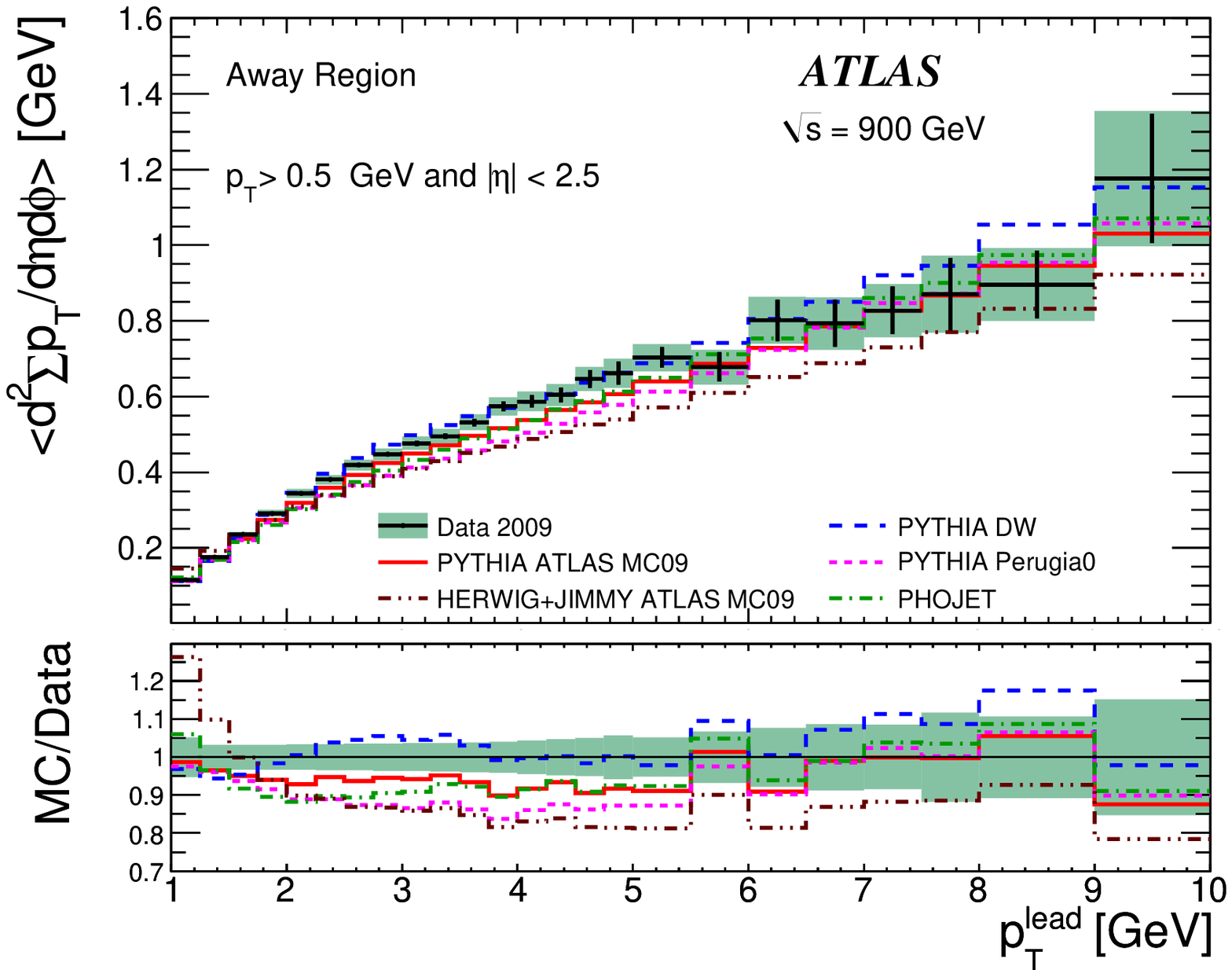}\hfill
   \includegraphics[width=.5\textwidth]{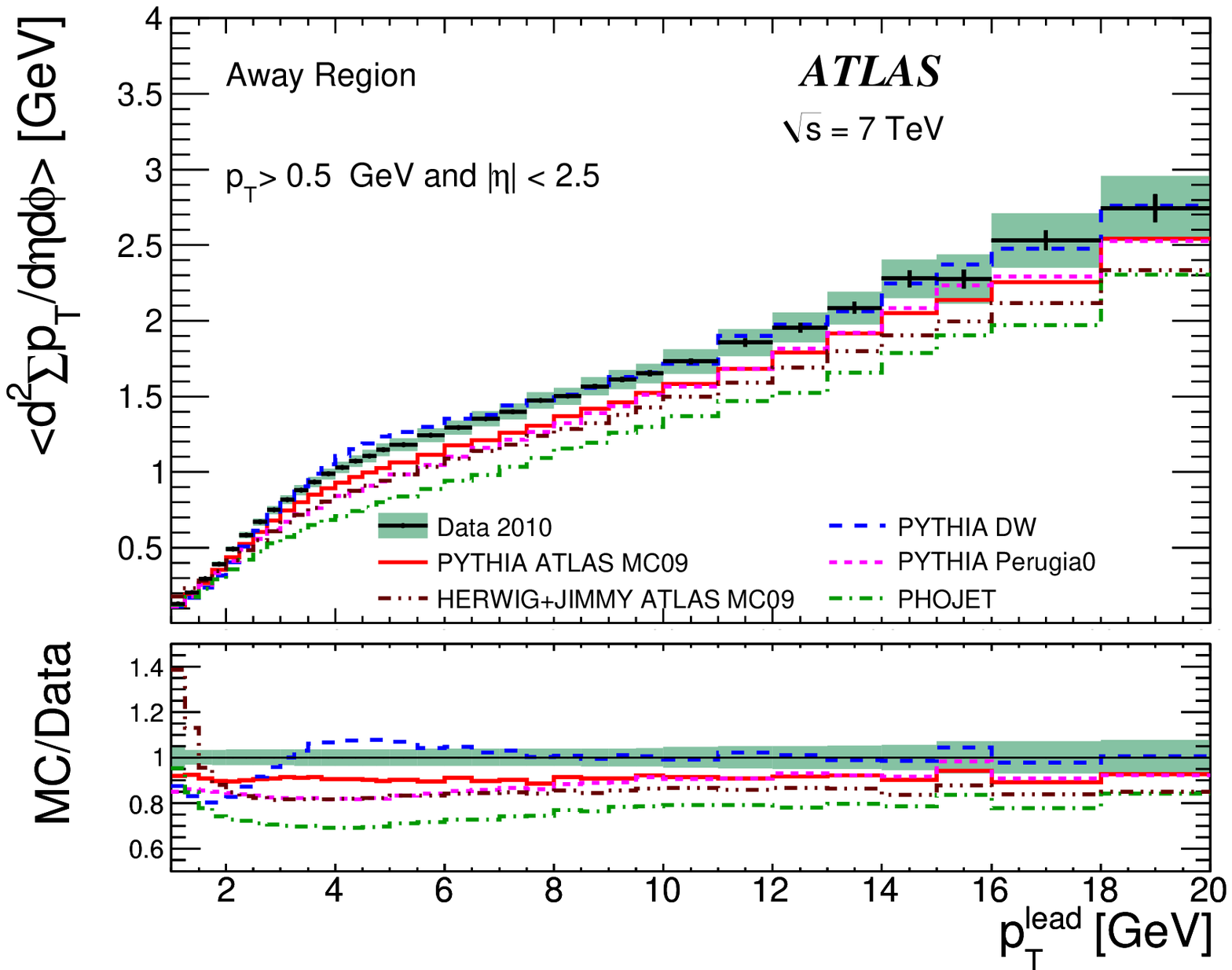}
 \end{center}
 \caption[]{ATLAS data at \unit{900}{\GeV} (left) and at \unit{7}{\TeV} (right)
   corrected back to particle level, showing the scalar $\ptsum$ density of the
   charged particles \dpTsumdetadphi with $\pt > \unit{0.5}{\GeV}$ and $\etamod
   < 2.5$, as a function of \ptlead.
   The data are compared with
   \Pythia ATLAS~MC09, DW and Perugia0 tunes, \Herwig+ \Jimmy ATLAS~MC09 tune, and \Phojet predictions.
   The top, middle and the bottom rows, respectively, show the transverse,
   toward and away regions defined by the leading charged particle.
   The error bars show the statistical uncertainty while the shaded
   area shows the combined statistical and systematic uncertainty.}
 \label{fig:cptsum}
\end{figure}

In \FigRef{fig:cptsum} the charged particle scalar \ptsum density,
in the kinematic range $\pt > \unit{0.5}{\GeV}$ and $\etamod < 2.5$,
is shown as a function of \ptlead at
$\sqrt{s} = \unit{900}{\GeV}$ and \unit{7}{\TeV}.


The summed charged particle \pt in the plateau characterises the mean
contribution of the underlying event to jet energies.  The higher number density
implies a higher \pt density as well.  All the MC tunes considered show 10--15\%
lower \ptsum than the data in the plateau part of the transverse region.  The
\Pythia DW tune is again seen to be the closest to data in the transverse
region, but it slightly overshoots the data in the toward and away
regions. \Phojet is again the model furthest from the data, particularly at
\unit{7}{\TeV}, and the strong deviation of \HerwigJimmy from the data at
low-\ptlead is again expected due to the range of validity of the model. The
value of \ptsum is seen to increase by slightly more than a factor of
two between \unit{900}{\GeV} and \unit{7}{\TeV} data, which is roughly
consistent with the increase predicted by the MC models.

In the toward and away regions jet-like rising profiles are observed,
in contrast to the plateau-like feature in the transverse region. The toward region includes
the leading charged particle, and has a higher \ptsum than the away region as
there is higher probability of high-\pt particles being produced in association
with the leading-\pt charged particle. In the toward region the highest fraction
of energy has been allocated to a single charged particle. This implicitly
reduces the number of additional charged particles in that region, since there
is less remaining energy to be partitioned. As a result the multiplicity of
charged particles is slightly lower in the toward region by comparison to the
away region for high-\ptlead.  The increase of the \pt densities in the toward
and away regions indicates the extent of the variation in the charged fraction
of the total energy in each region.

Multiplying the \ptsum density by the area associated with the toward region, the
\ptsum is nearly twice what it would be if the leading charged particle were
the only charged particle in the region. For the away region, the initial linear
rise corresponds to the region whose total \pt nearly balances that of the
leading charged particle alone.  The \unit{900}{\GeV} and \unit{7}{\TeV} data
show the same trend.

\subsection{Standard deviation of charged particle multiplicity and scalar \ptsum}
\label{sec:results:stddevs}

\begin{figure}[pbt]
 \begin{center}
   \includegraphics[width=.5\textwidth]{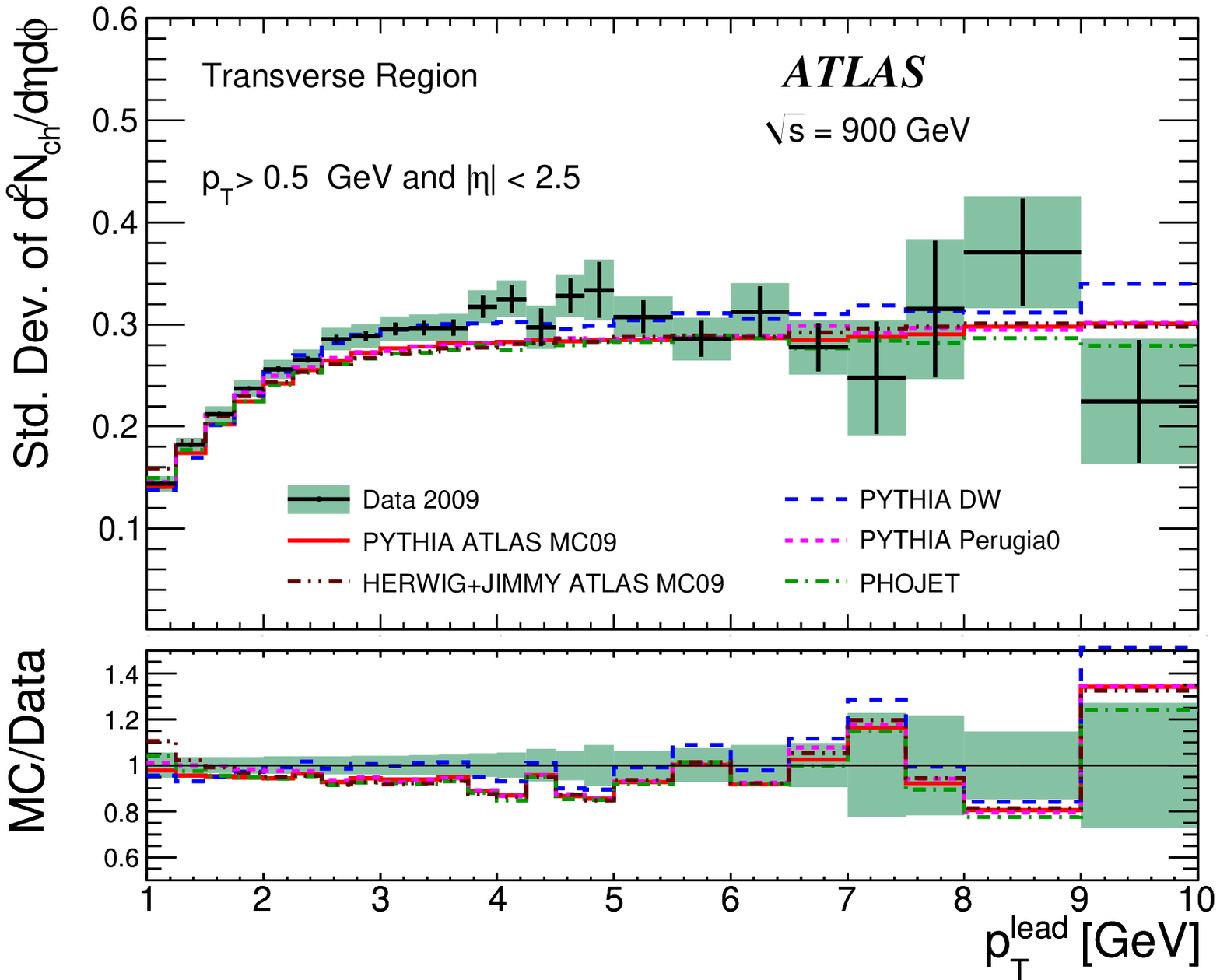}\hfill
   \includegraphics[width=.5\textwidth]{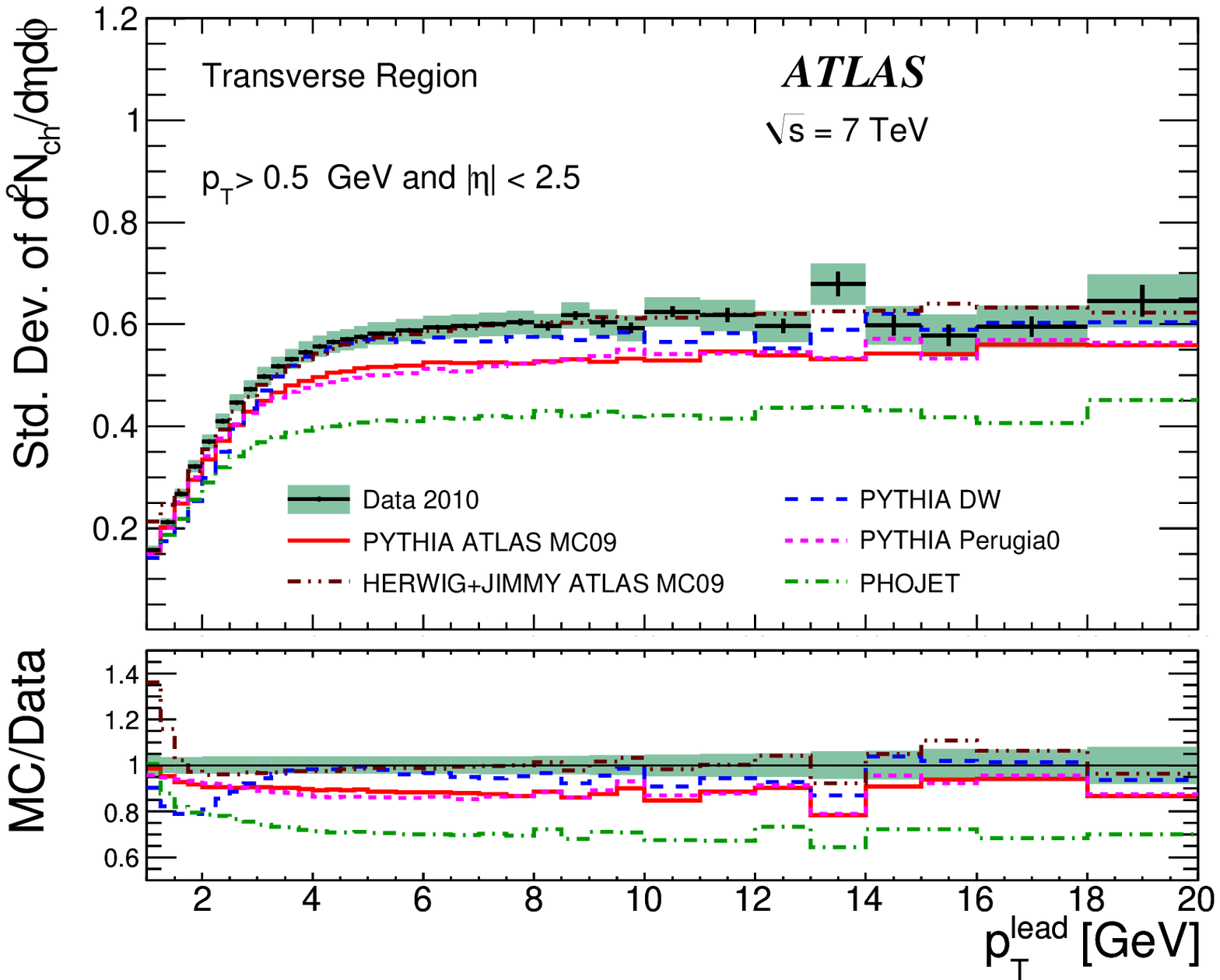}\hfill
   \includegraphics[width=.49\textwidth]{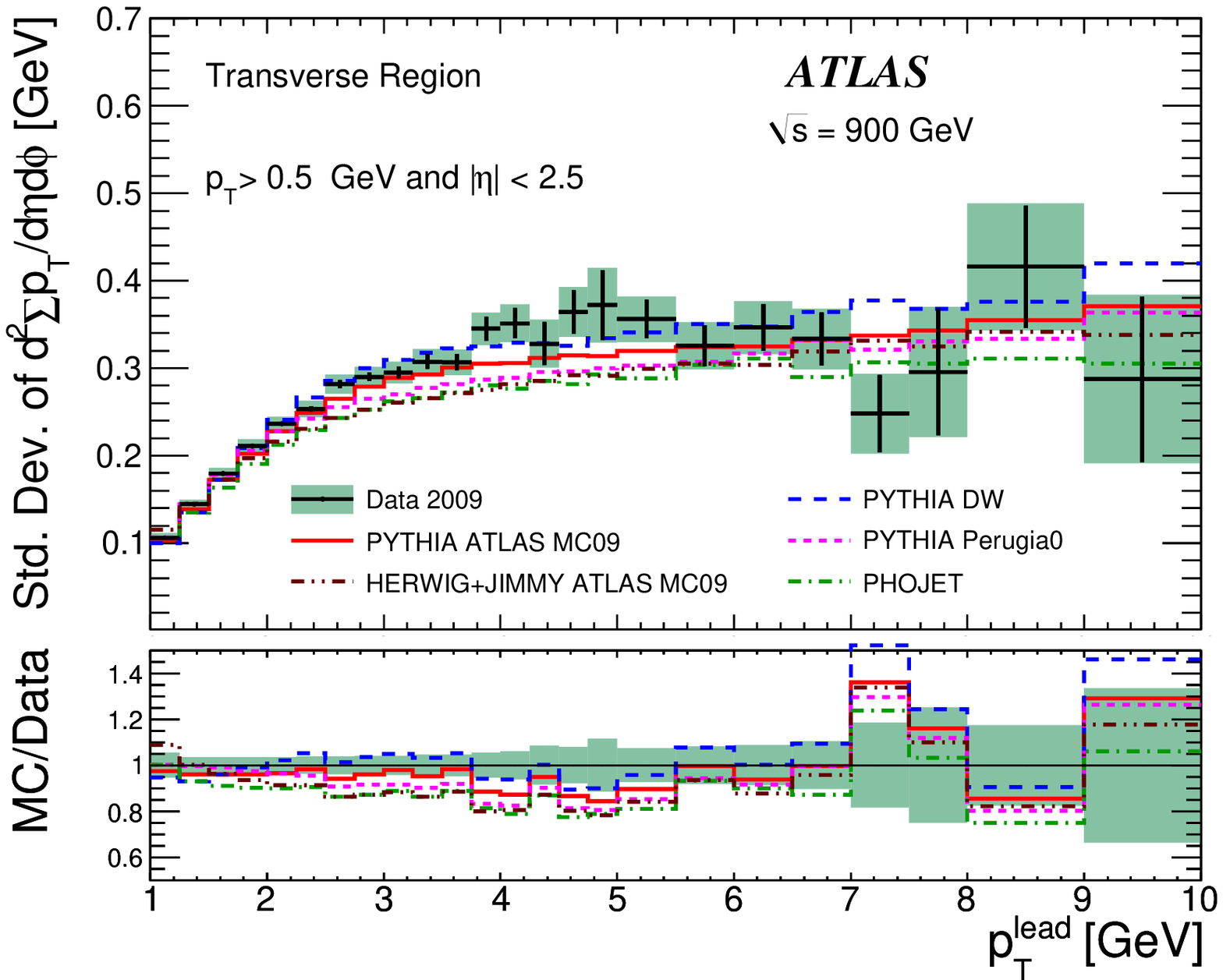} \hfill
   \includegraphics[width=.49\textwidth]{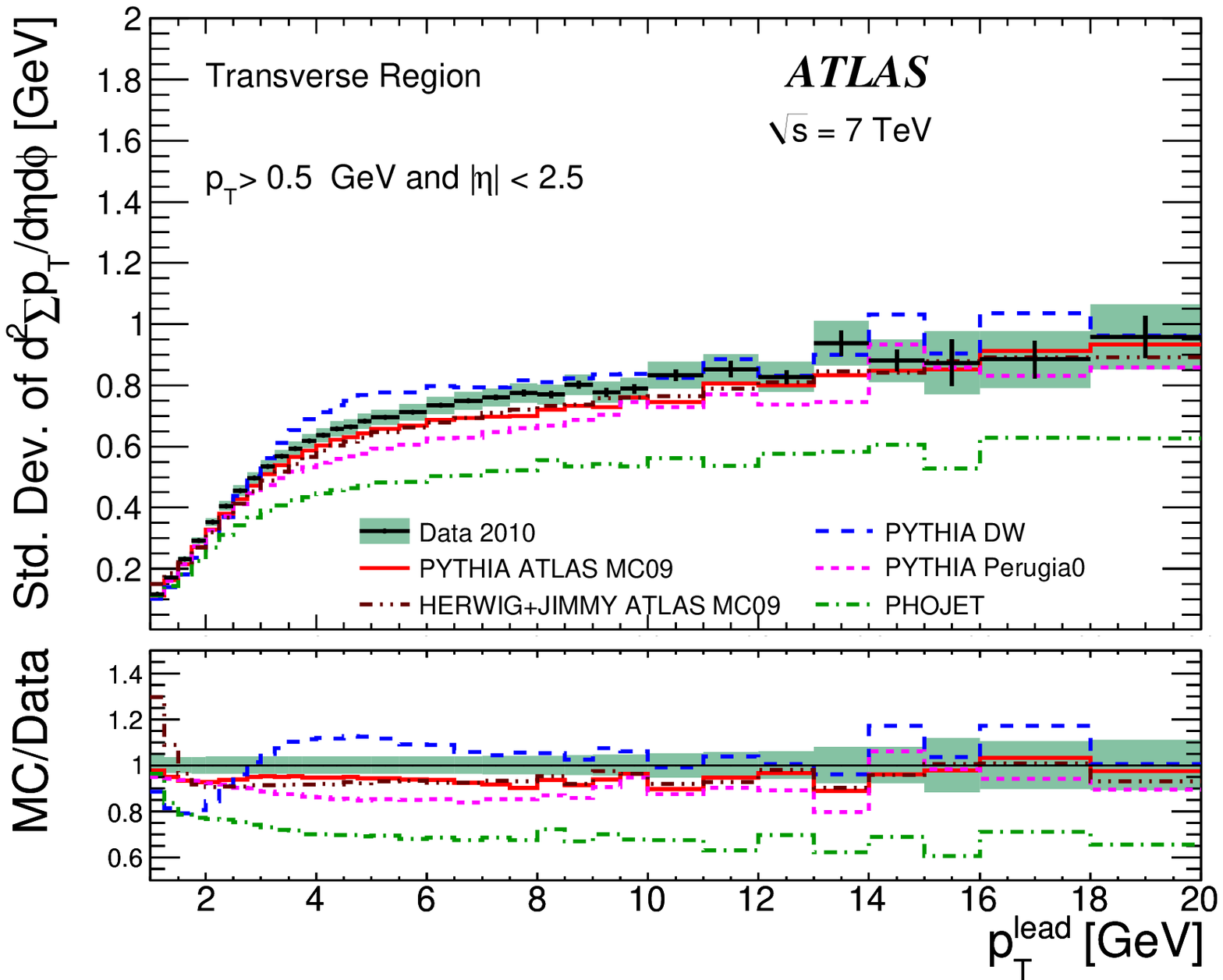}
 \end{center}
 \caption[]{ATLAS data at \unit{900}{\GeV} (left) and at \unit{7}{\TeV} (right)
   corrected back to the particle level, showing the standard deviation of the
   density of the charged particles \dNchgdetadphi (top row) and the standard
   deviation of the scalar $\ptsum$ density of charged particles
   \dpTsumdetadphi (bottom row) with $\pt > \unit{0.5}{\GeV}$ and $\etamod <
   2.5$, as a function of \ptlead, for the transverse region
   defined by the leading charged particle and compared with \Pythia ATLAS~MC09, DW and
   Perugia0 tunes, \HerwigJimmy ATLAS~MC09 tune, and \Phojet predictions. The error bars show
   the statistical uncertainty while the shaded area shows the combined
   statistical and systematic uncertainty.}
 \label{fig:SD}
\end{figure}

In \FigRef{fig:SD}, the standard deviation 
of the charged particle multiplicity and charged particle scalar \ptsum densities,
in the kinematic range $\pt > \unit{0.5}{\GeV}$ and $\etamod < 2.5$,
are shown against the leading charged particle \pt at $\sqrt{s} = \unit{900}{\GeV}$ and \unit{7}{\TeV}
(for the transverse region only).

The mean and standard deviation of the \pt
density in the transverse region characterize a range of additional energy that
jets might acquire if the underlying event were uniformly distributed.
As the error formula is neither trivial nor particularly standard, we
reproduce it here: for each bin, the sample variance of the variance of the
observable $x
\in \{\Nchg, \ptsum\}$ is $\text{var}(\text{var}(x)) = m_4(x) - 4 \,
m_3(x) \,
m_1(x) - m_2(x)^2 + 8 \, m_2(x) \, m_1(x)^2 - 4 \, m_1(x)^4$, where
$m_N(x) = \langle x^N \rangle$ is
the order $N$ moment of the distribution. This is then translated
into the standard error on the standard deviation of $x$ via error propagation
with a single derivative, giving symmetric errors of size
$\sqrt{\text{var}(\text{var}(x))/(n-2)} \, \big/ \,
2\!\sqrt{\text{var}(x)}$, where $n$ is
the number of entries in the bin.
The \unit{900}{\GeV} and \unit{7}{\TeV} data show the same trend.


The confirmation that the magnitude of the standard deviations of the
distributions are comparable to the magnitudes of the mean values indicates
that a subtraction of the underlying event from jets should be done on an
event by event basis, rather than by the subtraction of an invariant average value.
These distributions also
provide an additional constraint on generator models and tunes: the discrepancy
between models is much stronger at \unit{7}{\TeV} than at \unit{900}{\GeV}, with
\HerwigJimmy giving the best description and \Phojet in particular severely
undershooting the data at \unit{7}{\TeV}.


\subsection{Charged particle mean \pt}

\begin{figure}[pbt]
 \begin{center}
   \includegraphics[width=.5\textwidth]{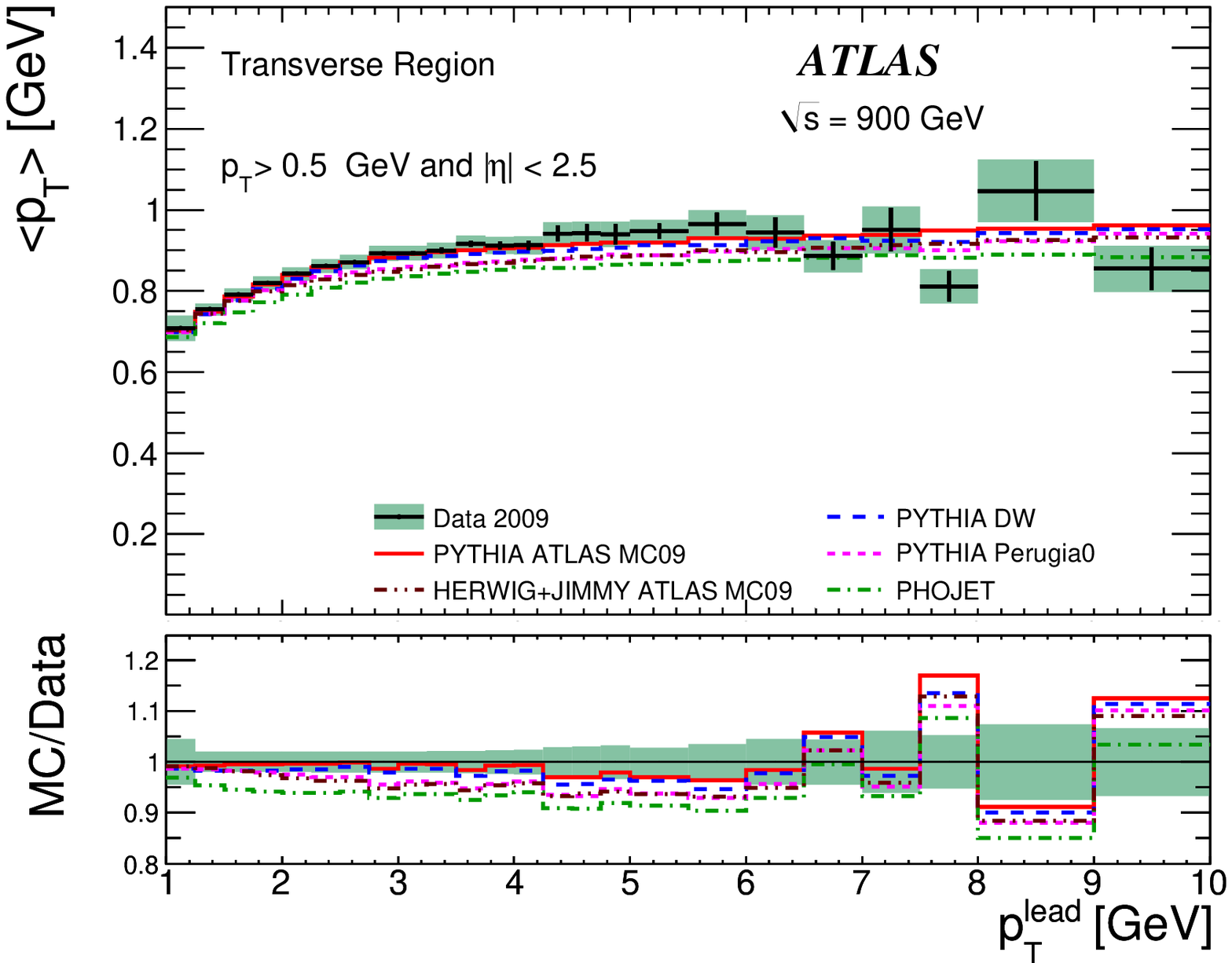}\hfill
   \includegraphics[width=.5\textwidth]{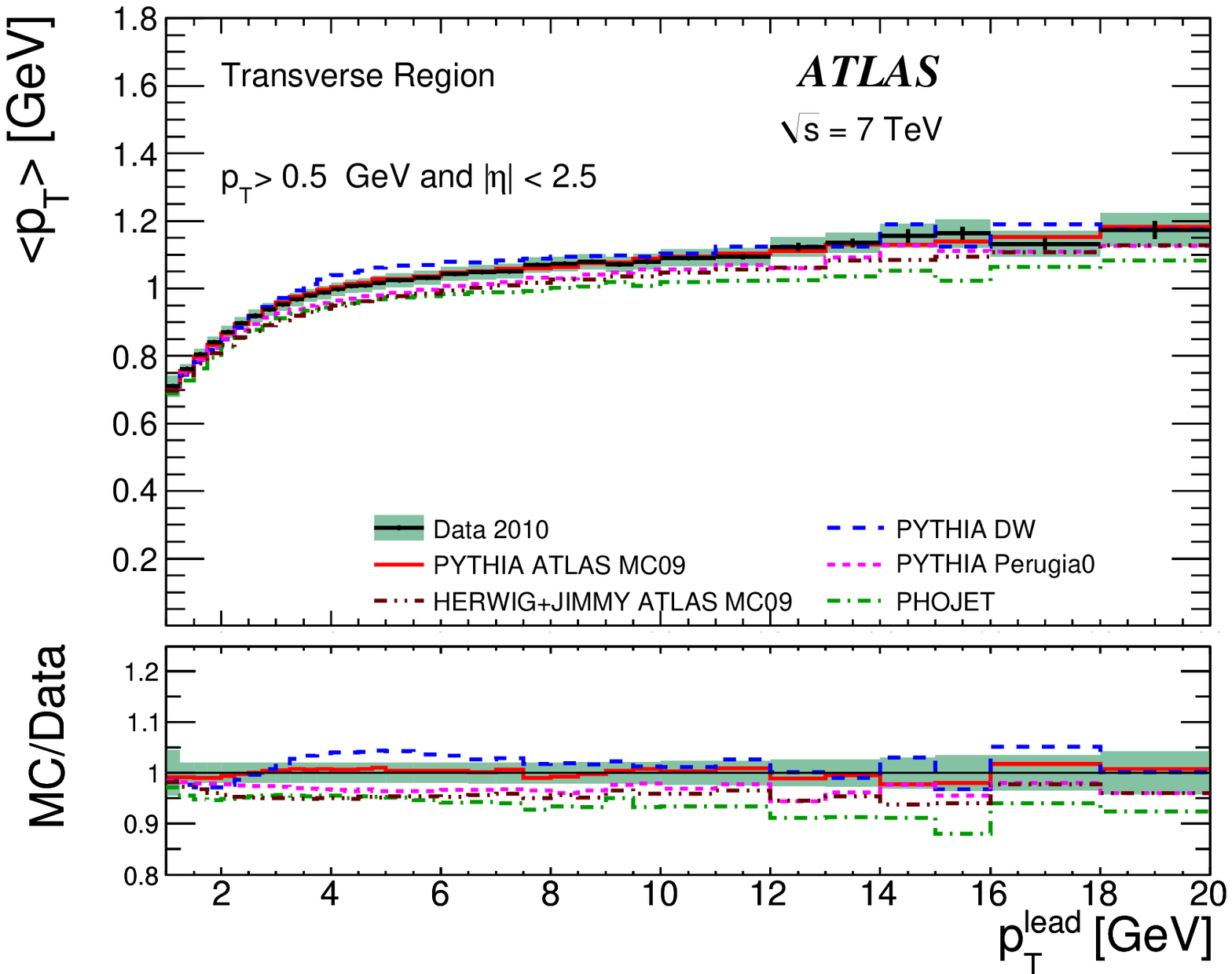}\hfill
   \includegraphics[width=.5\textwidth]{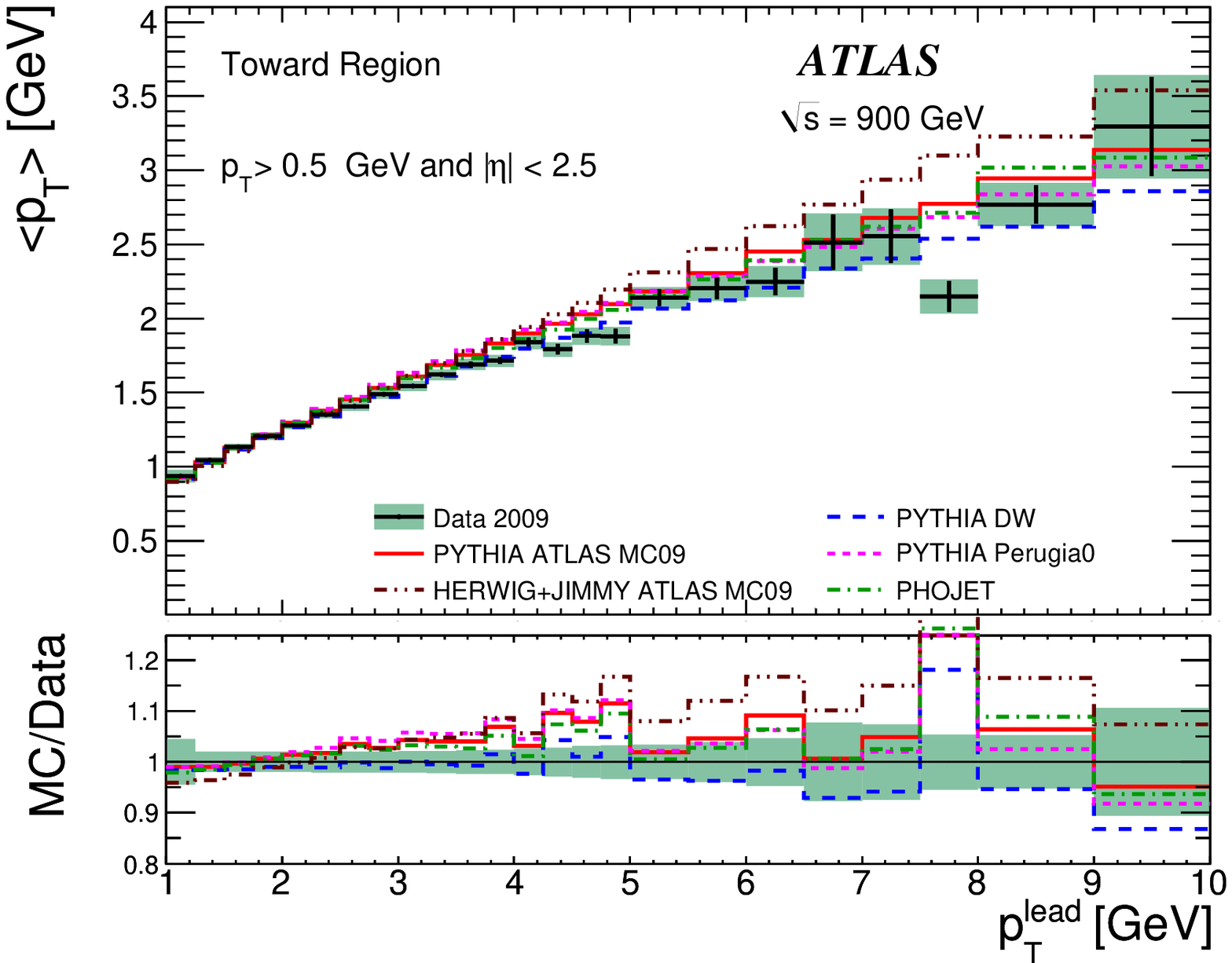}\hfill
   \includegraphics[width=.5\textwidth]{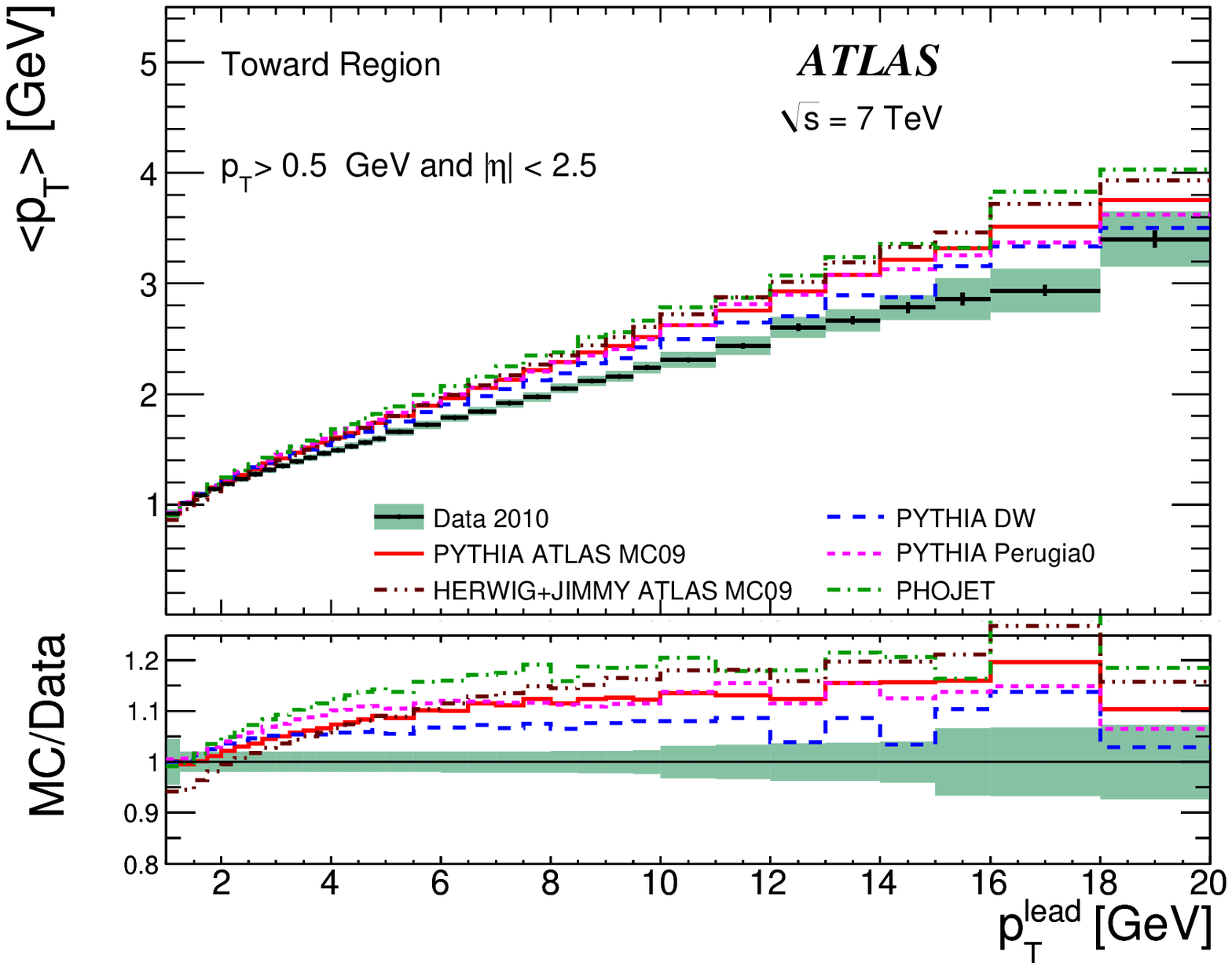}\hfill
   \includegraphics[width=.5\textwidth]{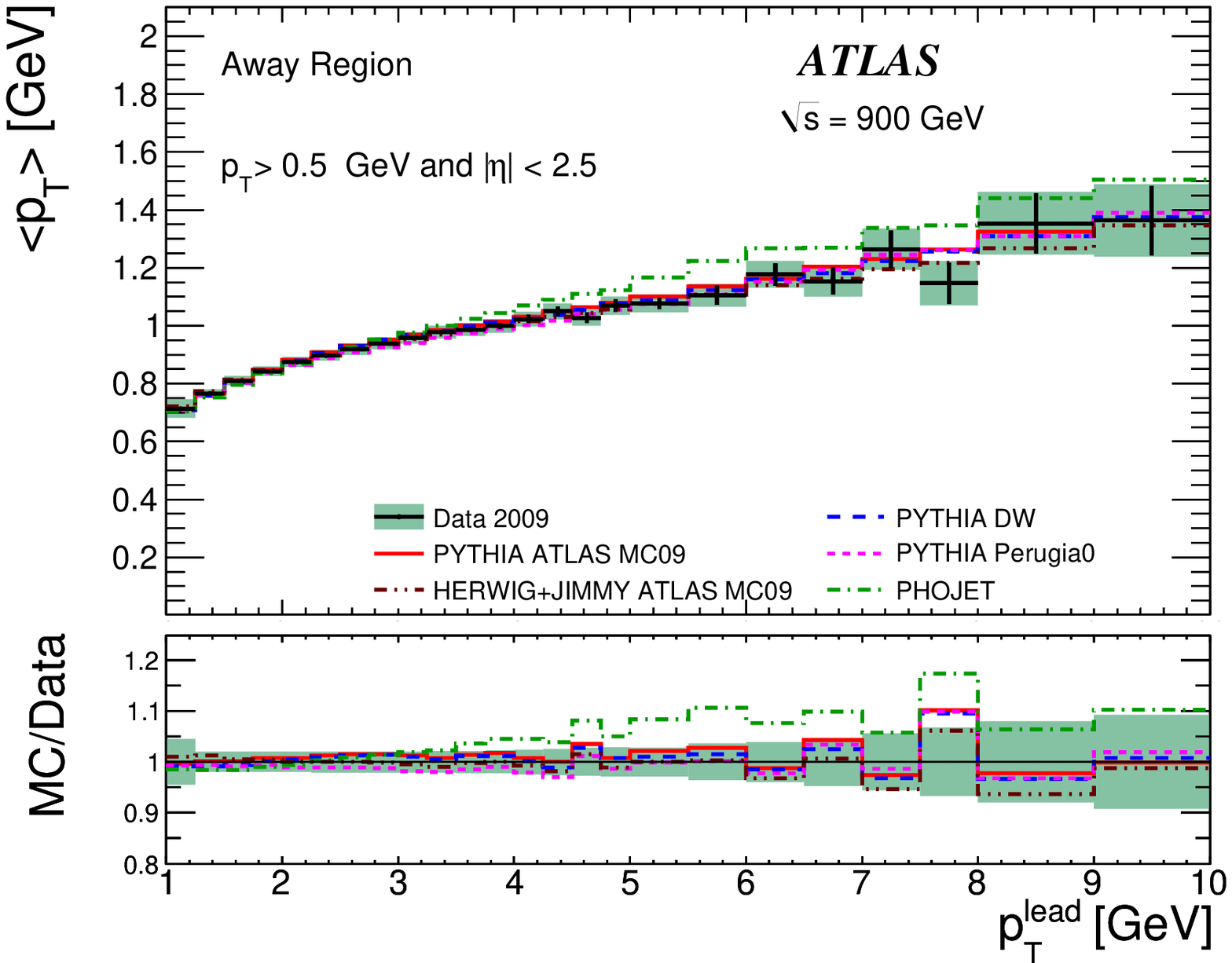}\hfill
   \includegraphics[width=.5\textwidth]{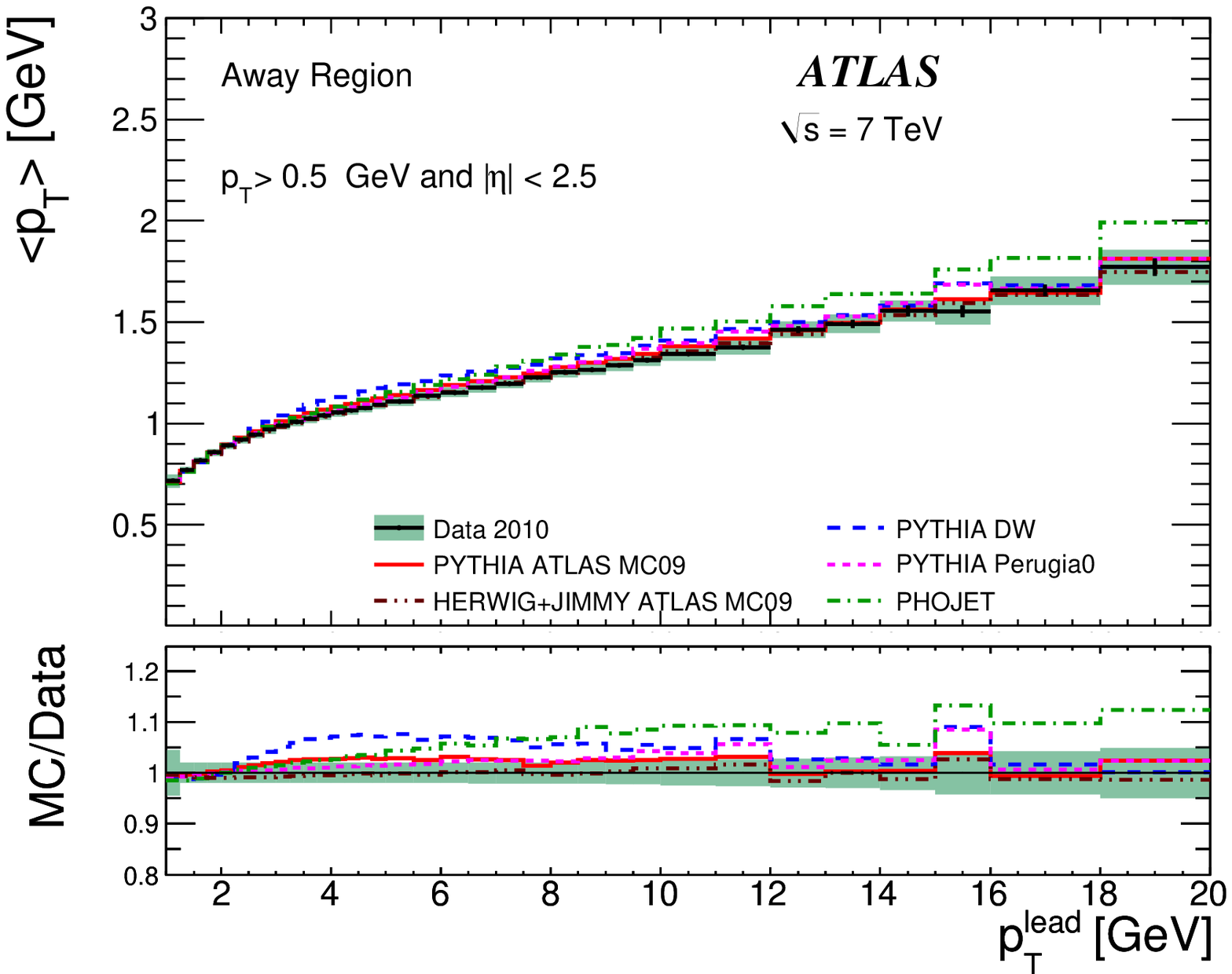}\
 \end{center}
 \caption[]{ATLAS data at \unit{900}{\GeV} (left) and at \unit{7}{\TeV} (right)
   corrected back to particle level, showing the mean \pt of the charged
   particles with $\pt > \unit{0.5}{\GeV}$ and $\etamod < 2.5$, as a function of \ptlead.
   The data are compared with
   \Pythia ATLAS~MC09, DW and Perugia0 tunes, \Herwig+ \Jimmy ATLAS~MC09 tune, and \Phojet predictions.
   The top, middle and the bottom rows, respectively, show the transverse,
   toward and away regions defined by the leading charged particle.
   The error bars show the statistical uncertainty while the shaded area shows the combined
   statistical and systematic uncertainty.}
 \label{fig:cptavg}
\end{figure}

In \FigRef{fig:cptavg} the average charged particle \ptsum,
in the kinematic range $\pt > \unit{0.5}{\GeV}$ and $\etamod < 2.5$,
is shown as a function of \ptlead at
$\sqrt{s} = \unit{900}{\GeV}$ and \unit{7}{\TeV}. These plots were constructed on an event-by-event
basis by dividing the total charged particle \pt in each region by the number of
charged particles in that region,
requiring at least one charged particle in the considered region.


All the MC tunes, except \Pythia tune DW, show somewhat lower mean \pt than the data in
the plateau part of the transverse region and overestimate the data in the toward and away regions.
The underlying event \ptmean is seen
to increase by about $20\%$ going from $\sqrt{s}=$\unit{900}{\GeV} to \unit{7}{\TeV},
again described by the MC models. There is relatively little
discrimination between MC models for this observable, all predictions are
within $\sim 10\%$ of the data values. The toward and away regions are dominated
by the jet-like rising profiles, in contrast to the plateau in the transverse
region. The toward region has a higher mean \pt than the away region since there
is higher probability of higher \pt particles being produced in association with
the leading charged particle.  The \unit{900}{\GeV} and \unit{7}{\TeV} data show
the same trend.



\subsection{Charged particle mean \pt and multiplicity correlations}
\label{sec:results:ptnch}

\begin{figure}[pbt]
 \begin{center}
   \includegraphics[width=.5\textwidth]{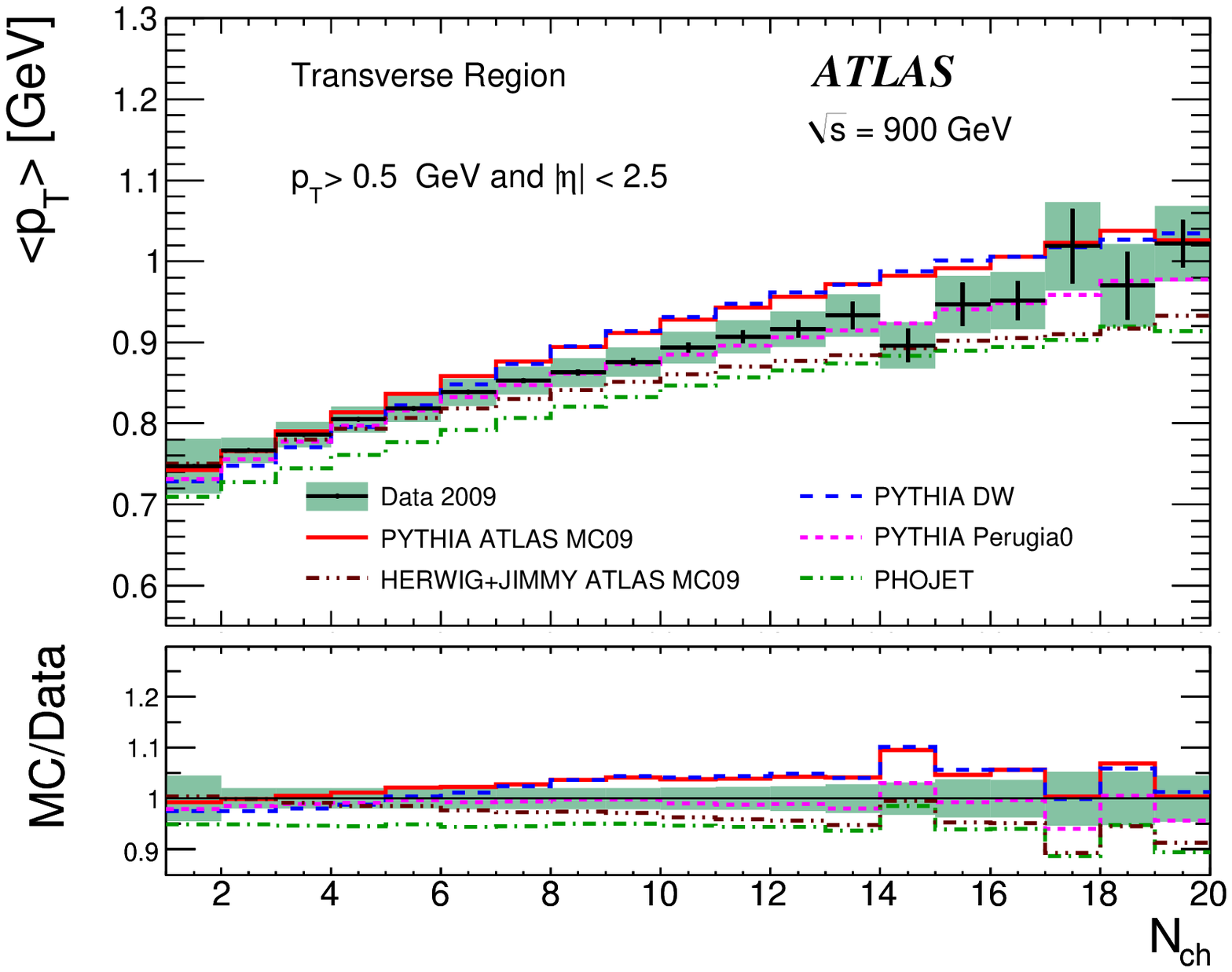}\hfill
   \includegraphics[width=.5\textwidth]{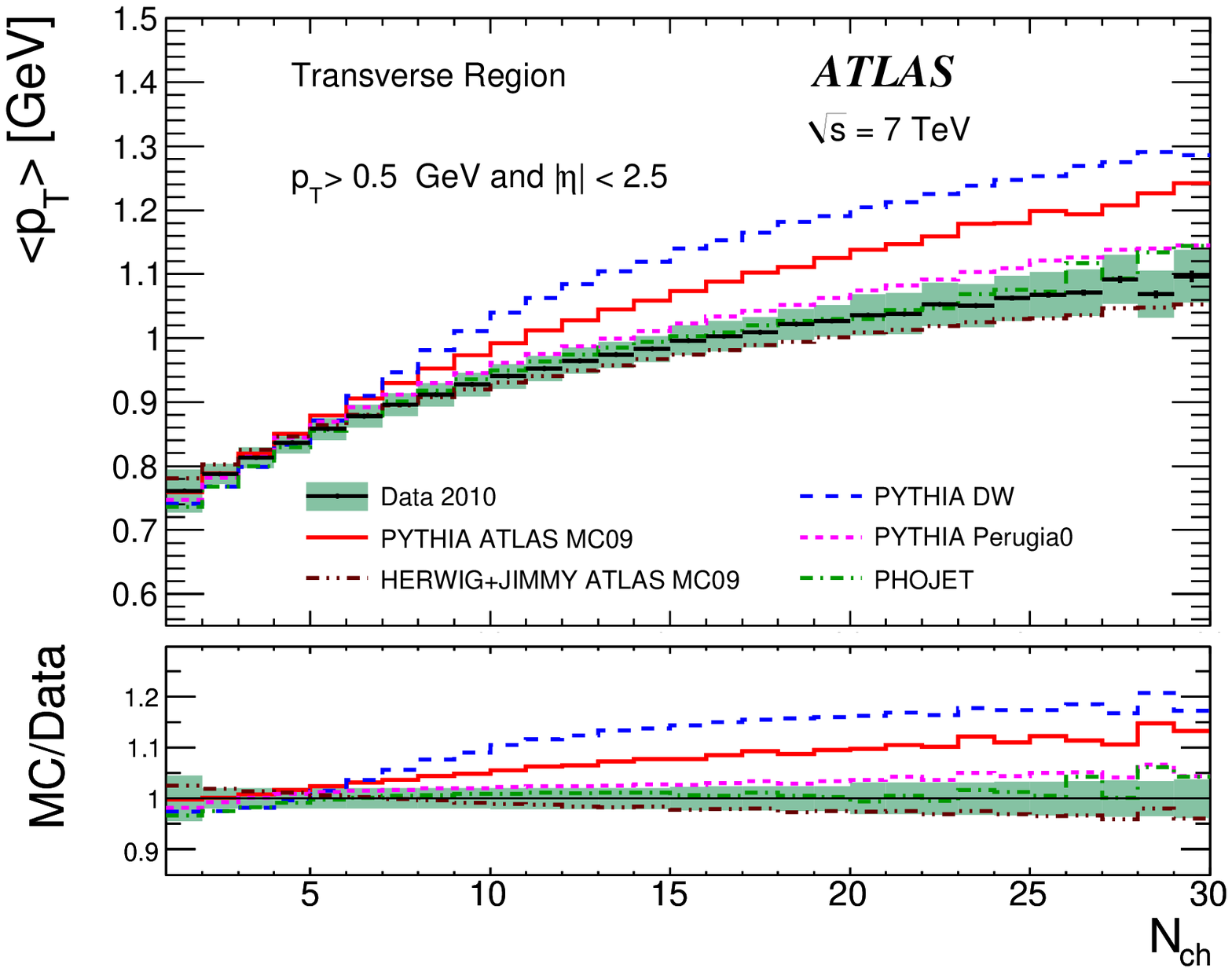}\hfill
   \includegraphics[width=.5\textwidth]{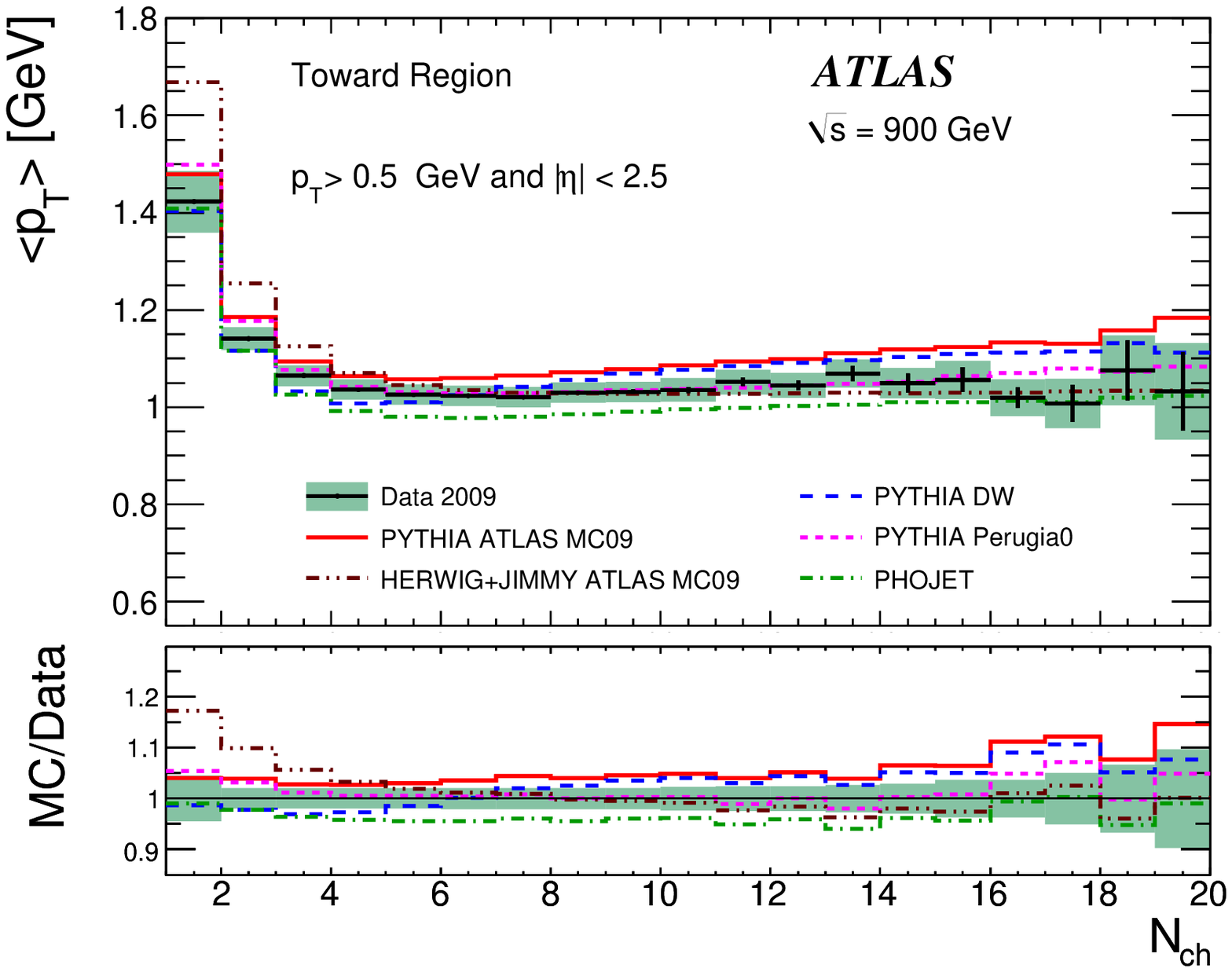}\hfill
   \includegraphics[width=.5\textwidth]{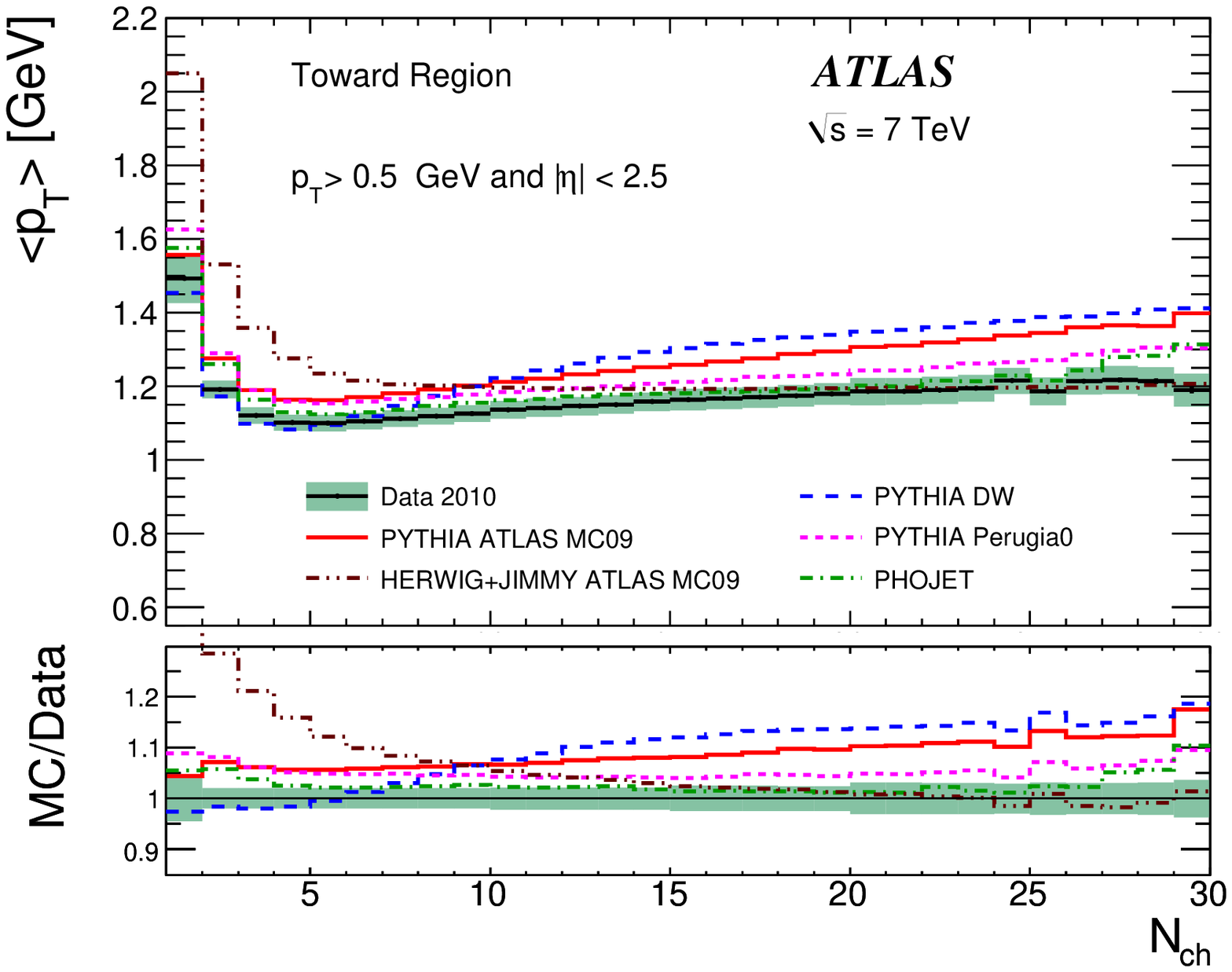}\hfill
   \includegraphics[width=.5\textwidth]{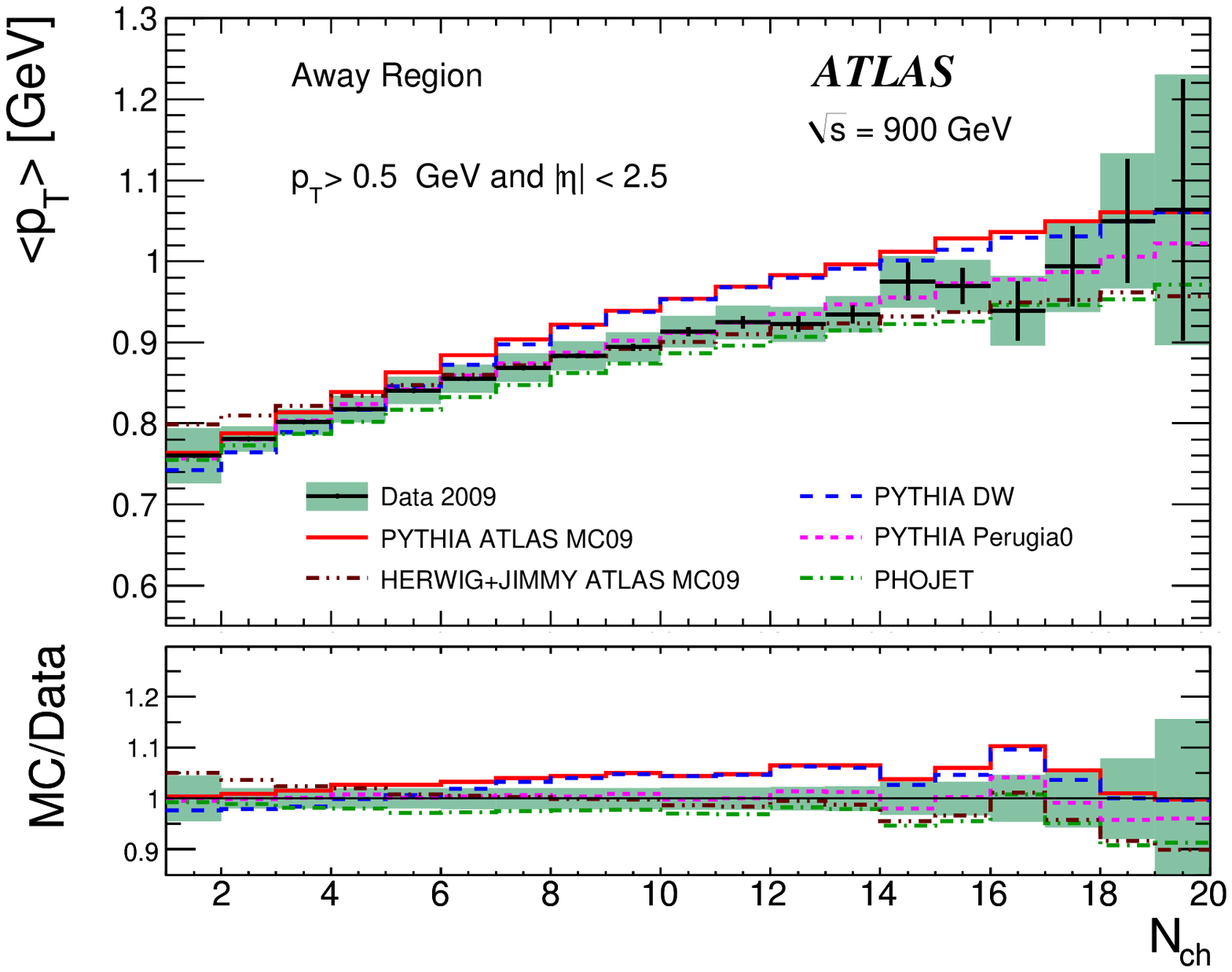}\hfill
   \includegraphics[width=.5\textwidth]{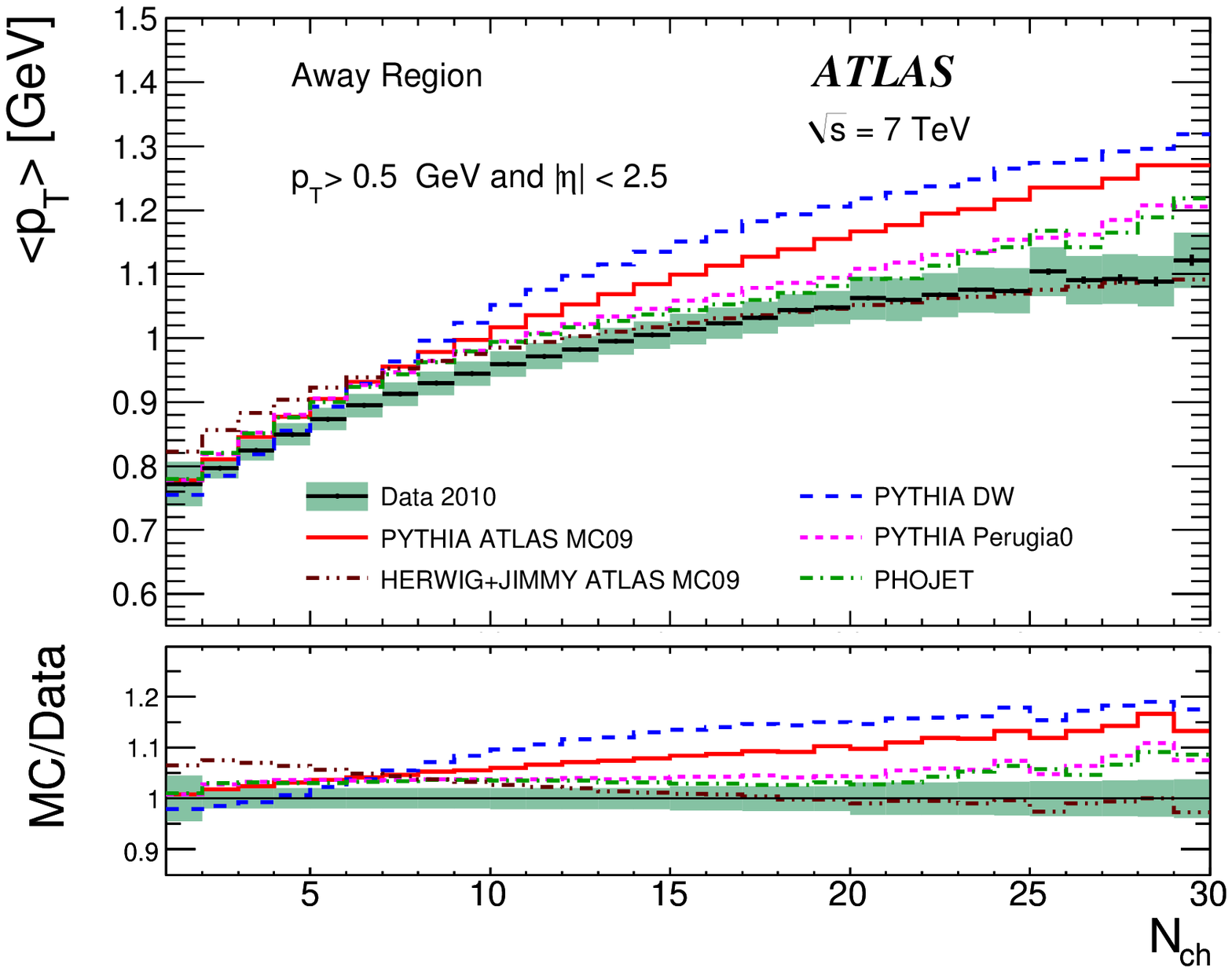}\
 \end{center}
 \caption[]{ATLAS data at \unit{900}{\GeV} (left) and at \unit{7}{\TeV} (right)
   corrected back to particle level, showing the mean \pt of the charged
   particles against the charged multiplicity, for charged particles with $\pt >
   \unit{0.5}{\GeV}$ and $\etamod < 2.5$.  The data are compared with \Pythia
   ATLAS~MC09, DW and Perugia0 tunes, \Herwig+ \Jimmy ATLAS~MC09 tune, and
   \Phojet predictions.  The top, middle and the bottom rows, respectively, show
   the transverse, toward and away regions defined by the leading charged
   particle.  The error bars show the statistical uncertainty while the shaded
   area shows the combined statistical and systematic uncertainty.}
 \label{fig:corr}
\end{figure}

The correlation between the mean \pt of charged particles and the charged
particle multiplicity in each region is sensitive to the amount of hard (perturbative QCD)
versus soft (non-perturbative QCD) processes contributing to the UE.
This has previously been measured for inclusive minimum bias events by
CDF~\cite{Aaltonen:2009ne} and ATLAS~\cite{Collaboration:2010ir}.
We present this quantity in \FigRef{fig:corr} for
each of the azimuthal regions in the kinematic range $\pt > \unit{0.5}{\GeV}$ and $\etamod < 2.5$.


The profiles in the transverse and away regions are very similar, showing a
monotonic increase of \ptmean with \Nchg. The profile of the toward region is
different, as it is essentially determined by the requirement of a
track with $\pt > \unit{1}{\GeV}$. For $N_{ch} =1$, it
contains only the leading charged particle and as $N_{ch}$ is increased by
inclusion of soft charged particles the average is reduced. However, for $N_{ch} > 5$
jet-like structure begins to form, and a weak rise of the mean \pt is observed.
 The \unit{900}{\GeV} and \unit{7}{\TeV} data show the same trend.
Comparing the \unit{900}{\GeV} and \unit{7}{\TeV} data, it is seen that the mean
charged particle \pt vs. \Nchg profiles are largely independent of the energy scale of
the collisions.

The MC models again show most differentiation for the \unit{7}{\TeV} data, and
it is interesting to see that the \HerwigJimmy model describes the data
well at this center-of-mass energy -- better than either the DW or ATLAS~MC09
\Pythia tunes (which both substantially overshoot at \unit{7}{\TeV}) and
comparably to the Perugia0 \Pythia tune. \Phojet gives the best description at
\unit{7}{\TeV}. However, both \HerwigJimmy and \Phojet undershoot the
transverse region data at \unit{900}{\GeV}, so no robust conclusion can be drawn
about the relative qualities of the models.



\subsection{Angular distributions}

\begin{figure}[pbt]
 \begin{center}
   \includegraphics[width=.7\textwidth]{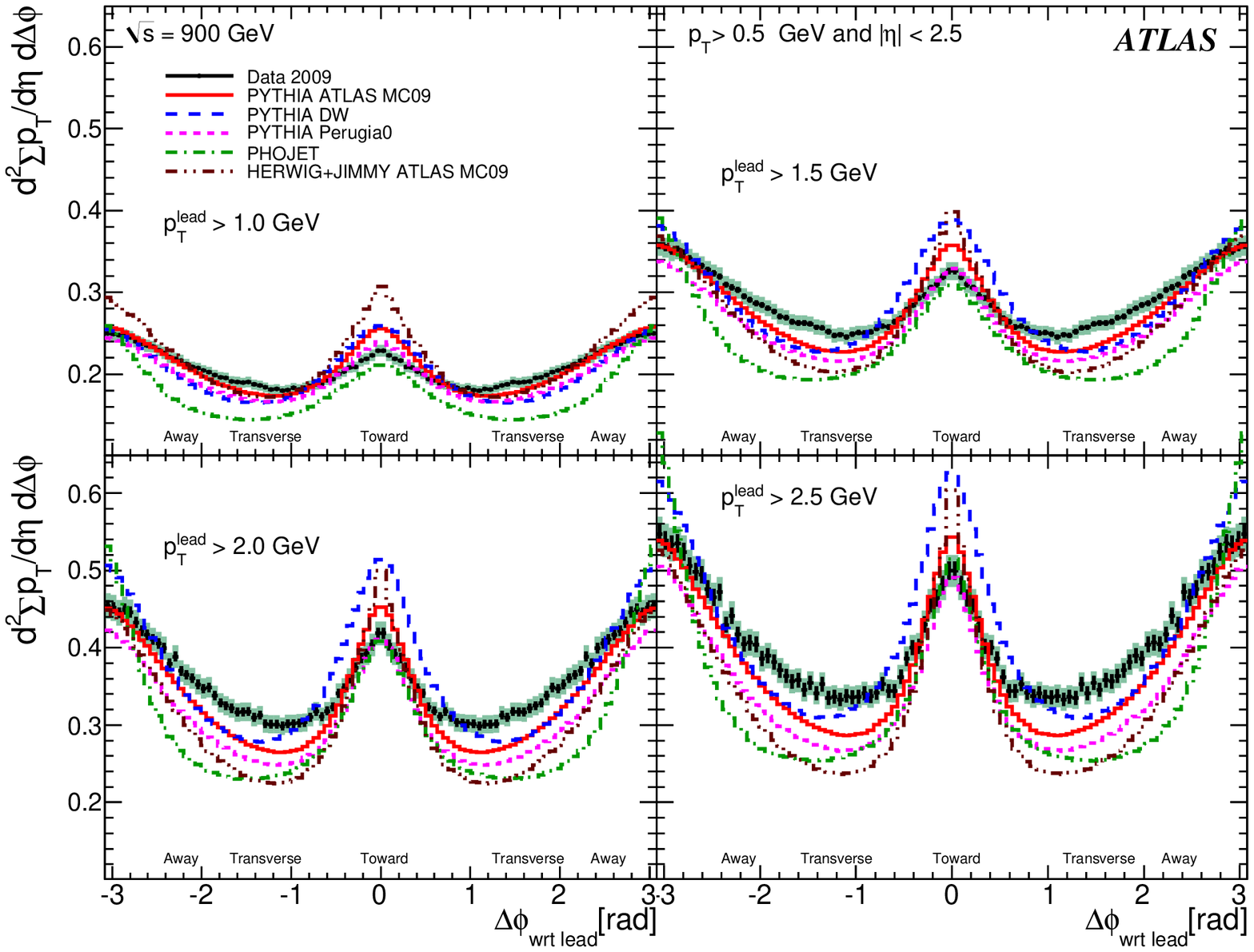}
   \includegraphics[width=.7\textwidth]{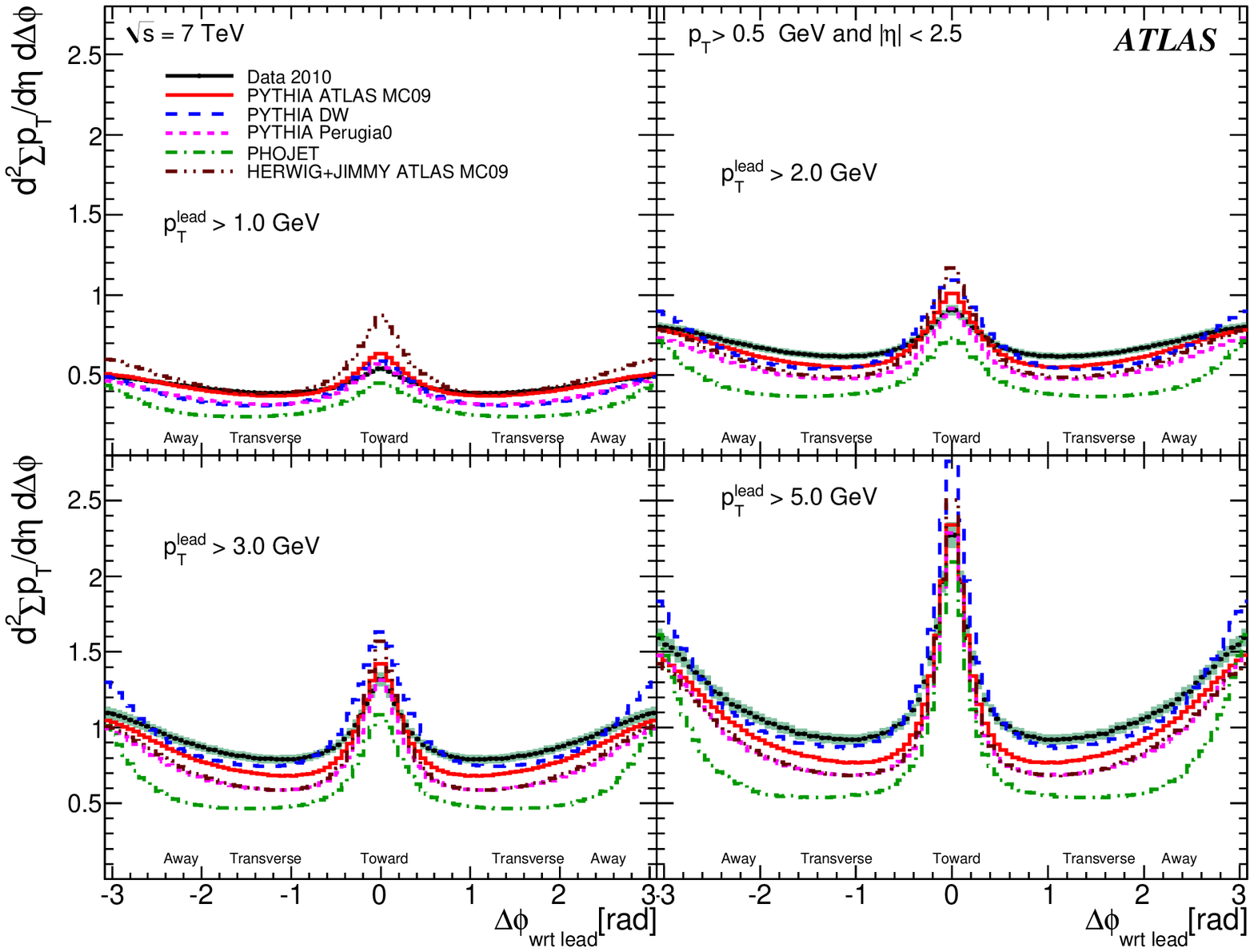}
 \end{center}
 \caption[]{ATLAS data at \unit{900}{\GeV} (top) and at \unit{7}{\TeV} (bottom)
   corrected back to the particle level, showing the $\phi$ distribution of
   charged particle \pt densities $\mathrm{d}^2 \pt
   /\mathrm{d}\eta\,\mathrm{d}\Delta \phi$ with respect
   to the leading charged particle (at $\Delta\phi = 0$), for $\pt >
   \unit{0.5}{\GeV}$ and $\etamod < 2.5$. The leading charged particle is
   excluded. The data are compared to MC predictions by the \Pythia ATLAS~MC09,
   DW and Perugia0 tunes, the \Herwig+\Jimmy ATLAS~MC09 tune, and \Phojet.
   The distributions obtained by restricting the
   minimum leading charged particle \pt to different values are shown separately.
   The plots have been symmetrized by reflecting them about $\Delta\phi=0$.
   The error bars show the statistical uncertainty while the shaded
   areas show the combined statistical and systematic uncertainty
   corresponding to each \pt lower cut value.}
 \label{fig:deltaphipt}
\end{figure}

The angular distributions with respect to the leading charged particle of the charged
particle number and \ptsum densities at $\sqrt{s} =
\unit{900}{\GeV}$ and \unit{7}{\TeV}, with charged particle $\pt > \unit{0.5}{\GeV}$, are
plotted in \FigRef[s]{fig:deltaphi} and \ref{fig:deltaphipt}. The leading charged particle taken to be at $\Delta\phi = 0$ has
been excluded from the distributions. The data are shown for four different
lower cut values in leading charged particle \pt. These distributions are constructed by
reflecting $|\Delta\phi|$ about zero, i.e. the region $-\pi \le \Delta\phi < 0$ is an exact
mirror image of the measured $|\Delta\phi|$ region shown in $0 \le \Delta\phi
\le \pi$.


\begin{figure}[pbt]
 \begin{center}
   \includegraphics[width=.7\textwidth]{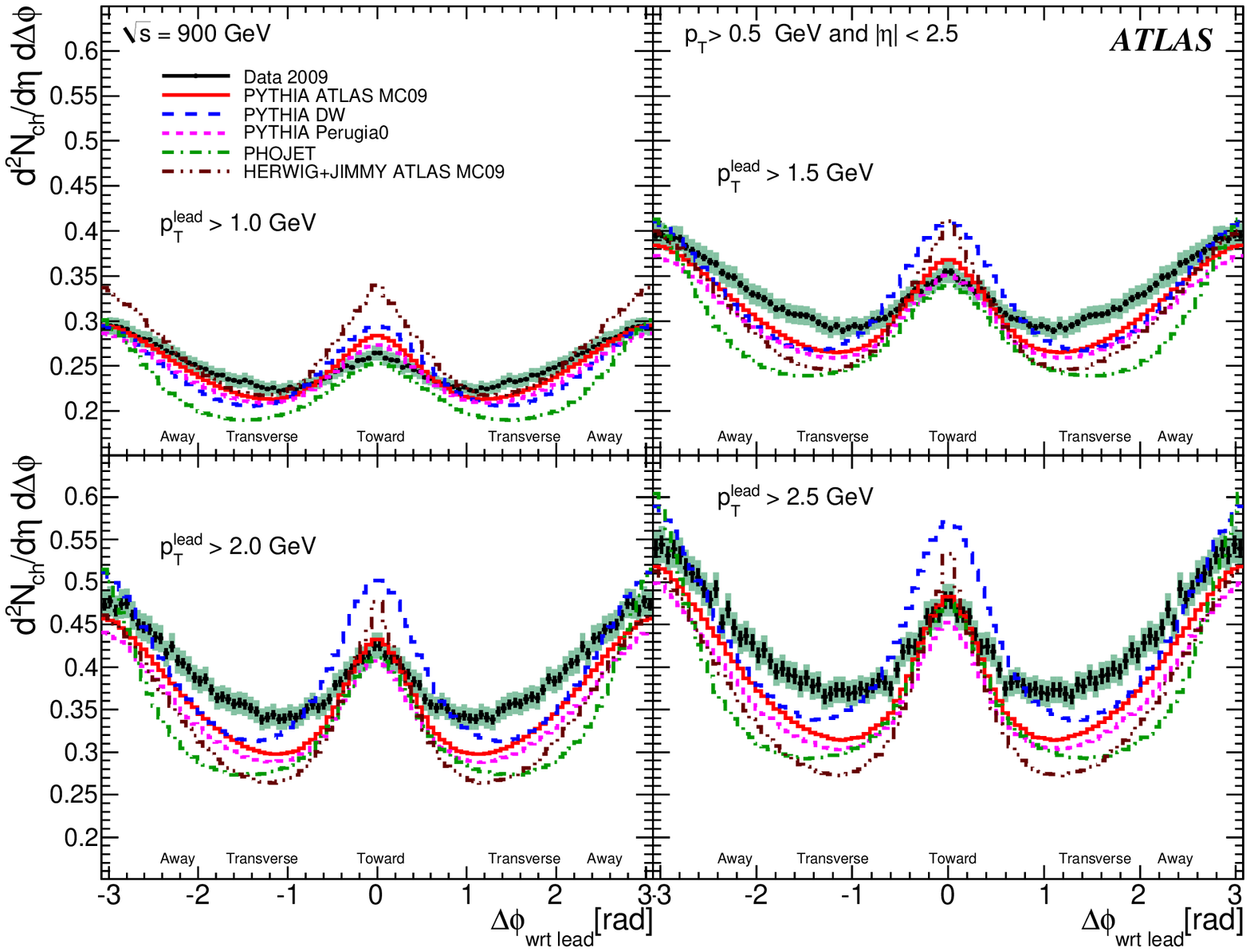}
   \includegraphics[width=.7\textwidth]{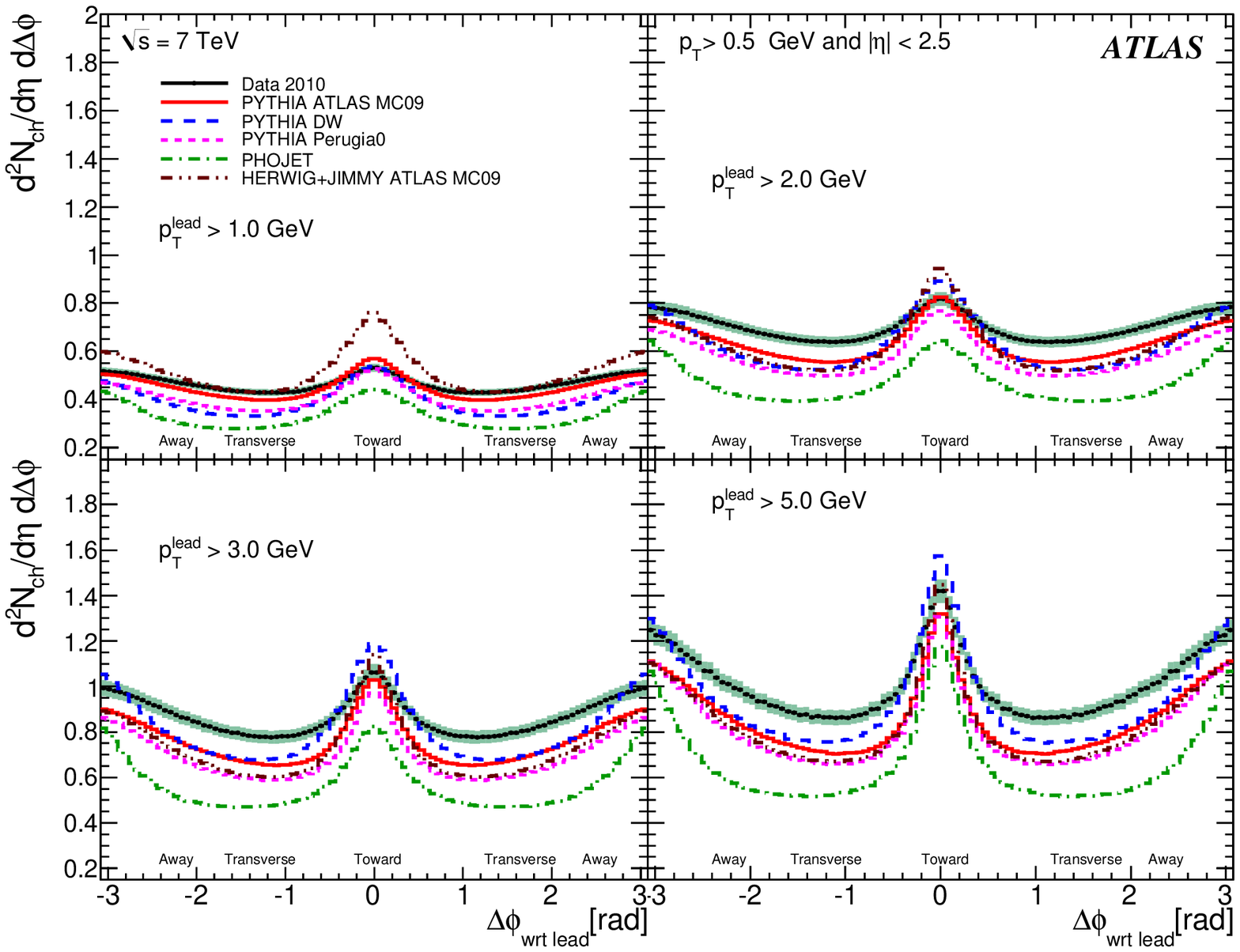}
 \end{center}
 \caption[]{ATLAS data at \unit{900}{\GeV} (top) and at \unit{7}{\TeV} (bottom)
   corrected back to the particle level, showing the $\phi$ distribution of
   charged particle densities
   $\mathrm{d}^2N_\text{ch}/\mathrm{d}\eta\,\mathrm{d}\Delta \phi$ with respect
   to the leading charged particle (at $\Delta\phi = 0$), for $\pt >
   \unit{0.5}{\GeV}$ and $\etamod < 2.5$. The leading charged particle is
   excluded.  The data are compared to MC predictions by the \Pythia ATLAS~MC09,
   DW and Perugia0 tunes, the \Herwig+\Jimmy ATLAS~MC09 tune, and \Phojet. The
   distributions obtained by restricting the minimum leading charged particle
   \pt to different values are shown separately.  The plots have been symmetrized by
   reflecting them about $\Delta\phi=0$.  The error bars show the statistical
   uncertainty while the shaded areas show the combined statistical and
   systematic uncertainty corresponding to each \pt lower cut value.}
 \label{fig:deltaphi}
\end{figure}



These distributions show a significant difference in shape between data and MC
predictions.  With the increase of the leading charged particle \pt, the development of
jet-like structure can be observed, as well as the corresponding sharper rise in
transverse regions compared to the MC.  The saturation at higher \pt indicates
the plateau region seen in \FigRef[s]{fig:nchg} and \ref{fig:cptsum}. \Pythia
tunes essentially predict a stronger correlation than is seen in the data, and
this discrepancy in the toward region associated particle density was also observed
at CDF~\cite{:RDFlead}.

\subsection{Charged particle multiplicity and scalar \ptsum for lower \pt cut}
\label{sec:results:lowpt}

\begin{figure}[pbt]
 \begin{center}
   \includegraphics[width=.5\textwidth]{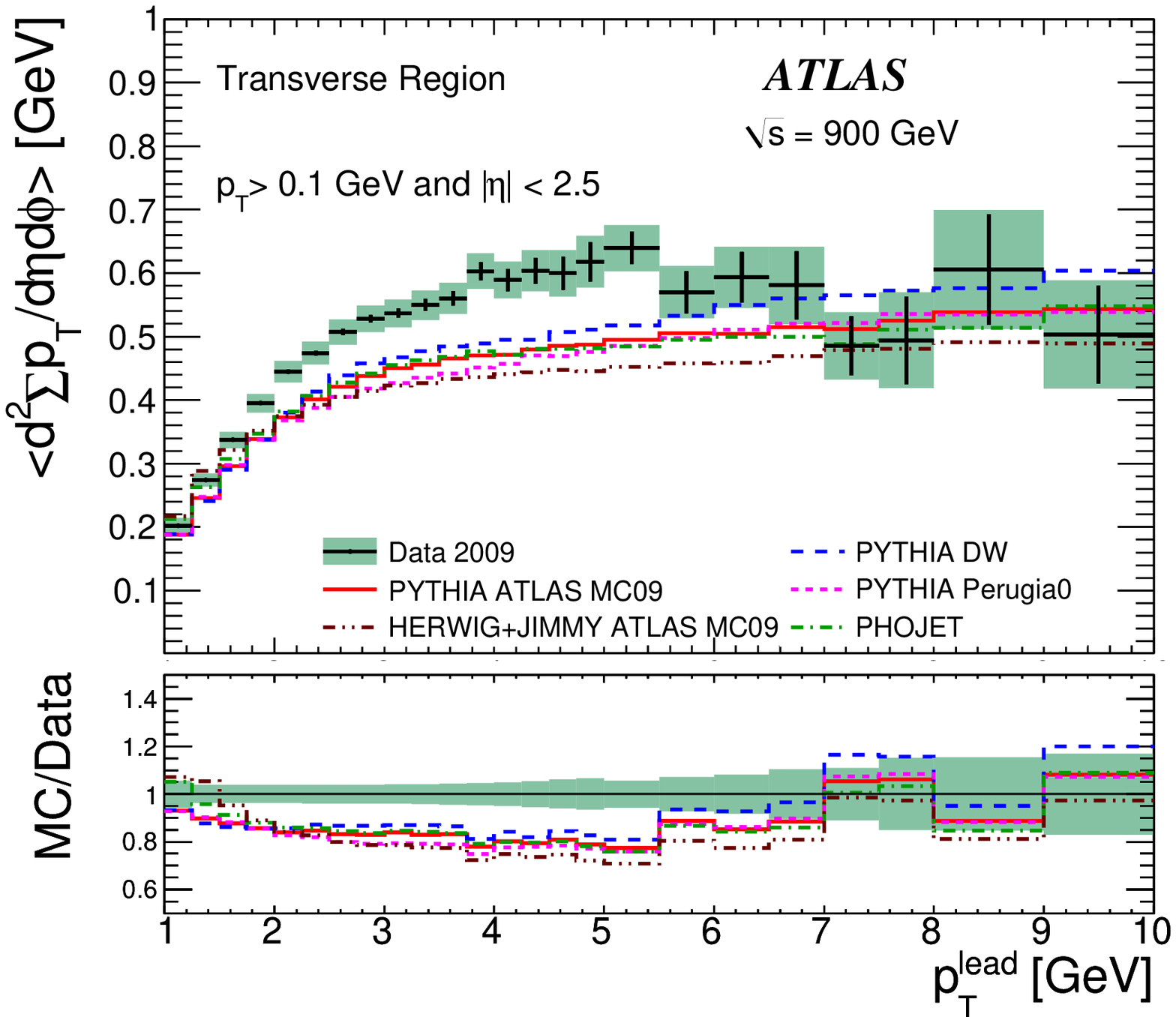}\hfill
   \includegraphics[width=.5\textwidth]{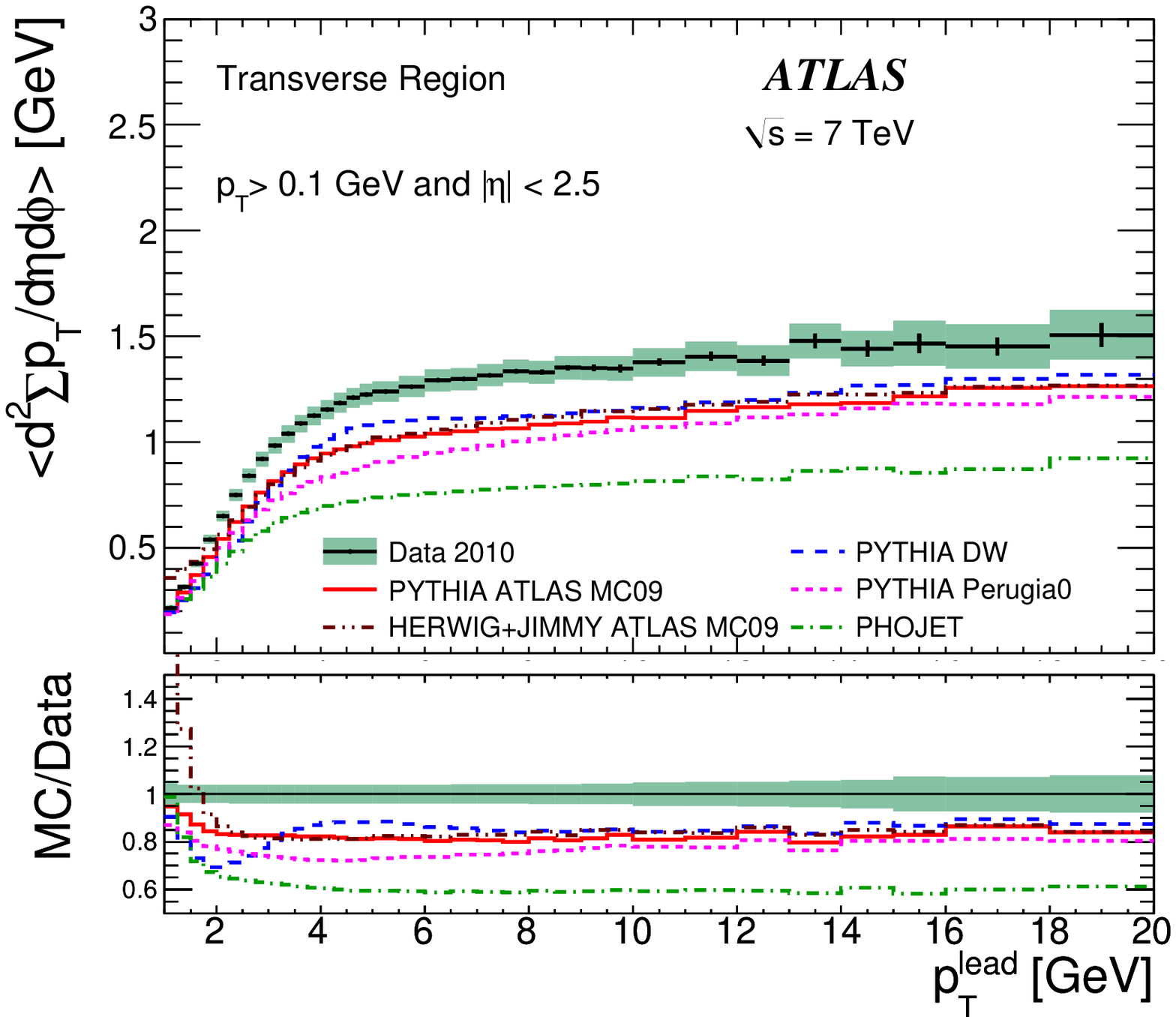}\hfill
   \includegraphics[width=.5\textwidth]{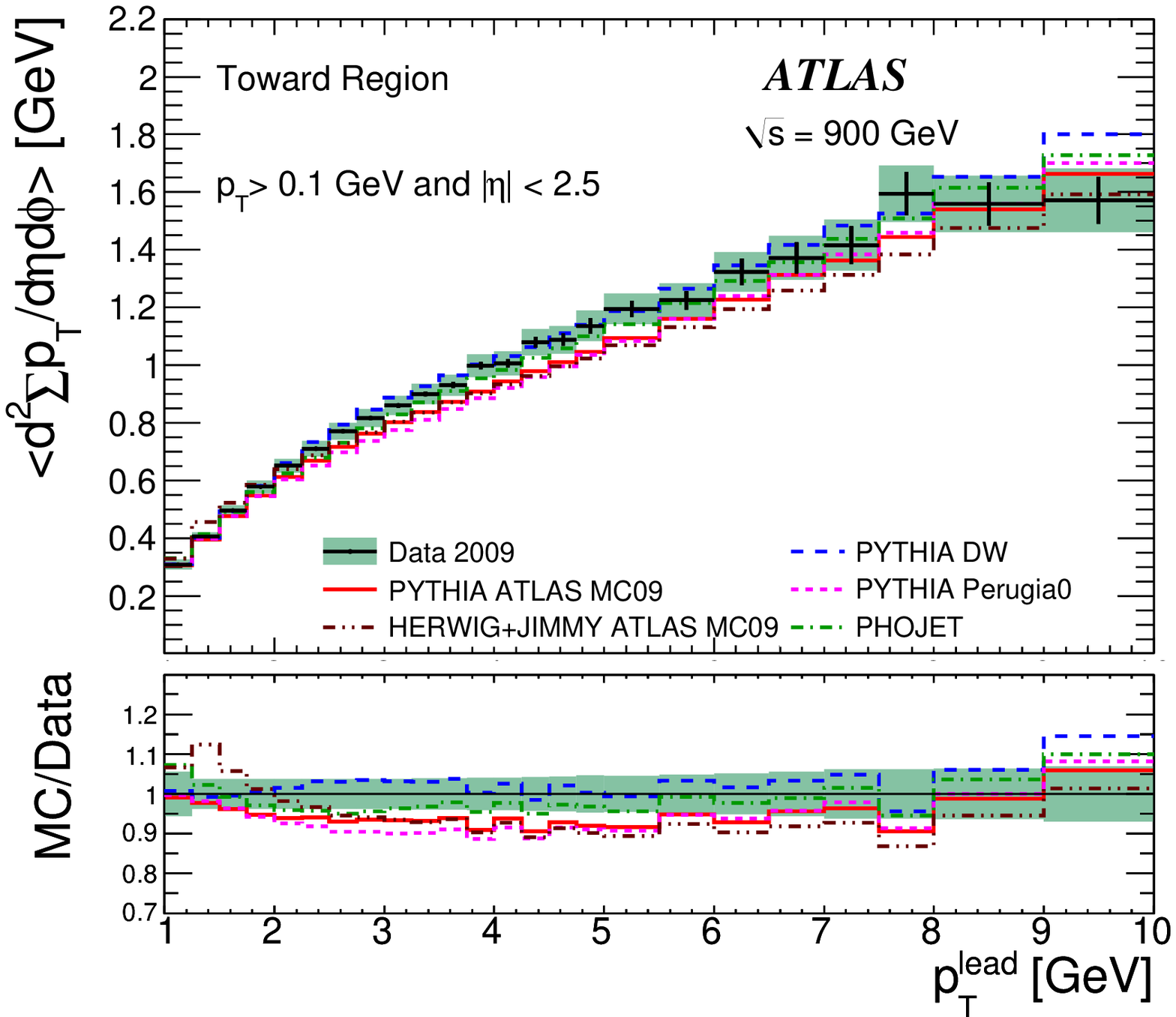}\hfill
   \includegraphics[width=.5\textwidth]{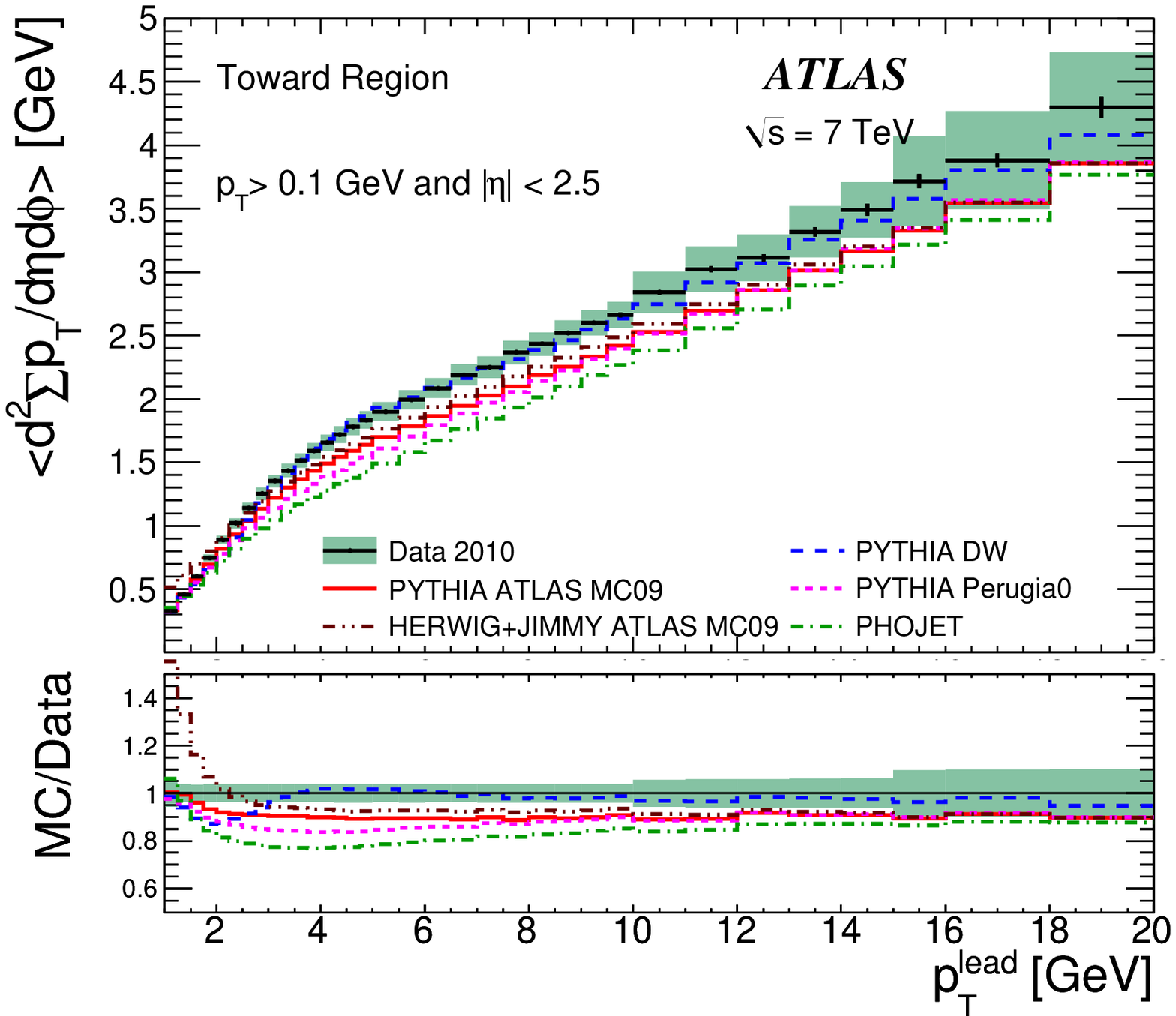}\hfill
   \includegraphics[width=.5\textwidth]{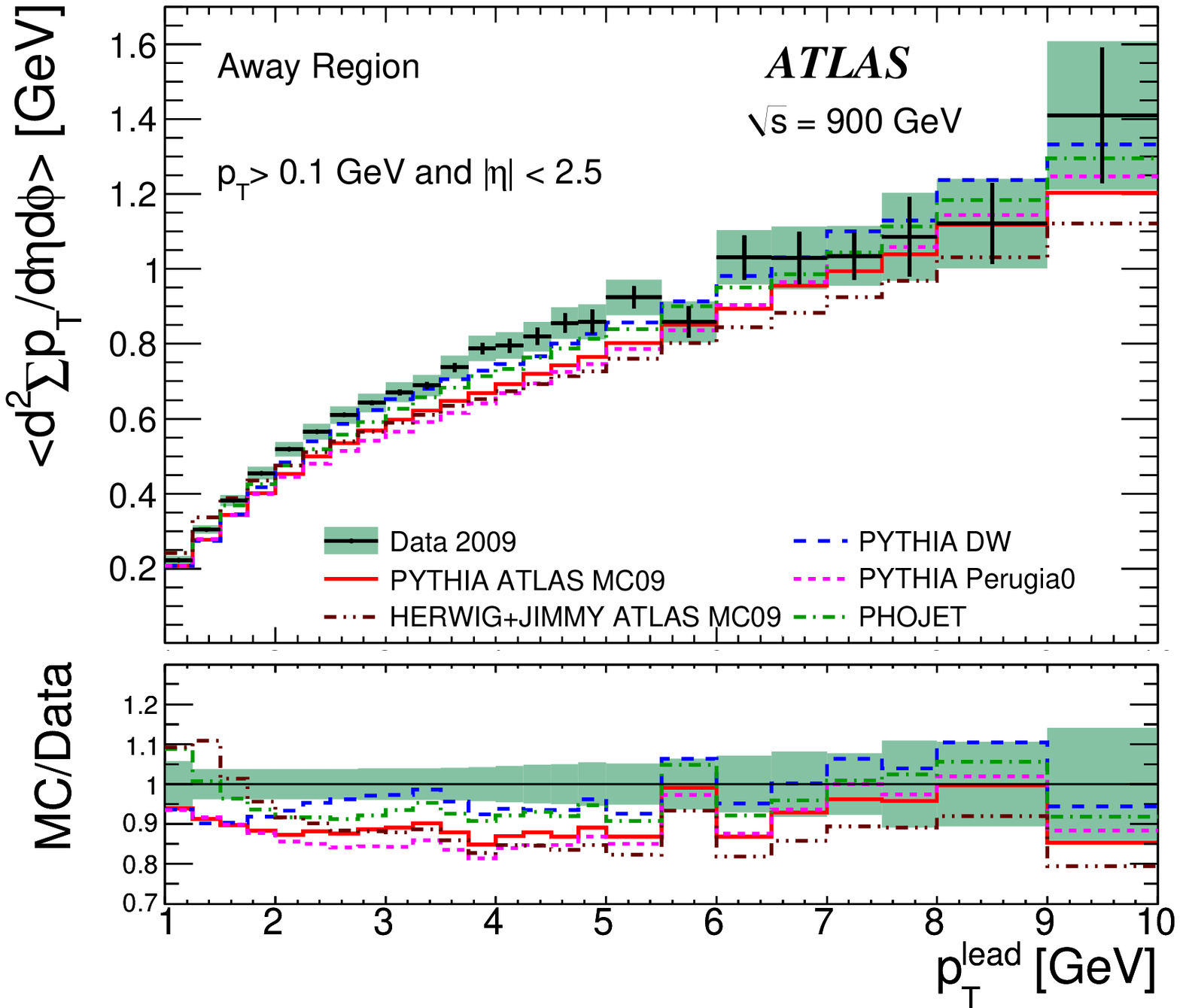}\hfill
   \includegraphics[width=.5\textwidth]{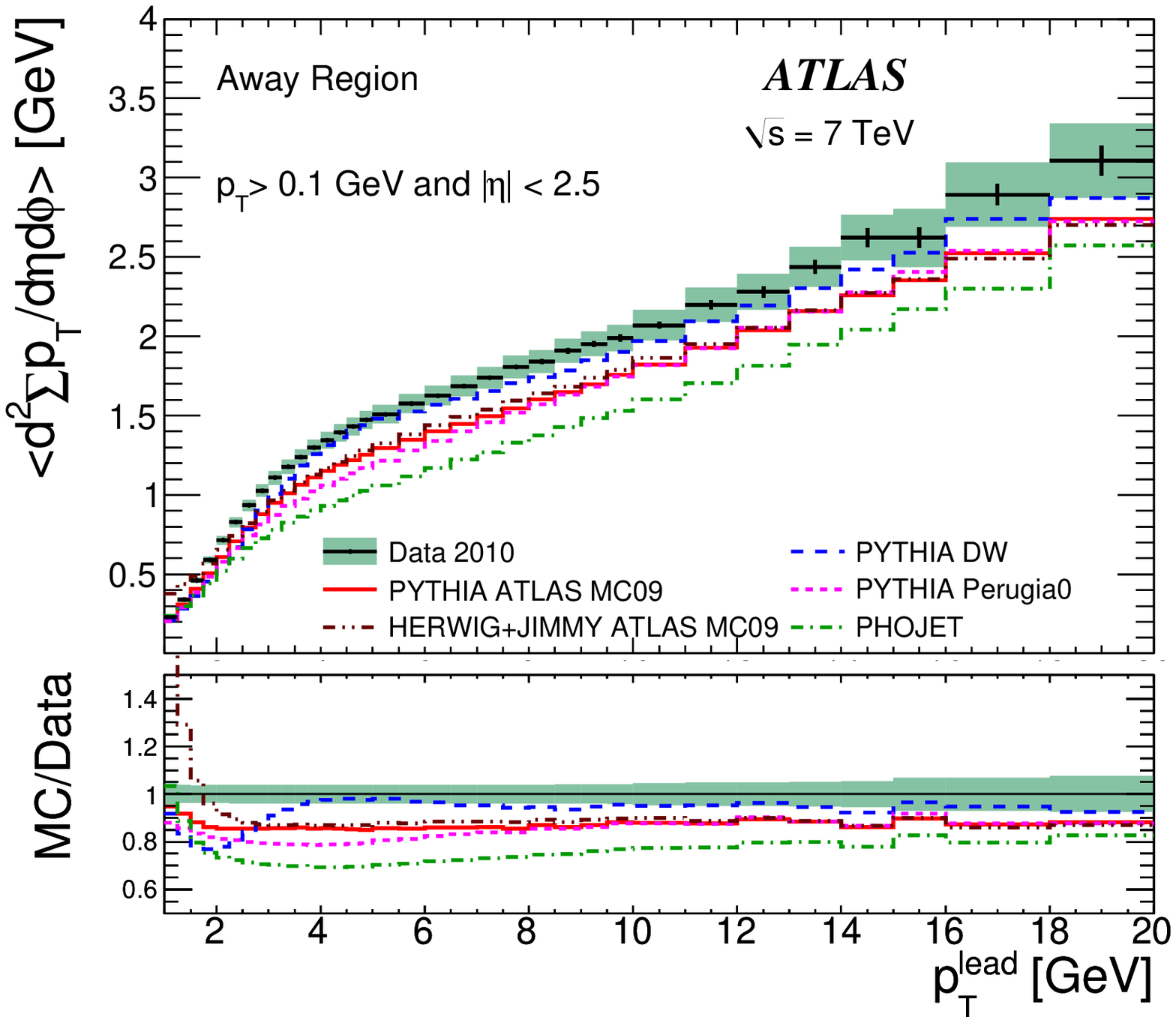}
 \end{center}
 \caption[]{ATLAS data at \unit{900}{\GeV} (left) and at \unit{7}{\TeV} (right)
   corrected back to particle level, showing the scalar $\ptsum$ density of the
   charged particles \dpTsumdetadphi with $\pt > \unit{0.1}{\GeV}$ and $\etamod
   < 2.5$, as a function of \ptlead.
   The data are compared with
   \Pythia ATLAS~MC09, DW and Perugia0 tunes, \Herwig+ \Jimmy ATLAS~MC09 tune and \Phojet predictions.
   The top, middle and the bottom rows, respectively, show the transverse,
   toward and away regions defined by the leading charged particle.
   The error bars show the statistical uncertainty while the shaded
   area shows the combined statistical and systematic uncertainty.}
 \label{fig:lowpt-cptsum}
\end{figure}

In \FigRef[s]{fig:lowpt-nchg} and \ref{fig:lowpt-cptsum}, the charged particle multiplicity density and charged
particle scalar \ptsum density are shown against the leading charged particle
\pt at $\sqrt{s} = \unit{900}{\GeV}$ and \unit{7}{\TeV}. This time
a lower \pt cut-off of \unit{0.1}{\GeV} is applied for the charged particles in $\etamod < 2.5$.


\begin{figure}[pbt]
 \begin{center}
   \includegraphics[width=.5\textwidth]{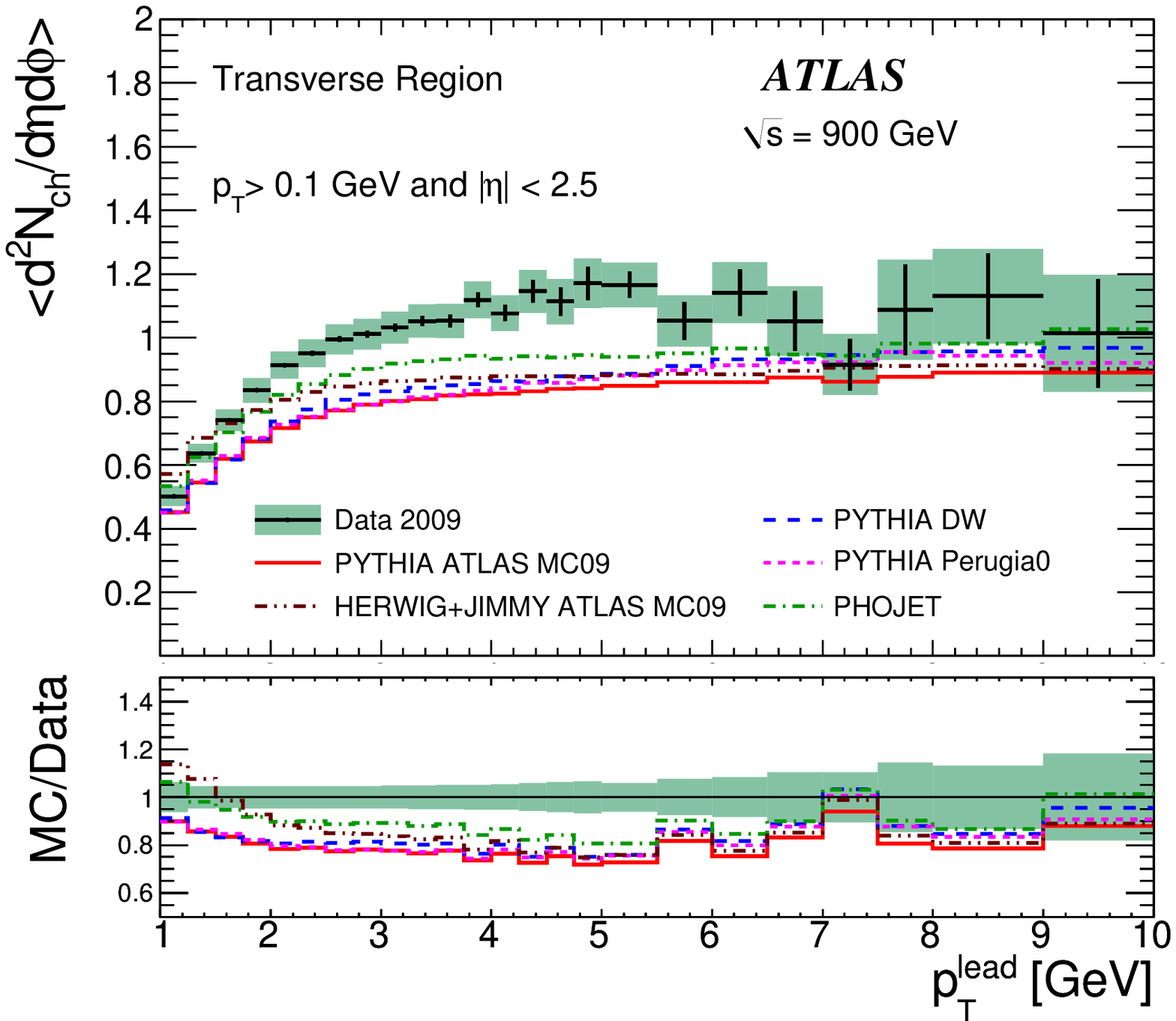}\hfill
   \includegraphics[width=.5\textwidth]{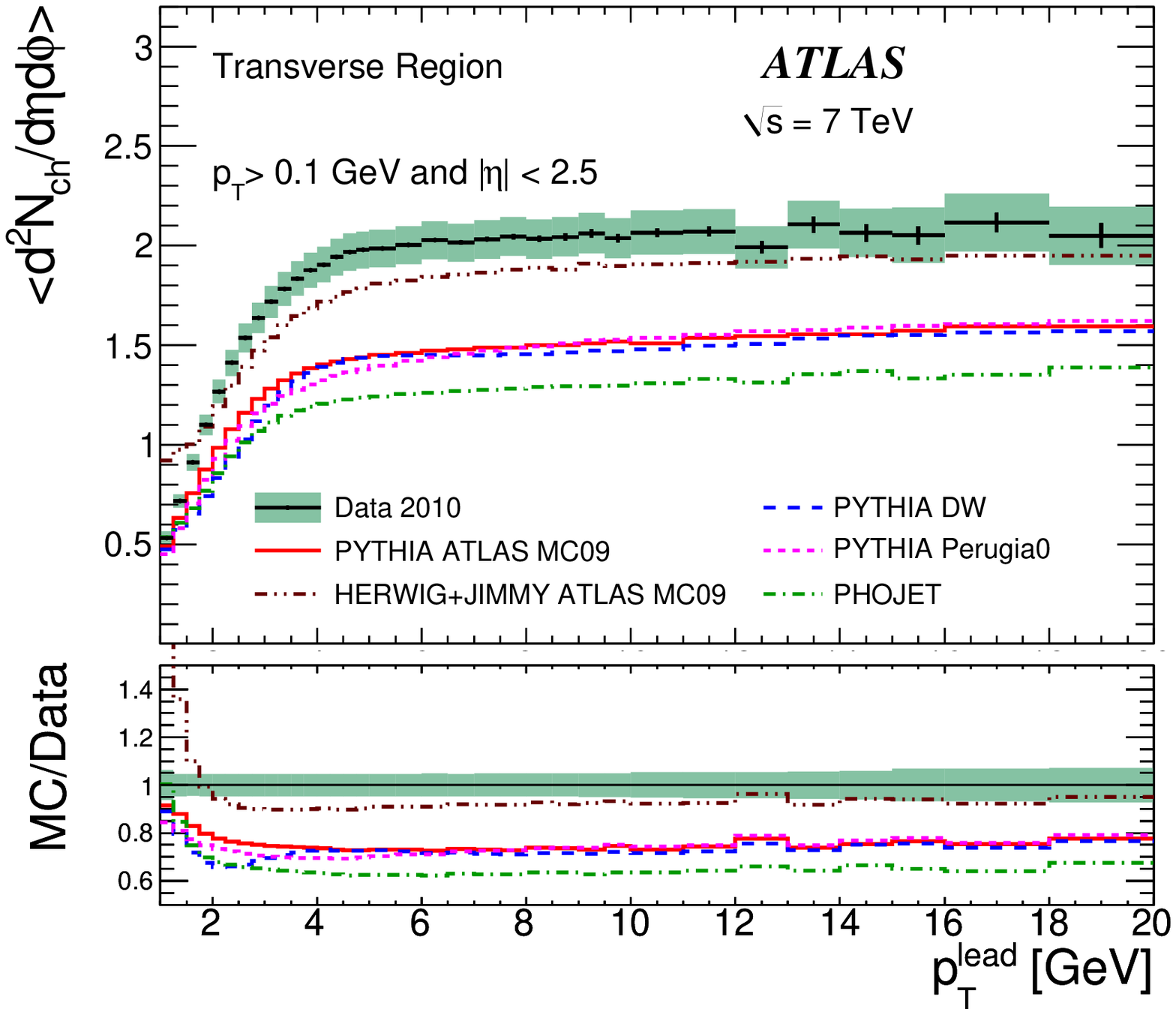}\hfill
   \includegraphics[width=.5\textwidth]{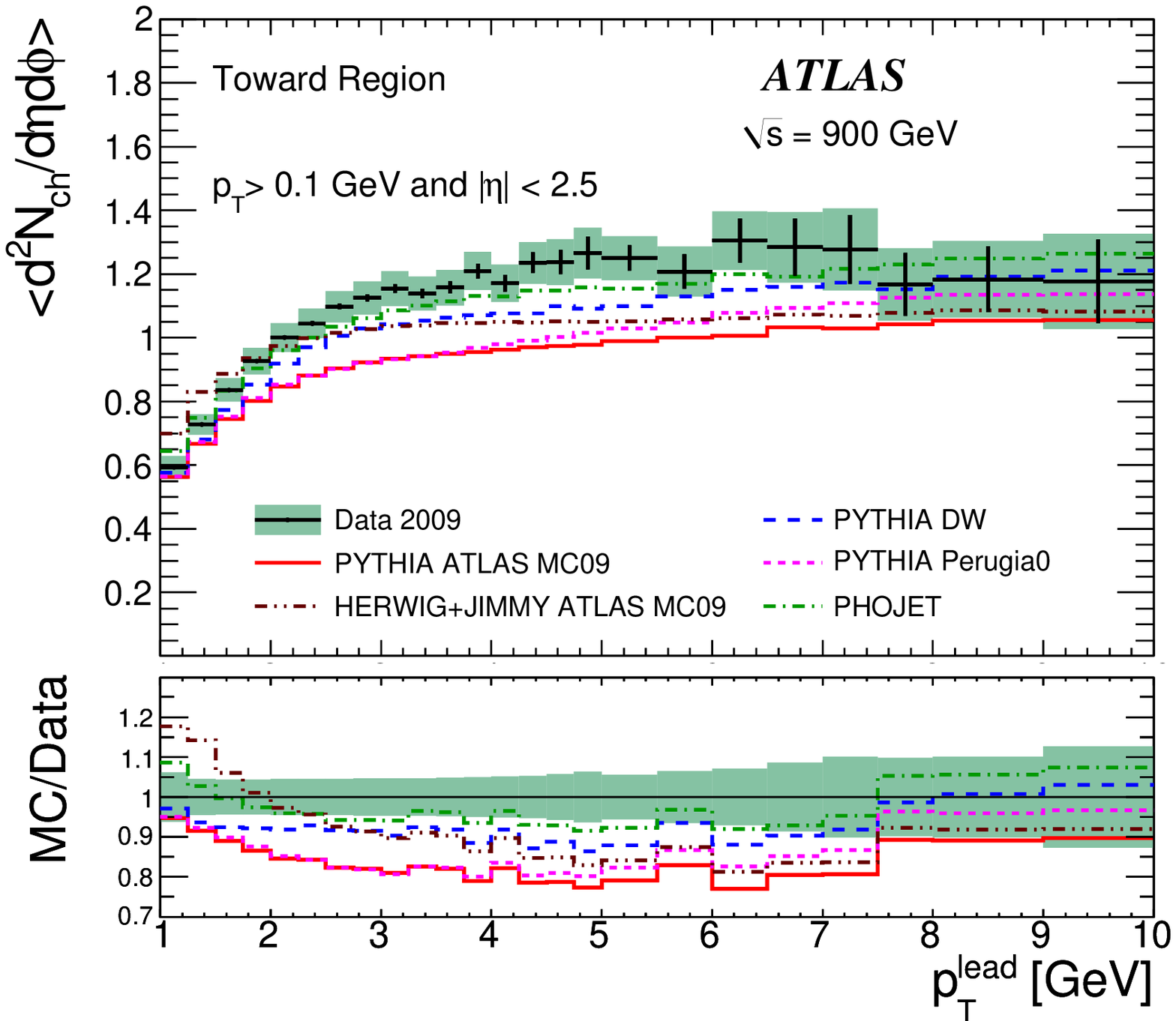}\hfill
   \includegraphics[width=.5\textwidth]{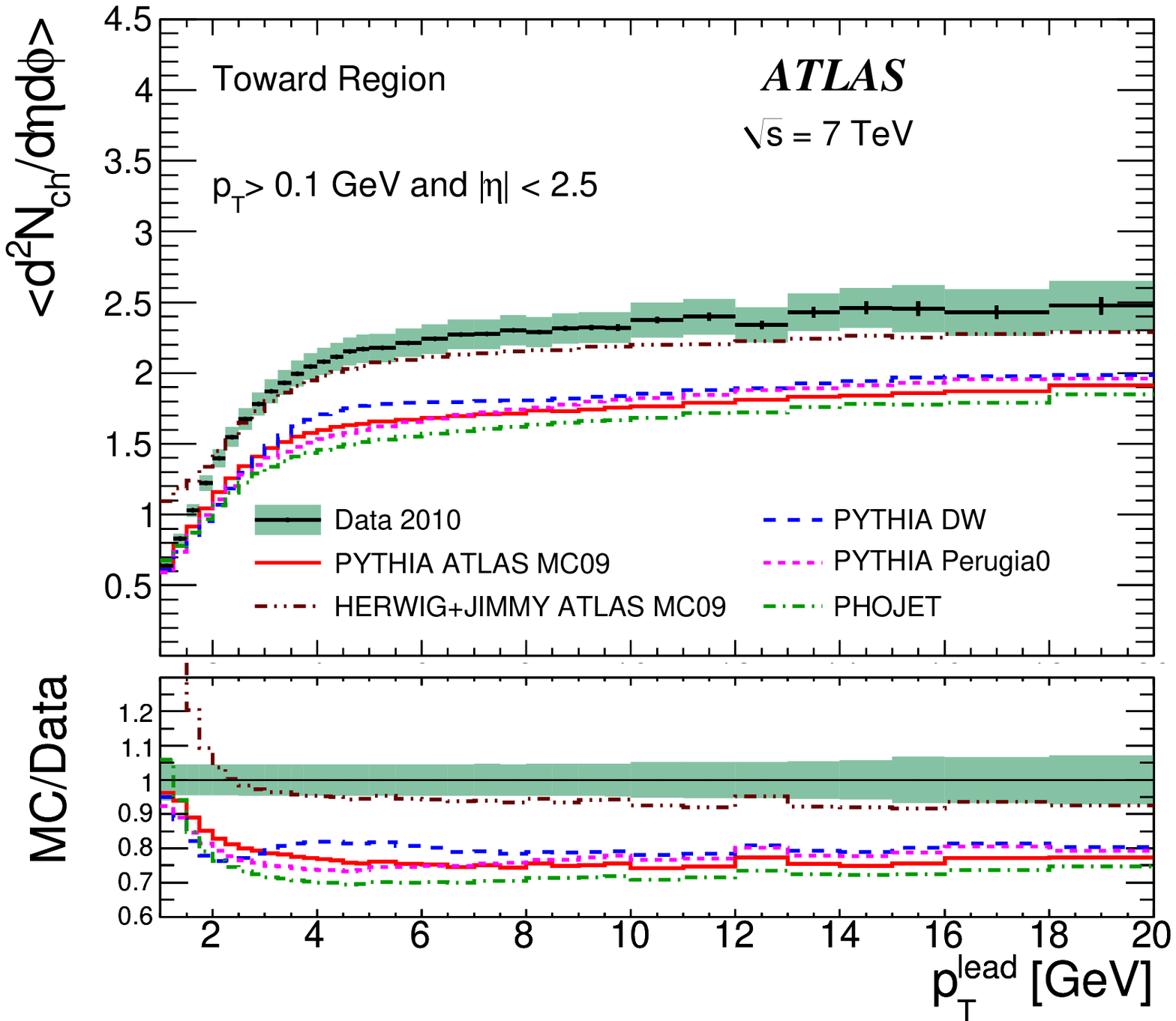}\hfill
   \includegraphics[width=.49\textwidth]{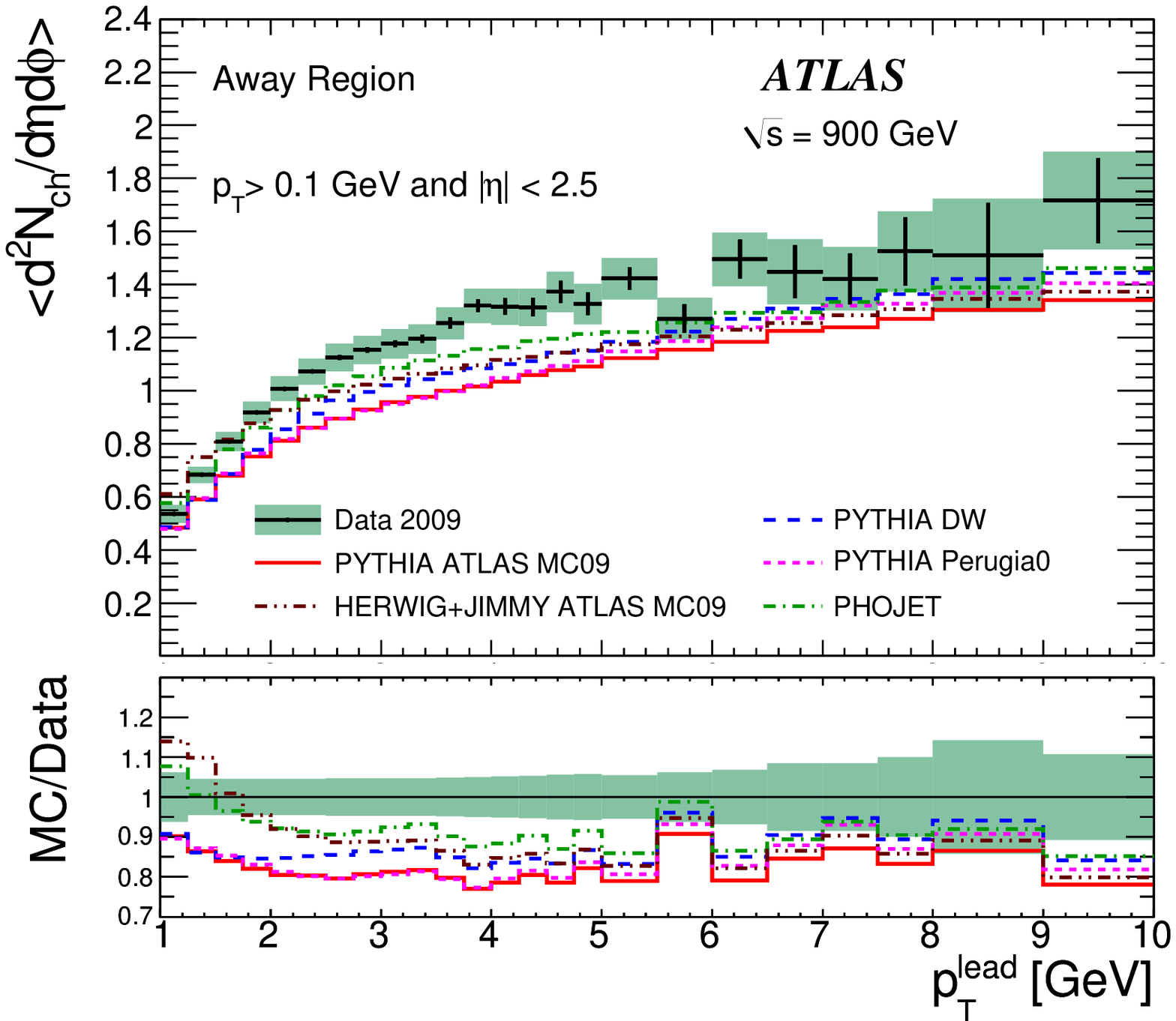} \hfill
   \includegraphics[width=.49\textwidth]{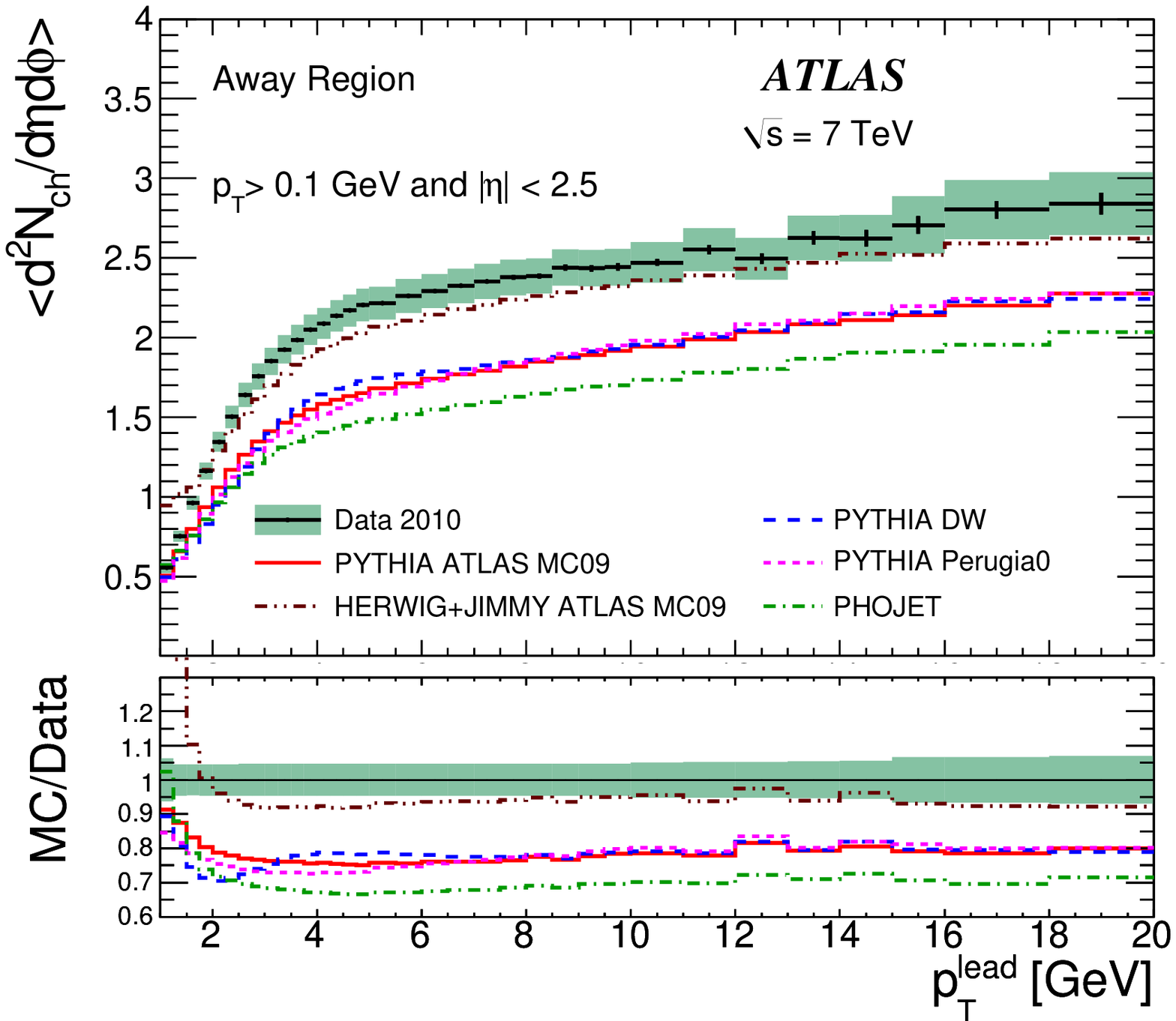}
 \end{center}

 \caption[]{ATLAS data at \unit{900}{\GeV} (left) and at \unit{7}{\TeV} (right)
   corrected back to particle level, showing the density of the charged
   particles \dNchgdetadphi with $\pt > \unit{0.1}{\GeV}$ and $\etamod < 2.5$, as
   a function of \ptlead.
   The data are compared with
   \Pythia ATLAS~MC09, DW and Perugia0 tunes, \Herwig+ \Jimmy ATLAS~MC09 tune, and \Phojet predictions.
   The top, middle and the bottom rows, respectively, show the transverse,
   toward and away regions defined by the leading charged particle.
   The error bars show the statistical uncertainty while the shaded area shows the
   combined statistical and systematic uncertainty.}
 \label{fig:lowpt-nchg}
\end{figure}



Compared to the previous plots with $\pt > \unit{500}{\MeV}$ (\FigRef[s]{fig:nchg} and \ref{fig:cptsum}, almost a twofold
increase in multiplicity is observed, but the scalar \ptsum stays very similar.
Again, the pre-LHC MC tunes show lower activity than the data in the plateau
part of the transverse region, except for \HerwigJimmy which predicts the
charged particle multiplicity density better than other models, but does not do
better for the \ptsum density. As this distinction of MC models is not seen for
the $\pt > \unit{500}{\MeV}$ \Nchg profile in \SecRef{sec:results:nch}, it can be
seen that \HerwigJimmy produces more particles between \unit{100}{\MeV}
and \unit{500}{\MeV} than the other MC models. A similar effect may be observed
in the \ptmean vs. \Nchg observable of \SecRef{sec:results:ptnch}.



\subsection{Charged particle multiplicity and scalar \ptsum vs. $\etamod$ of the leading charged particle}

\begin{figure}[tp]
 \begin{center}
   \includegraphics[width=.49\textwidth]{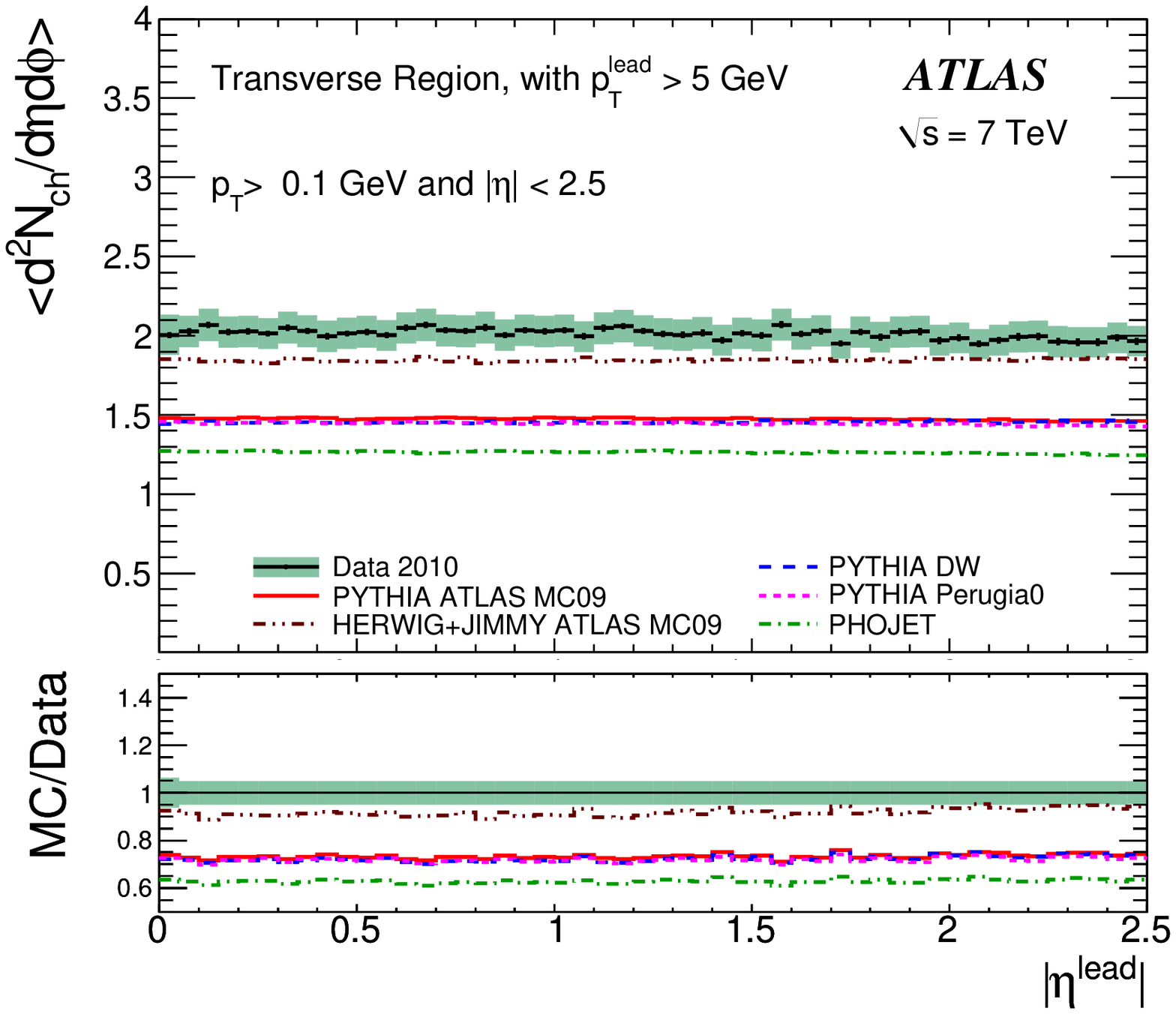}\hfill
   \includegraphics[width=.49\textwidth]{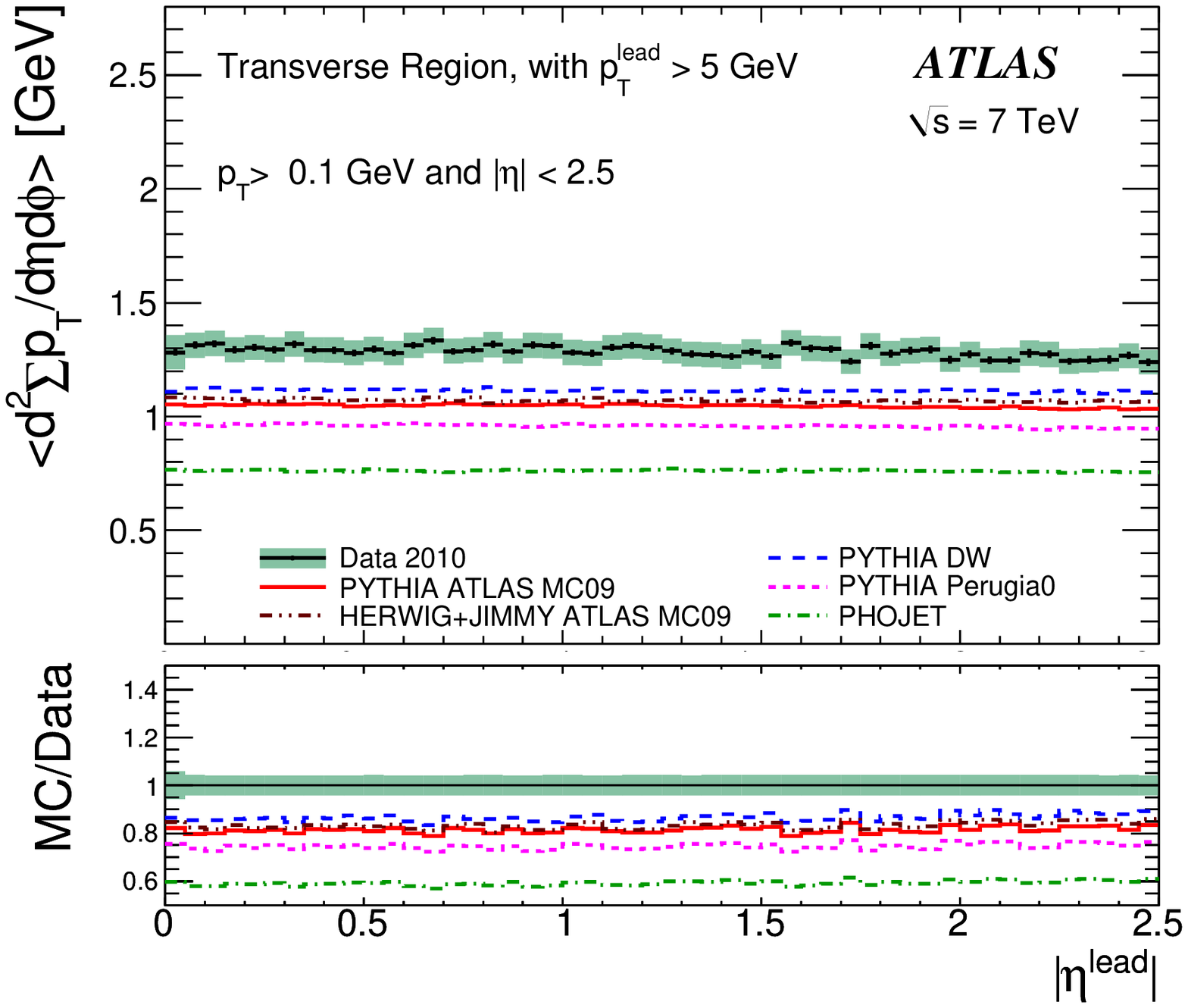}
 \end{center}
 \caption[]{ATLAS data at \unit{7}{\TeV}
   corrected back to the particle level, showing the density of the charged
   particles \dNchgdetadphi (left) and the scalar $\ptsum$ density of
   charged particles \dpTsumdetadphi (right) with $\pt > \unit{0.1}{\GeV}$
   and $\etamod < 2.5$, as a function of the leading charged particle $\etamod$, for the
   transverse region plateau ($\ptlead > \unit{5}{\GeV}$), defined by the leading
   charged particle and compared with \Pythia ATLAS~MC09, DW and Perugia0 tunes, and
   \HerwigJimmy ATLAS~MC09 tune, and \Phojet predictions. The error bars show the statistical
   uncertainty while the shaded area shows the combined statistical and
   systematic uncertainty.}
 \label{fig:leadeta}
\end{figure}

\FigureRef{fig:leadeta} shows the charged particle multiplicity density and \ptsum density
 in the kinematic range $\pt > \unit{0.1}{\GeV}$ and $\etamod
< 2.5$, for $p_T^{lead} > 5$ GeV, against the leading charged particle pseudorapidity for
$\sqrt{s}=\unit{7}{\TeV}$. As this observable is
composed only from events on the low-statistics transverse region plateau, the
available statistics were not sufficient at $\sqrt{s}= \unit{900}{\GeV}$ for a
robust analysis. However, the same behavior is seen as for \unit{7}{\TeV}.


It has been proposed that the dependence of the event characteristics on the (pseudo)rapidity
can be a useful test of the centrality of the events~\cite{Frankfurt:2010ea}.
In \FigRef{fig:leadeta}, the multiplicity
and \ptsum are seen to be independent of $\etamod$ for the transverse region
plateau,
suggesting that the average impact parameters in $pp$ collisions do not depend strongly on $\eta$ of the leading particle for a given \pT.

\FloatBarrier


\section{Conclusions}
\label{sec:conclusions}

Measurements of underlying event structure with the ATLAS detector have been
presented, using the data delivered by the LHC during 2009 and 2010 at
center-of-mass energies of \unit{900}{\GeV} and \unit{7}{\TeV}. This is the
first underlying event analysis at \unit{7}{\TeV}, and the first such analysis
at \unit{900}{\GeV} to be corrected for detector-specific effects.

The data have been corrected with minimal model-dependence and are provided as
inclusive distributions at the particle level. The selected phase-space and the
precision of this analysis highlight significant differences between Monte Carlo
models and the measured distributions. The same trend was observed for the ATLAS
inclusive charged particle multiplicity measurement \cite{Collaboration:2010rd,
Collaboration:2010ir}.  \Phojet, \HerwigJimmy and all pre-LHC MC tunes of \Pythia
predict less activity in the transverse region (i.e in the underlying event)
than is actually observed, for both center-of-mass energies and for charged
particle minimum \pt requirements of both \unit{100}{\MeV} and
\unit{500}{\MeV}. The charged particle multiplicity in the plateau of the
transverse region distribution was found to be about two times higher than that
of minimum bias particle density in the overall event.

One of the goals of this analysis is to provide data that can be used to test
and improve Monte Carlo models in preparation for other physics studies at the
LHC. The underlying event observables presented here are particularly important
for constraining the energy evolution of multiple partonic interaction models,
since the plateau heights of the UE profiles are highly correlated
to multiple parton interaction activity. As MC models of soft physics are least
predictive when modeling diffractive processes, it is particularly useful
that the UE profiles are largely insensitive to contributions from soft
diffraction models: the \Pythia soft diffraction model indicates that these are
constrained to the lowest bins in \ptlead. However, the sensitivity to more
complete diffraction models with a hard component, such as implemented in
\Pythia~8 \cite{Sjostrand:2007gs} or \Phojet, has not yet been fully
ascertained.

The data at \unit{7}{\TeV} are particularly important for MC tuning, since measurements are
needed with at least two energies to constrain the energy evolution of MPI
activity. While measurements from CDF exist at \unit{630}{\GeV}, \unit{1800}{\GeV} and \unit{1960}{\GeV},
in addition to these ATLAS measurements at \unit{900}{\GeV} and \unit{7}{\TeV},
there is a tension between the CDF and ATLAS measurements: the ATLAS analyses
indicate higher levels of activity, as evidenced by the failure of MC tunes to
CDF data to match the ATLAS data\footnote{The CDF measurements are within
  $\etamod < 1$, but ATLAS measurements restricted to that $\eta$ range show the
  same discrepancy as seen for the $\etamod < 2.5$ results presented here.}. Hence,
ATLAS UE measurement at two energies provides the best tuning data for MC predictions of ATLAS UE at higher energies.
While the \Pythia DW tune fits the ATLAS UE profile data closer than any other
current tune, it fails to describe other data -- as highlighted in the shape
of the distribution of \ptmean vs. \Nchg (\FigRef{fig:corr}). The increase of initial state radiation activity (and different shower models)
in tune DW may be responsible for this behavior. There is therefore no current
standard MC tune which adequately describes all the early ATLAS data.  However,
using diffraction-limited minimum bias distributions and the plateau regions of the
underlying event distributions presented here, ATLAS has developed a new \Pythia
tune AMBT1 (ATLAS Minimum Bias Tune 1) and a new \HerwigJimmy tune AUET1
(ATLAS Underlying Event Tune 1) which model the \pt and charged multiplicity
spectra significantly better than the pre-LHC tunes of those
generators~\cite{Collaboration:2010ir, :auet}.



\section{Acknowledgements}

We wish to thank CERN for the efficient commissioning and operation of the LHC
during this initial high-energy data-taking period as well as the support staff
from our institutions without whom ATLAS could not be operated efficiently.

We acknowledge the support of ANPCyT, Argentina; YerPhI, Armenia; ARC,
Australia; BMWF, Austria; ANAS, Azerbaijan; SSTC, Belarus; CNPq and FAPESP,
Brazil; NSERC, NRC and CFI, Canada; CERN; CONICYT, Chile; CAS, MOST and NSFC,
China; COLCIENCIAS, Colombia; MEYS (MSMT), MPO and CCRC, Czech Republic; DNRF,
DNSRC and Lundbeck Foundation, Denmark; ARTEMIS, European Union; IN2P3-CNRS,
CEA-DSM/IRFU, France; GNAS, Georgia; BMBF, DFG, HGF, MPG and AvH Foundation,
Germany; GSRT, Greece; ISF, MINERVA, GIF, DIP and Benoziyo Center, Israel; INFN,
Italy; MEXT and JSPS, Japan; CNRST, Morocco; FOM and NWO, Netherlands; RCN,
Norway; MNiSW, Poland; GRICES and FCT, Portugal; MERYS (MECTS), Romania; MES of
Russia and ROSATOM, Russian Federation; JINR; MSTD, Serbia; MSSR, Slovakia; ARRS
and MVZT, Slovenia; DST/NRF, South Africa; MICINN, Spain; SRC and Wallenberg
Foundation, Sweden; SER, SNSF and Cantons of Bern and Geneva, Switzerland; NSC,
Taiwan; TAEK, Turkey; STFC, the Royal Society and Leverhulme Trust, United
Kingdom; DOE and NSF, United States of America.

The crucial computing support from all WLCG partners is acknowledged gratefully,
in particular from CERN and the ATLAS Tier-1 facilities at TRIUMF (Canada), NDGF
(Denmark, Norway, Sweden), CC-IN2P3 (France), KIT/GridKA (Germany), INFN-CNAF
(Italy), NL-T1 (Netherlands), PIC (Spain), ASGC (Taiwan), RAL (UK) and BNL (USA)
and in the Tier-2 facilities worldwide.




%

\clearpage



\end{document}